\documentclass
[prd,a4paper,superscriptaddress,fleqn,nofootinbib,preprint,final,twocolumn,10pt,amsfonts,amssymb,tightenlines,nobibnotes]{revtex4}
\ifx\msidvipdf\msiundefined
\usepackage{graphicx}
\else%
\usepackage[dvipdfm]{graphicx}
\fi
\usepackage{amsfonts}
\usepackage{amsmath}
\usepackage{amssymb}
\usepackage{bm}
\usepackage{accents}
\usepackage{color}%
\providecommand{\U}[1]{\protect\rule{.1in}{.1in}}
\providecommand{\U}[1]{\protect\rule{.1in}{.1in}}
\providecommand{\U}[1]{\protect\rule{.1in}{.1in}}
\ifx\pdfoutput\relax\let\pdfoutput=\undefined\fi
\newcount\msipdfoutput
\ifx\pdfoutput\undefined\else
\ifcase\pdfoutput\else
\ifx\paperwidth\undefined\else
\ifdim\paperheight=0pt\relax\else\pdfpageheight\paperheight\fi
\ifdim\paperwidth=0pt\relax\else\pdfpagewidth\paperwidth\fi
\fi\fi\fi
\ifx\msidvipdf\msiundefined\else
\AtBeginDvi{}
\fi
\begin{document}
\preprint{KEK Preprint 2019-2. CHIBA-EP-226}
\title{Confinement/deconfinement phase transition and dual Meissner effect \\in SU(3) Yang-Mills theory }
\author{Akihiro Shibata}
\affiliation{Computing Research Center, High Energy Accelerator Research Organization
(KEK), Tsukuba 305-0801, Japan }
\affiliation{SOUKENDAI (The Graduate University for Advanced Studies), Tsukuba
305-0801,Japan }
\author{Kei-Ichi Kondo}
\affiliation{Department of Physics, Graduate School of Science, Chiba University, Chiba
263-8522, Japan}
\author{Seikou Kato}
\affiliation{Oyama National College of Technology, Oyama, Tochigi 323-0806, Japan}
\keywords{dual Meissner effect, confinement, phase transition}
\pacs{PACS number}

\begin{abstract}
We investigate the confinement-deconfinement phase transition at finite
temperature of the $SU(3)$ Yang-Mills theory on the lattice from a viewpoint
of the dual superconductor picture based on the novel reformulation of the
Yang-Mills theory. In particular, we compare the conventional Abelian dual
superconductor picture with the non-Abelian dual superconductor picture
proposed in our previous works as the mechanism of quark confinement in the
$SU(3)$ Yang-Mills theory. For the $SU(3)$ Yang-Mills theory, the
reformulation allows two possible options called maximal and minimal. The
maximal option corresponds to the manifestly gauge-invariant extension of the
Abelian projection scheme, while the minimal option is really new to give the
non-Abelian dual superconductor picture due to non-Abelian magnetic monopoles.
Keeping these differences in mind, we present the numerical evidences that the
confinement/deconfinement phase transition is caused by
appearance/disappearance of the dual Meissner effects. First, we measure the
Polyakov loop average at various temperatures to determine the critical
temperature separating the low-temperature confined phase and the
high-temperature deconfined phase. Second, we measure the static
quark-antiquark potential at various temperatures. Third, we measure the
chromoelectric and chromomagnetic flux created by a pair of quark and
antiquark at temperatures below and above the critical temperature. We observe
no more squeezing of the chromoelectric flux tube at high-temperatures above
the critical temperature. Finally, we measure the associated
magnetic--monopole current induced around the chromo-flux tube and observe
that the confinement/deconfinement phase transition is associated with the
appearance/disappearance of the induced magnetic--monopole current,
respectively. We confirm that these results are also obtained by the
restricted field alone in both options, indicating the restricted field
dominance in quark confinement at finite temperature.

\end{abstract}
\volumeyear{year}
\volumenumber{number}
\issuenumber{number}
\eid{identifier}
\maketitle

\section{Introduction}

The \textit{dual superconductivity} is one of the most promising mechanisms
for quark confinement \cite{dualsuper}. To establish the \textit{dual
superconductor picture}, we must show that magnetic monopoles play a dominant
role in quark confinement. For this purpose, we have constructed a new
formulation \cite{KMS05} of the $SU(N)$ Yang-Mills theory on the continuum
space-time which enables us to define the gauge-invariant magnetic monopoles
in the gauge-independent way. Subsequently, we have implemented the new
formulation to the $SU(N)$ Yang-Mills theory on the lattice
\cite{KKMSSI06,KSSMKI08,SKS10}, which enables us to perform the numerical
simulations to obtain non-perturbative results. The new reformulation is
feasible for decomposing the gauge field into the two pieces, i.e., the
restricted field and the remaining one in the gauge covariant way
so that the restricted field is identified with the dominant mode for quark
confinement in the gauge independent way.
See \cite{KKSS15} for a review.

The conventional method called the \textit{Abelian projection} \cite{tHooft81}
allows us to extract specific magnetic monopoles called \textit{Abelian
magnetic monopoles} in the pure Yang-Mills theory without matter fields, which
are however possible only in special Abelian gauges such as the maximal
Abelian (MA) gauge \cite{KLSW87} and the Laplacian Abelian gauge. In fact, the
Abelian projection is a kind of gauge fixing to explicitly break the gauge
symmetry, which also breaks the color symmetry (global symmetry). Therefore,
the Abelian magnetic monopoles are not gauge-independent objects. This fact
casts a doubt on the validity of the results obtained under the Abelian projection.

The new formulation can overcome the criticism raised for the Abelian
projection method to extract magnetic monopoles in the pure Yang-Mills theory
without matter fields. The new formulation of the $SU(3)$ Yang-Mills theory
has two possible options for choosing the fundamental field variables which we
call the minimal and maximal options. The respective option is discriminated
by a \textit{maximal stability subgroup} $\tilde{H}$, a subgroup of the gauge
group $G$.

In the \textit{minimal option}, the maximal stability group is the non-Abelian
group $\tilde{H}=U(2)\cong SU(2)\times U(1)$ and the restricted field is used
to extract \textit{non-Abelian magnetic monopoles} yielding
\textit{non-Abelian dual superconductivity}. In the preceding works, indeed,
we have provided numerical evidences of the non-Abelian dual superconductivity
using the minimal option for the $SU(3)$ Yang-Mills theory on a lattice. The
minimal option is suggested from the non-Abelian Stokes theorem
\cite{Kondo08,KS08} for the Wilson loop operator in the fundamental
representation. We have found that both the restricted field variable and the
extracted non-Abelian magnetic monopole dominantly reproduce the string
tension in the linear potential of the $SU(3)$ Yang-Mills theory
\cite{KSSK11}. In this way we have demonstrated the gauge-independent
\textit{restricted field dominance} (corresponding to the conventional Abelian
dominance): the string tension $\sigma_{V}$ calculated from the restricted
field reproduces the string tension $\sigma_{full}$ of the original Yang-Mills
field: $\sigma_{V}/\sigma_{full}=93\pm16\%$. Moreover, we have also
demonstrated the gauge-independent \textit{non-Abelian magnetic monopole
dominance}: the string tension $\sigma_{V}$ was reproduced by $\sigma_{mon}$
calculated from the (non-Abelian) magnetic monopole part extracted from the
restricted field: $\sigma_{mon}/\sigma_{V}=94\pm9\%$, see
\cite{KSSMKI08,lattice2008,lattice2009,lattice2010,KSSK11}.

The dual superconductivity has been established by demonstrating the existence
of chromoelectric flux connecting a quark and antiquark and the associated
magnetic-monopole current induced around the flux tube \cite{KSSK11} where
both the chromoelectric flux and the magnetic-monopole current are
gauge-invariant objects. Moreover, the investigation of the chromoelectric
flux tube leads to the surprising conclusion that the vacuum of the $SU(3)$
Yang-Mills vacuum is the type I dual superconductor, which is a novel feature
obtained by the numerical simulations \cite{SKKS13}. This is sharp contrast to
the proceeding studies: the border of type I and type II or of week type I
\cite{flusx:AP}. In the $SU(3)$ case, there are many works on chromo flux by
using Wilson line/loop operator, see e.g.,
\cite{Cardaci2011,Cardso,CCP12,CCCP14}.

These results should be compared with another option called the
\textit{maximal option} which was first constructed by Cho \cite{SUN-decomp1}
and Faddeev and Niemi \cite{SUN-decomp2} by extending the
Cho-Duan-Ge-Faddeev-Niemi (CDGFN) decomposition for the $SU(2)$ case
\cite{CFNS-C}. The maximal stability group in the maximal option of $SU(3)$ is
an Abelian group $\tilde{H}=U(1)\times U(1)$, the maximal torus subgroup of
$SU(3)$. Therefore, the restricted field in the maximal option involves only
the Abelian magnetic monopole, which is indeed detected on the lattice
\cite{lattce2007,CCLL14}. The maximal option in the new formulation gives a
gauge invariant extension of the Abelian projection in the maximal Abelian
(MA) gauge \cite{GIS12,SS14}.

The similar results are also obtained for $SU(2)$ Yang-Mills theory on the
lattice \cite{IKKMSS06, SKKMSI07, lattice2009k,Cea1995, Cea2012a14}. The
restricted field corresponding to the stability group $\tilde{H}=U(1)$ of
$SU(2)$ reproduces the dual Meissner effect \cite{KKS14,NKSSK2018}. For
$SU(2)$, there is a unique option which is regarded as a gauge-invariant
version of the Abelian projection in the MA gauge \cite{SY90,SS94,SNW94}.

Furthermore, the dual superconductivity for the Wilson loop in higher
representations is investigated by using the extension of the new formulation
of the $SU(N)$ Yang-Mills theory on the lattice. The restricted field
reproduces the string tension in the linear potential of the $SU(3)$
Yang-Mills theory for the Wilson loops in higher representations
\cite{MSKK2019}.


The purpose of this paper is to investigate the confinement/deconfinement
phase transition at finite temperature of the $SU(3)$ Yang-Mills theory from a
viewpoint of the dual superconductor picture,
as an investigation subsequent to quark confinement due to dual
superconductivity in the $SU(3)$ Yang-Mills theory at zero temperature, see
\cite{lattice2012,lattice2013,lattice2014,lattice2015,lattice2016,SCGT15,confinementX,confinement2018}
for preliminary results. In particular, we compare the conventional Abelian
dual superconductor picture with the non-Abelian dual superconductor picture
proposed as the mechanism of quark confinement in the $SU(3)$ Yang-Mills
theory in our previous works based on the novel reformulation of the
Yang-Mills theory. We examine the dual Meissner effect at finite temperature
by measuring the distribution of the chromoelectric field strength (or chromo
flux) generated from a static quark-antiquark pair and the associated
magnetic-monopole current induced around it. We present the numerical
evidences for them at finite temperature by using the gauge link decomposition
for extracting the magnetic monopole in the gauge invariant way. In
particular, we discuss the role of magnetic monopoles in
confinement/deconfinement phase transition. We focus on whether there are
distinctions for the physics of quark confinement between the maximal and
minimal options in the new reformulation of the $SU(3)$ Yang-Mills theory.

This paper is organized as follows. In section II, we review the new
formulations of the $SU(3)$ Yang-Mills theory on the lattice to see the
differences between the minimal and maximal options.\quad In section III, we
give the results of the numerical simulations on the lattice. First, we
measure the Polyakov loop average at various temperatures to determine the
critical temperature $T_{c}$ separating the low-temperature confined phase and
the high-temperature deconfined phase. We also measure the correlation
functions of the Polyakov loops which are defined for both the original
Yang-Mills gauge field and the restricted field to examine the restricted
field dominance in the Polyakov loop average at finite temperature. Second, we
measure the string tension for the restricted field of both minimal and
maximal option in comparison with the string tension for the original gauge
field. Third, we investigate the dual Meissner effect by measuring the
distribution of the chromoelectric flux and chromomagnetic flux created by a
pair of static quark and antiquark at finite temperature below and above the
critical temperature to investigate its relevance to the phase transition. We
observe no more squeezing of the chromoelectric flux tube in the
high-temperature deconfinement phase above the critical temperature $T_{c}$.
Finally, we measure the associated magnetic--monopole current induced around
the chromo-flux tube and observe that the confinement/deconfinement phase
transition is associated with the appearance/disappearance of the induced
magnetic--monopole current, respectively. We observe that these results are
also obtained by the restricted field alone, confirming the restricted field
dominance at finite temperature. These results are the numerical evidences
that the confinement/deconfinement phase transition is caused by
appearance/disappearance of the non-Abelian dual superconductivity. The final
section is devoted to conclusion and discussion.

\section{New formulation of lattice gauge theory}

\subsection{Gauge link decompositions}

We introduce a new formulation of the lattice Yang-Mills theory, which enables
one to extract the dominant mode for quark confinement in the $SU(N)$
Yang-Mills theory \cite{KKMSSI06,KSSMKI08,SKS10}.
We decompose the gauge link variable $U_{x,\mu}$ into the product of the two
variables $V_{x,\mu}$ and $X_{x,\mu}$:
\[
U_{x,\mu}=X_{x,\mu}V_{x,\mu}\in G=SU(N),
\]
in such a way that the new variable $V_{x,\mu}$ is transformed by the full
$SU(3)$ gauge transformation $\Omega_{x}$ as the gauge link variable
$U_{x,\mu}$, while $X_{x,\mu}$ transforms as the site variable:
\begin{subequations}
\label{eq:gaugeTransf}%
\begin{align}
U_{x,\mu}  &  \longrightarrow U_{x,\nu}^{\prime}=\Omega_{x}U_{x,\mu}%
\Omega_{x+\mu}^{\dag},\\
V_{x,\mu}  &  \longrightarrow V_{x,\nu}^{\prime}=\Omega_{x}V_{x,\mu}%
\Omega_{x+\mu}^{\dag},\\
X_{x,\mu}  &  \longrightarrow X_{x,\nu}^{\prime}=\Omega_{x}X_{x,\mu}\Omega
_{x}^{\dag}.
\end{align}
From the physical point of view, $V_{x,\mu}$ could be identified with the
dominant mode for quark confinement, while $X_{x,\mu}$ is the remaining part.

For the $SU(3)$ Yang-Mills theory, we have two possible options discriminated
by the \textit{stability subgroup} $\tilde{H}$ of the gauge group $G$, which
we call the minimal and maximal options.

\subsection{Minimal option}

The \textit{minimal option} is obtained for the choice of the stability
subgroup $\tilde{H}=U(2)=SU(2)\times U(1)\subset SU(3)$. In the minimal
option, we introduce a single \textit{color field} $\bm{h}_{x}$ taking the
value in the Lie algebra as a site variable
\end{subequations}
\begin{equation}%
\bm{h}%
_{x}=\xi_{x}\frac{\lambda^{8}}{2}\xi_{x}^{\dag}\in Lie[SU(3)/U(2)],
\end{equation}
with $\lambda^{8}$ being the last diagonal Gell-Mann matrix and $\xi_{x}$ an
$SU(3)$ group element. Then the decomposition is obtained by solving the
\textit{defining equations}:
\begin{subequations}
\label{eq:def-min}%
\begin{align}
&  D_{\mu}^{\varepsilon}[V]%
\bm{h}%
_{x}:=\frac{1}{\varepsilon}\left[  V_{x,\mu}%
\bm{h}%
_{x+\mu}-%
\bm{h}%
_{x}V_{x,\mu}\right]  =0\text{ },\label{eq:def1-min}\\
&  g_{x}:=e^{i2\pi q/3}\exp\left(  -ia_{x}^{0}%
\bm{h}%
_{x}-i\sum\nolimits_{j=1}^{3}a_{x}^{(j)}\mathbf{u}_{x}^{(j)}\right)  \text{
}.\label{eq:def2-min}%
\end{align}
These defining equations can be solved exactly \cite{SKS10}, and the solution
is given by
\end{subequations}
\begin{subequations}
\label{eq:decomp-min}%
\begin{align}
V_{x,\mu} &  =X_{x,\mu}^{\dag}U_{x,\mu}=g_{x}\widehat{L}_{x,\mu}U_{x,\mu},\\
X_{x,\mu} &  =\widehat{L}_{x,\mu}^{\dag}\det(\widehat{L}_{x,\mu})^{1/3}%
g_{x}^{-1},\\
\widehat{L}_{x,\mu} &  :=\left(  L_{x,\mu}L_{x,\mu}^{\dag}\right)
^{-1/2}L_{x,\mu},\\
\text{\ }L_{x,\mu} &  :=\frac{5}{3}\mathbf{1}+\frac{2}{\sqrt{3}}(%
\bm{h}%
_{x}+U_{x,\mu}%
\bm{h}%
_{x+\mu}U_{x,\mu}^{\dag})\nonumber\\
&  +8%
\bm{h}%
_{x}U_{x,\mu}%
\bm{h}%
_{x+\mu}U_{x,\mu}^{\dag}.
\end{align}
Here the variable $g_{x}$ is the $U(2)$ part which is undetermined from
Eq.(\ref{eq:def1-min}) alone, $\mathbf{u}_{x}^{(j)}$ 's are $su(2)$-Lie
algebra valued, and \thinspace$q$ is an integer, see \cite{SKS10} for the
details. In what follows, we chose $g_{x}=\mathbf{1}\,\ $.

Note that the above defining equations (\ref{eq:def1-min}) and
(\ref{eq:def2-min}) correspond to the continuum version: $D_{\mu}[%
\bm{V}%
]%
\bm{h}%
(x)=0$ and $\mathrm{tr}(%
\bm{X}%
_{\mu}(x)%
\bm{h}%
(x))$ $=0$, respectively. By taking the naive continuum limit, indeed, we can
reproduce the decomposition in the continuum theory \cite{KMS05}:
\end{subequations}
\begin{subequations}
\label{eq:minimal_cont}%
\begin{align}
\bm{A}_{\mathbf{\mu}}(x)  &  =\bm{V}_{\mu}(x)+\bm{X}_{\mu}(x),\\
\bm{X}_{\mu}(x)  &  =ig^{-1}\frac{4}{3}\left[  \partial_{\mu}%
\bm{h}(x),\bm{h}(x)\right] \nonumber\\
&  +\frac{4}{3}\left[  \left[  \bm{A}_{\mathbf{\mu}}(x),\bm{h}(x)\right]
,\bm{h}(x)\right]  ,\\
\bm{V}_{\mu}(x)  &  =\bm{A}_{\mathbf{\mu}}(x)-ig^{-1}\frac{4}{3}\left[
\partial_{\mu}\bm{h}(x),\bm{h}(x)\right] \nonumber\\
&  -\frac{4}{3}\left[  \left[  \bm{A}_{\mathbf{\mu}}(x),\bm{h}(x)\right]
,\bm{h}(x)\right]  ,
\end{align}

In this way, the decomposition is uniquely determined according to
(\ref{eq:decomp-min}),
once a set of color fields $\{\bm{h}_{x}\}$ are given. To determine the
configuration $\{\bm{h}_{x}\}$ of color fields, we adopt the procedure of
minimizing the functional:
\end{subequations}
\begin{equation}
F_{\text{red}}[\bm{h}_{x}]=\sum_{x,\mu}\mathrm{tr}\left\{  (D_{\mu
}^{\varepsilon}[U_{x,\mu}]\bm{h}_{x})^{\dag}(D_{\mu}^{\varepsilon}[U_{x,\mu
}]\bm{h}_{x})\right\}  , \label{eq:reduction-min}%
\end{equation}
which yields the condition to be satisfied for the color field, which we call
the \textit{reduction condition}. Thus we obtain the reformulated Yang-Mills
theory written in terms of the new variables ($X_{x,\mu}$,$V_{x,\mu}$), which
is equipollent to the original Yang-Mills theory.

%

\begin{table*}[bht] \centering
\
\begin{tabular}
[c]{|c|c|c||c|c|c||c|c|c|}\hline
& \multicolumn{2}{|c}{} & \multicolumn{3}{||c||}{$N_{T}=6$} &
\multicolumn{3}{||c|}{$N_{T}=8$}\\\hline
$\beta$ & $\varepsilon\sqrt{\sigma}$ & $\varepsilon$ [fm] \  & $T$%
[$\sqrt{\sigma}$] & $T$[MeV] & $T/T_{c}$ & $T$[$\sqrt{\sigma}$] & $T$[MeV] &
$\ T/T_{c}$ \ \\\hline
$5.75$ & $[0.3544]$ & $0.159$ & $0.1231$ & $54.14$ & $0.7483$ & $-$ & $-$ &
$-$\\\hline
$5.8$ & $[0.3209]$ & $0.144$ & $0.5194$ & $228.8$ & $0.8264$ & $0.3896$ &
$171.4$ & $0.6198$\\\hline
$5.85$ & $0.2874(7)$ & $0.129$ & $0.5799$ & $255.2$ & $0.9228$ & $-$ & $-$ &
$-$\\\hline
$5.875$ & $[0.2763]$ & $0.124$ & $0.6032$ & $265.4$ & $0.9598$ & $-$ & $-$ &
$-$\\\hline
$5.9$ & $[0.2652]$ & $0.119$ & $0.6285$ & $276.5$ & $1.000$ & $0.4714$ &
$207.4$ & $0.7500$\\\hline
$5.925$ & $[0.25415]$ & $0.114$ & $0.6558$ & $288.5$ & $1.034$ & $-$ & $-$ &
$-$\\\hline
$5.95$ & $[0.2431]$ & $0.109$ & $0.6856$ & $301.7$ & $1.091$ & $-$ & $-$ &
$-$\\\hline
$6.0$ & $0.2209(23)$ & $0.0989$ & $0.7545$ & $332.0$ & $1.201$ & $0.5659$ &
$249.0$ & $0.9005$\\\hline
$6.1$ & $[0.1905]$ & $0.085$ & $0.8748$ & $385..0$ & $1.392$ & $0.6561$ &
$288.7$ & $1.044$\\\hline
$6.2$ & $0.1610(9)$ & $0.0721$ & $1.035$ & $455.5$ & $1.647$ & $0.7763$ &
$341.6$ & $1.235$\\\hline
$6.3$ & $[0.1412]$ & $0.0632$ & $1.180$ & $519.3$ & $1.878$ & $0.8850$ &
$389.4$ & $1.409$\\\hline
$6.4$ & $0.1214(12)$ & $0.0543$ & $1.373$ & $604.1$ & $2.185$ & $1.030$ &
$453.1$ & $1.639$\\\hline
$6.5$ & $0.1068(9)$ & $0.0478$ & $1.561$ & $686.6$ & $2.483$ & $1.171$ &
$515.1$ & $1.862$\\\hline
\end{tabular}
\caption{
The summary of  temperature $T$ for various $\beta
$ and $N_T$ for the data set I .
The relation between
$\beta$ and  $ \varepsilon\sqrt{\sigma}$ is  obtaein from Ref \cite{Edward98},
where  values of   $\varepsilon\sqrt{\sigma}
$  in the squared brackets are lack of data
but  these are  obtained by linear  interpolation of measured values.
The lattice spacing is determined by using the relation
$\sigma_{\text{latt}} =\varepsilon^2 \sigma$
with the physical string tenstion defined by   $\sqrt{\sigma}=440 MeV$ .
The temperuture is obtained by using the relation $1/T = \varepsilon N_T $.
The critical temperature $T_c$ is  determined by using  the Polyakov loop average and the susceptibility as given
in Subsection \ref{sec:Ploop}.
}
\label{tbl:physscale}
\end{table*}%

%

\begin{table}[bht] \centering
\begin{tabular}
[c]{|l||l|l|l||l|l|l|}\hline
& \multicolumn{3}{|c||}{$\beta=6.2$} & \multicolumn{3}{||c|}{$\beta=6.0$%
}\\\hline
$N_{T}$ & $T$[$\sqrt{\sigma}$] & $T$[MeV] & $\ T/T_{c}$ \  & $T$[$\sqrt
{\sigma}$] & $T$[MeV] & $\ T/T_{c}$ \ \\\hline
\multicolumn{1}{|c||}{$24$} & \multicolumn{1}{||c|}{$0.2588$} &
\multicolumn{1}{|c|}{$113.9$} & \multicolumn{1}{|c||}{$0.4112$} &
\multicolumn{1}{||c|}{$0.1886$} & \multicolumn{1}{|c|}{$82.89$} &
\multicolumn{1}{|c|}{$0.3001$}\\\hline
\multicolumn{1}{|c||}{$16$} & \multicolumn{1}{||c|}{$0.3882$} &
\multicolumn{1}{|c|}{$170.8$} & \multicolumn{1}{|c||}{$0.6181$} &
\multicolumn{1}{||c|}{$0.2829$} & \multicolumn{1}{|c|}{$124.5$} &
\multicolumn{1}{|c|}{$0.4502$}\\\hline
\multicolumn{1}{|c||}{$14$} & \multicolumn{1}{||c|}{$0.4437$} &
\multicolumn{1}{|c|}{$195.2$} & \multicolumn{1}{|c||}{$0.7061$} &
\multicolumn{1}{||c|}{$0.3223$} & \multicolumn{1}{|c|}{$141.8$} &
\multicolumn{1}{|c|}{$0.5129$}\\\hline
\multicolumn{1}{|c||}{$12$} & \multicolumn{1}{||c|}{$0.5176$} &
\multicolumn{1}{|c|}{$227.7$} & \multicolumn{1}{|c||}{$0.8237$} &
\multicolumn{1}{||c|}{$0.3772$} & \multicolumn{1}{|c|}{$166.0$} &
\multicolumn{1}{|c|}{$0.6002$}\\\hline
\multicolumn{1}{|c||}{$10$} & \multicolumn{1}{||c|}{$0.6211$} &
\multicolumn{1}{|c|}{$273.3$} & \multicolumn{1}{|c||}{$0.9884$} &
\multicolumn{1}{||c|}{$0.4527$} & \multicolumn{1}{|c|}{$199.2$} &
\multicolumn{1}{|c|}{$0.7204$}\\\hline
\multicolumn{1}{|c||}{$8$} & \multicolumn{1}{||c|}{$0.7763$} &
\multicolumn{1}{|c|}{$341.6$} & \multicolumn{1}{|c||}{$1.235$} &
\multicolumn{1}{||c|}{$0.5659$} & \multicolumn{1}{|c|}{$249.0$} &
\multicolumn{1}{|c|}{$0.9005$}\\\hline
\multicolumn{1}{|c||}{$6$} & \multicolumn{1}{||c|}{$1.035$} &
\multicolumn{1}{|c|}{$455.5$} & \multicolumn{1}{|c||}{$1.647$} &
\multicolumn{1}{||c|}{$0.7545$} & \multicolumn{1}{|c|}{$332.0$} &
\multicolumn{1}{|c|}{$1.201$}\\\hline
\multicolumn{1}{|c||}{$4$} & \multicolumn{1}{||c|}{$1.553$} &
\multicolumn{1}{|c|}{$683.2$} & \multicolumn{1}{|c||}{$2.471$} &
\multicolumn{1}{||c|}{$1.132$} & \multicolumn{1}{|c|}{$498.1$} &
\multicolumn{1}{|c|}{$1.801$}\\\hline
\end{tabular}
\caption{
The summary of  temperature $T$ for various $\beta$ and $N_T$ for data set II.
The temperatures are determined in the same way as
Table \ref{tbl:physscale}
}
\label{tbl:physscale2}%
\end{table}%


\subsection{Maximal option}

The \textit{maximal option} is obtained for the choice of the stability
subgroup of the maximal torus subgroup of $G$: $\tilde{H}=U(1)\times
U(1)\subset SU(3)$. In the maximal option, we introduce two kinds of color
fields,
\begin{subequations}
\begin{align}
\bm{n}_{x}^{(3)} &  =\xi_{x}\frac{\lambda^{3}}{2}\xi_{x}^{\dag}\in
Lie[SU(3)/U(1)\times U(1)],\\
\bm{n}_{x}^{(8)} &  =\xi_{x}\frac{\lambda^{8}}{2}\xi_{x}^{\dag}\in
Lie[SU(3)/U(2)],
\end{align}
with $\lambda^{3}$, $\lambda^{8}$ being the two diagonal Gell-Mann matrices
and $\xi$ an $SU(3)$ group element. Then the decomposition is obtained by
solving the defining equations ($j=3,8$):
\end{subequations}
\begin{subequations}
\label{eq:Defeq-max}%
\begin{align}
&  D_{\mu}^{\varepsilon}[V]\bm{n}_{x}^{(j)}:=\frac{1}{\varepsilon}\left[
V_{x,\mu}\bm{n}_{x+\mu}^{(j)}-\bm{n}_{x}^{(j)}V_{x,\mu}\right]
=0,\label{eq:def-max-1}\\
&  g_{x}:=e^{i2\pi q/3}\exp(-ia_{x}^{3}\bm{n}_{x}^{(3)}-ia_{x}^{(8)}%
\bm{n}_{x}^{(8)})\text{ .}\label{eq:def-max-2}%
\end{align}
These defining equations can be solved exactly, and the solution is given by
\end{subequations}
\begin{subequations}
\label{eq:decomp_max}%
\begin{align}
V_{x,\mu} &  =X_{x,\mu}^{\dag}U_{x,\mu},\\
X_{x,\mu} &  =\widehat{K}_{x,\mu}^{\dag}\det(\widehat{K}_{x,\mu})^{1/3}%
g_{x}^{-1},\\
\widehat{K}_{x,\mu} &  :=\left(  K_{x,\mu}K_{x,\mu}^{\dag}\right)
^{-1/2}K_{x,\mu},\\
K_{x,\mu} &  :=\mathbf{1}+6(\bm{n}_{x}^{(3)}U_{x,\mu}\bm{n}_{x+\mu}%
^{(3)}U_{x,\mu}^{\dag})\nonumber\\
&  +6(\bm{n}_{x}^{(8)}U_{x,\mu}\bm{n}_{x+\mu}^{(8)}U_{x,\mu}^{\dag})\text{ }.
\end{align}
Here the variable $g_{x}$ is the $U(1)\times U(1)$ part which is undetermined
from Eq.(\ref{eq:def-max-1}) alone, and $q$ is an integer, see \cite{SKS10}
for the details. In what follows, we chose $g_{x}=1\,$.

Note that the above defining equations (\ref{eq:def-max-1}) and
(\ref{eq:def-max-2}) correspond to the continuum version: $D_{\mu
}[\bm{V}]\bm{n}^{(j)}(x)=0$ and $\mathrm{tr}(\bm{X}_{\mu}(x)\bm{n}^{(j)}(x))$
$=0$, respectively. By taking the naive continuum limit, we can reproduce the
decomposition in the continuum theory \cite{KMS05}:
\end{subequations}
\begin{subequations}
\begin{align}
\bm{A}_{\mu}(x)=  &  \bm{V}_{\mu}(x)+\bm{X}_{\mu}(x),\\
\bm{X}_{\mu}(x)=  &  ig^{-1}\sum_{j=3,8}[\partial_{\mu}\bm{n}^{(j)}%
(x),\bm{n}^{(j)}(x)]\nonumber\\
&  +\sum_{j=3,8}[[\bm{A}_{\mu}(x),\bm{n}^{(j)}(x)],\bm{n}^{(j)}(x)],\\
\bm{V}_{\mu}(x)=  &  \sum_{j=3,8}(\bm{A}_{\mu}(x)\cdot\bm{n}^{(j)}%
(x))\bm{n}^{(j)}(x)\nonumber\\
&  -ig^{-1}\sum_{j=3,8}[\partial_{\mu}\bm{n}^{(j)}(x),\bm{n}^{(j)}(x)].
\end{align}

To formulate the new theory written in terms of the new variables ($X_{x,\mu}%
$,$V_{x,\mu}$) which is equipollent to the original Yang-Mills theory, we must
determine the configuration of color fields $\{\bm{n}_{x}^{(3)},\bm{n}_{x}%
^{(8)}\}$. In the maximal option, the color fields $\{\bm{n}_{x}%
^{(3)},\bm{n}_{x}^{(8)}\}$ are obtained by minimizing the functional:
\end{subequations}
\begin{align}
&  F_{\text{red}}[\bm{n}_{x}^{(3)},\bm{n}_{x}^{(8)}]\nonumber\\
&  =\sum_{x,\mu}\sum_{j=3,8}\mathrm{tr}\left\{  (D_{\mu}^{\varepsilon
}[U_{x,\mu}]\bm{n}_{x}^{(j)})^{\dag}(D_{\mu}^{\varepsilon}[U_{x,\mu
}]\bm{n}_{x}^{(j)})\right\}  . \label{eq:reductionMAG}%
\end{align}
It should be noticed that the decomposition in the maximal option with the
reduction condition obtained from the functional Eq.(\ref{eq:reductionMAG})
gives the gauge invariant extension of the \textit{Abelian projection} in the
\textit{maximal Abelian (MA) gauge}.

\section{Numerical simulations on the lattice}

\subsection{Lattice setup}

We adopt the standard Wilson action with the inverse gauge coupling constant
$\beta=2N_{c}/g^{2}$ ($N_{c}=3)$. We prepare the gauge field configurations
(link variables) $\{U_{x,\mu}\}$ on the lattice of size $L^{3}\times N_{T}$
from the two different data sets at finite temperature according to the
following setups:

(Data set I) For the fixed spatial size and temporal size $L=24$, $N_{T}=6$,
$8$, the temperature varies by changing the coupling $\beta$ in the range
$5.8\leq\beta\leq6.5$ (see Table~\ref{tbl:physscale}). The lattice spacing
$\varepsilon$ and the physical volume vary with temperature $T$.

(Data set II) For the fixed spatial size and coupling constant $L=24$,
$\beta=6.0$, $6.2$, the temperature varies by changing the temporal size
$N_{T}$ (see Table \ref{tbl:physscale2}). The lattice spacing $\varepsilon$
and the physical volume are the same for each $\beta$.

Table~\ref{tbl:physscale} gives the dictionary of the data set I for obtaining
the temperature $T$ for a given lattice size $L^{3}\times N_{T}$ and the
coupling constant $\beta$ by making use of the lattice spacing $\varepsilon$.
Here the physical units or the lattice spacing is determined on the lattice at
zero temperature according to Ref.~\cite{Edward98}. Table \ref{tbl:physscale2}
gives the dictionary of the data set II for obtaining the temperature $T$ and
the temporal size $N_{T}$ \ for a given coupling constant $\beta$.

We generate the configurations by using the standard method, i.e., one-sweep
update is obtained by applying once the pseudo heat-bath method and
twelve-times the over-relaxation method. We thermalize 16000 sweeps with the
cold start, and prepare 1000 configurations every 400 sweeps.

We obtain the color field configuration $\{%
\bm{h}%
_{x}\}$ for the minimal option by minimizing the functional
Eq.(\ref{eq:reduction-min}) for each set of the gauge field configurations
$\{U_{x,\mu}\}$, \ while we obtain the color field configuration $\{%
\bm{n}^{(3)}%
_{x}\}$ and $\{%
\bm{n}^{(8)}%
_{x}\}$ for the maximal option by minimizing the functional
Eq.(\ref{eq:reductionMAG}) for each set of the gauge field configurations
$\{U_{x,\mu}\}.$ Then we perform the decomposition of the gauge link variable
$U_{x,\mu}=X_{x,\mu}V_{x,\mu}$ by using the formula given in the previous
section. In the measurement of the Polyakov loop average and the Wilson loop
average defined below, we apply the APE smearing technique \cite{Albanese87}
to reduce noises.

\subsection{Polyakov-loop average at the confinement/ deconfinement
transition}

\label{sec:Ploop}%

\begin{figure*}  [thb] \centering
\includegraphics[height=52mm, viewport=0 0 360  252, clip]
{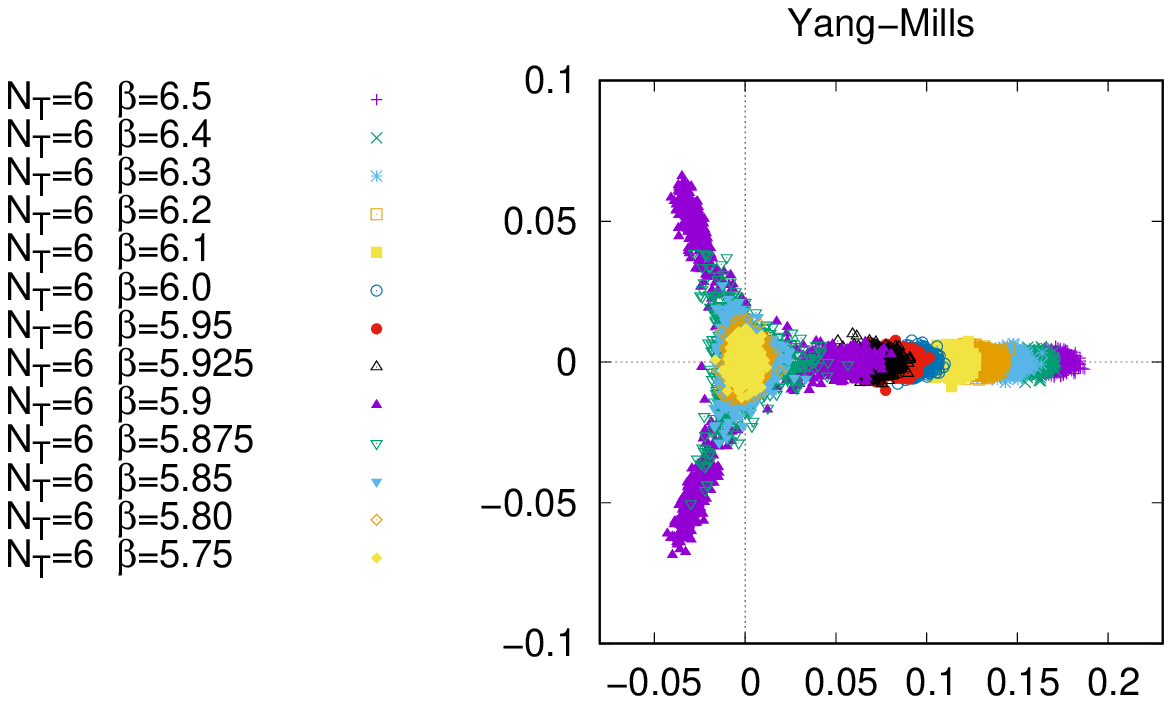}%
\includegraphics[height=52mm, viewport=115 0 360 252 , clip,]
{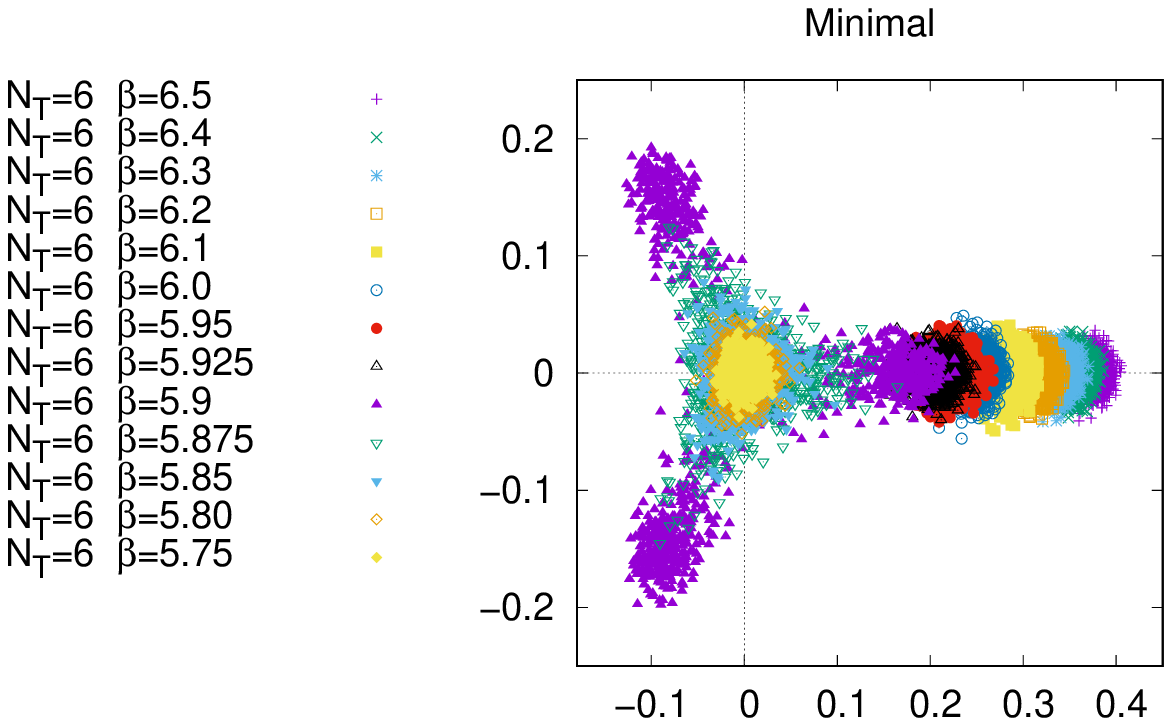}%
\includegraphics[height=52mm, viewport= 115 0 3600 252 ,clip]
{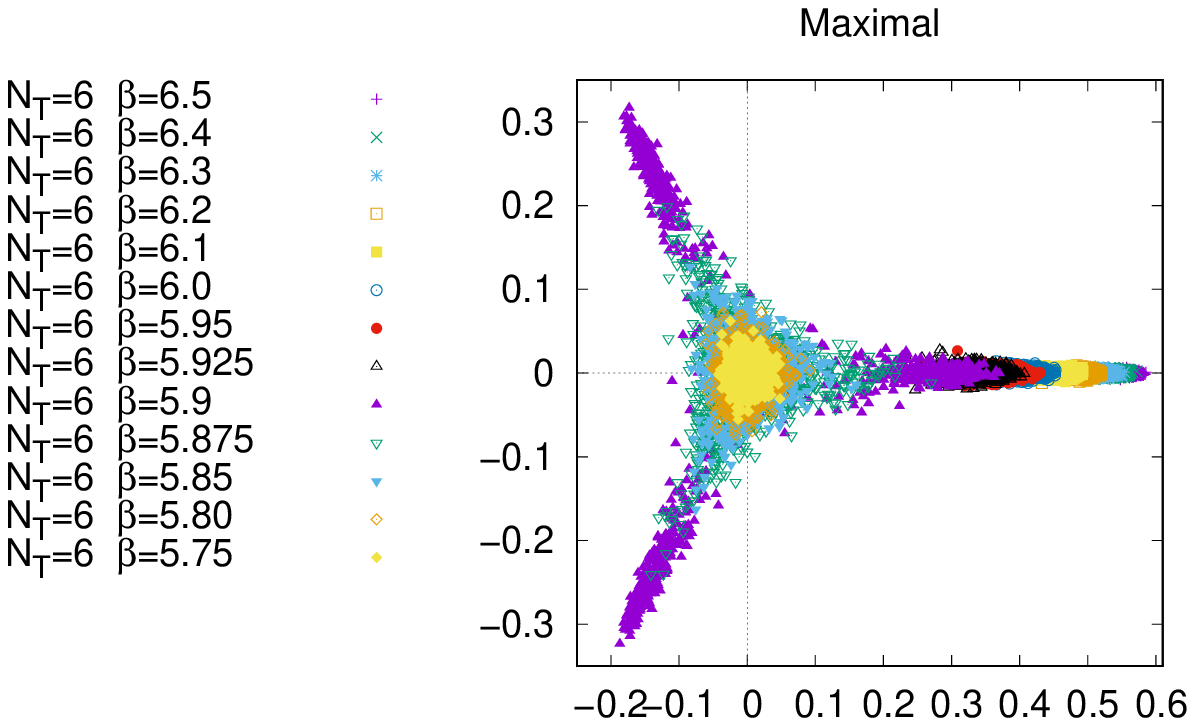}%
\linebreak
\includegraphics[height=52mm, viewport=0 0 360  252, clip]
{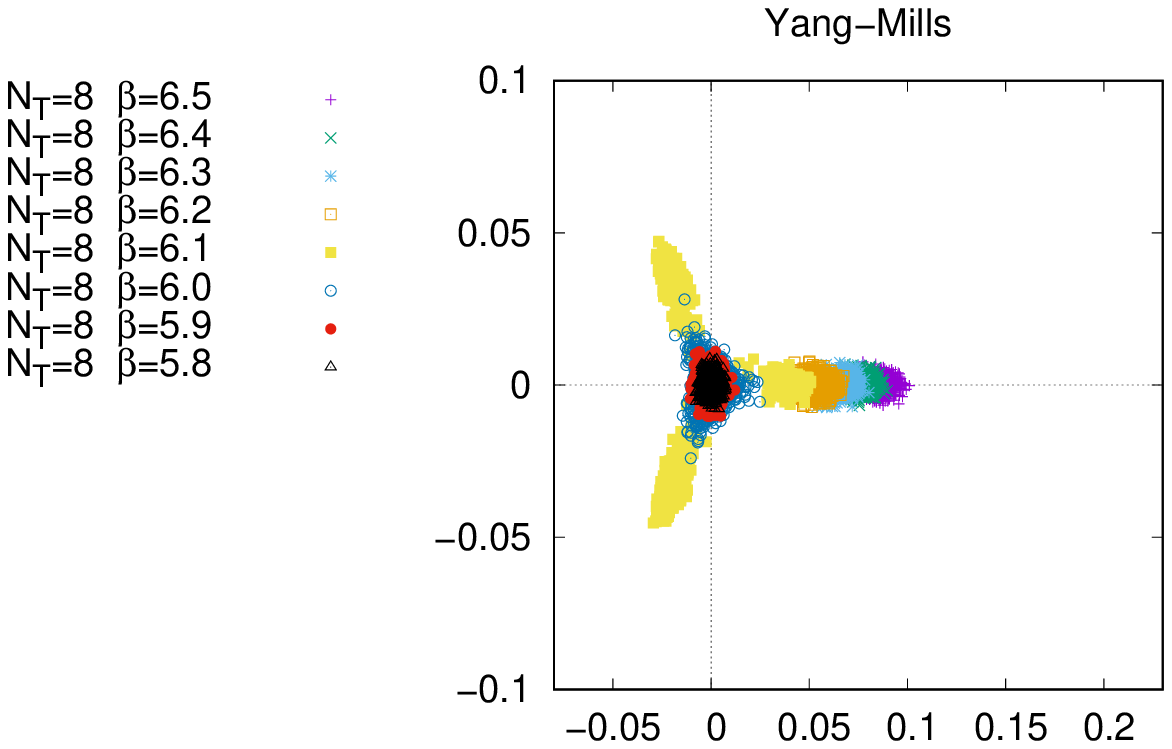}%
\includegraphics[height=52mm, viewport=115 0 360 252 , clip,]
{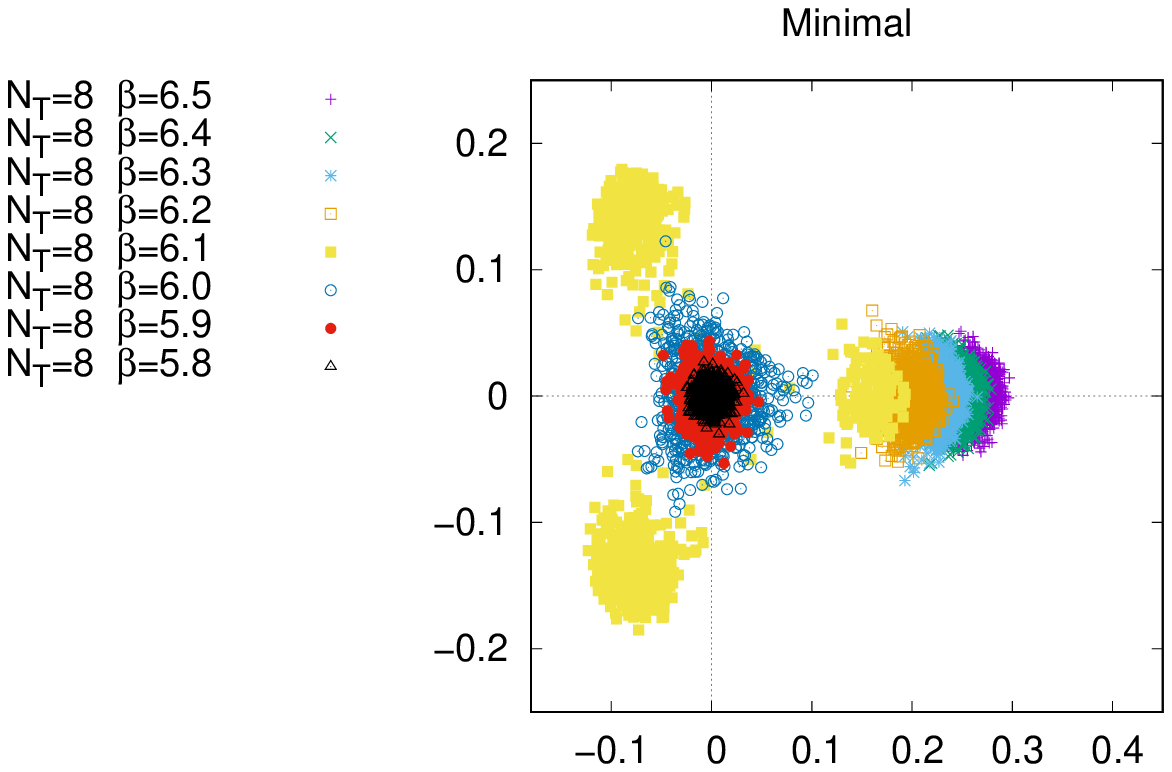}%
\includegraphics[height=52mm, viewport= 115 0 3600 252 ,clip]
{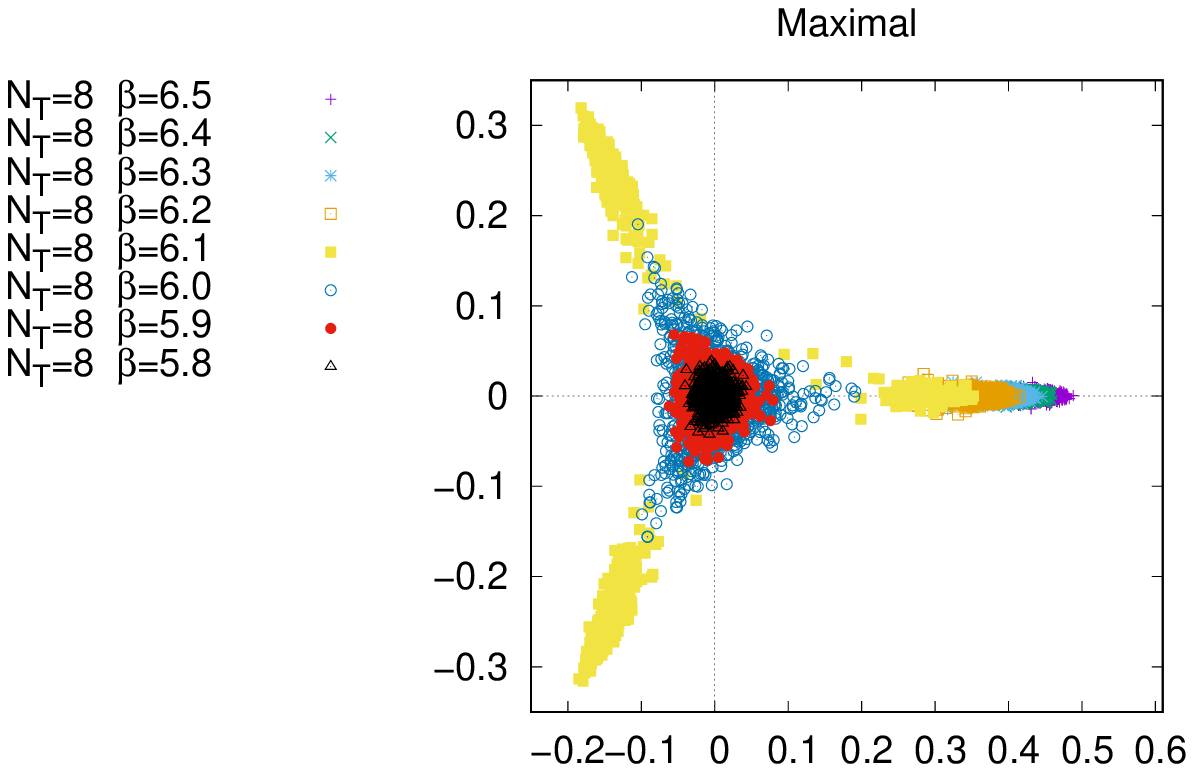}%
\caption{
Distributions of the space-averaged Polyakov loop on the complex plane for the data set I:
(upper three panels) for $N_T=6$  and $\beta=5.75,5.80,5.85,5.875,5.9,5.925,5.95,6.0,6.1,6.2,6.3,6.4,6.5$
(lower three panels) $N_T=8$ and $\beta=5.8,5.9,6.0,6.1,6.2,6.3,6.4,6.5$.
Plots are given from left to right for the original Yang-Mills field $P_{U}^{YM}$, the restricted field in the minimal option  $P_{V}^{\text{min}}$ and the restricted field  in  the maximal option  $P_{V}^{\text{max}}$.
} \label{Fig:PLP}%
\end{figure*}%
%

\begin{figure*}[thb] \centering
\includegraphics[height=52mm, viewport=0 0 360  252, clip]
{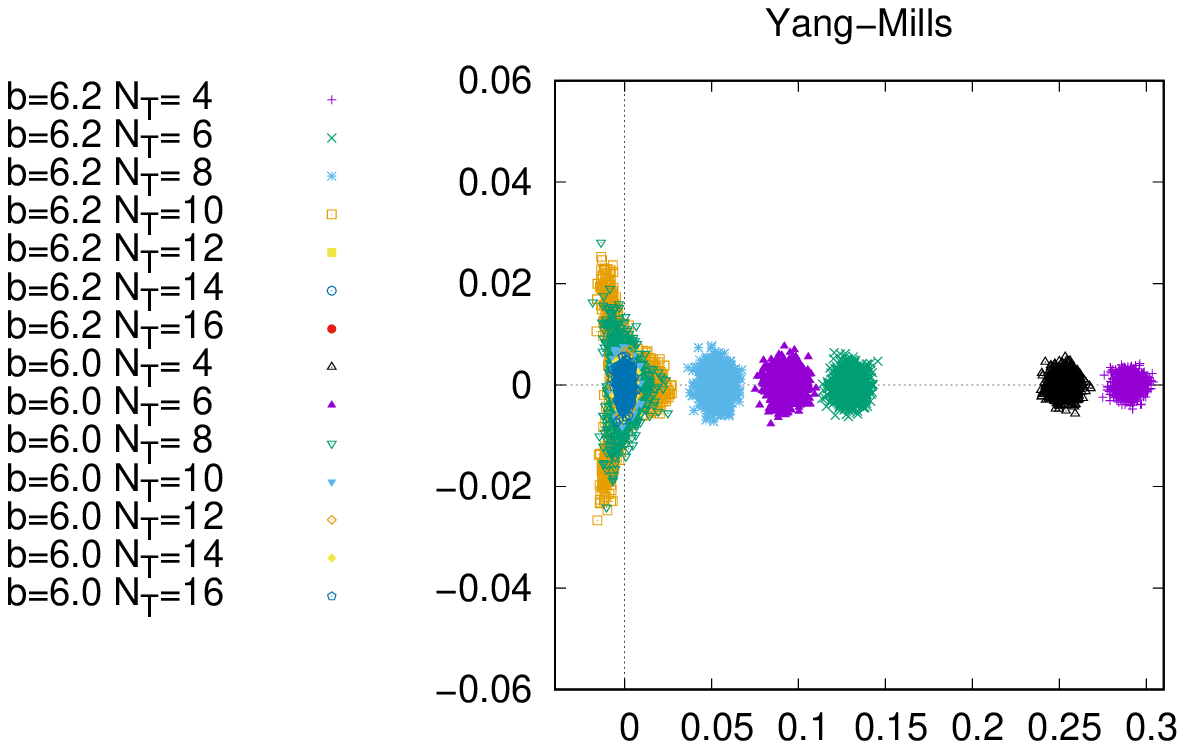}%
\includegraphics[height=52mm, viewport=115 0 360 252 , clip,]
{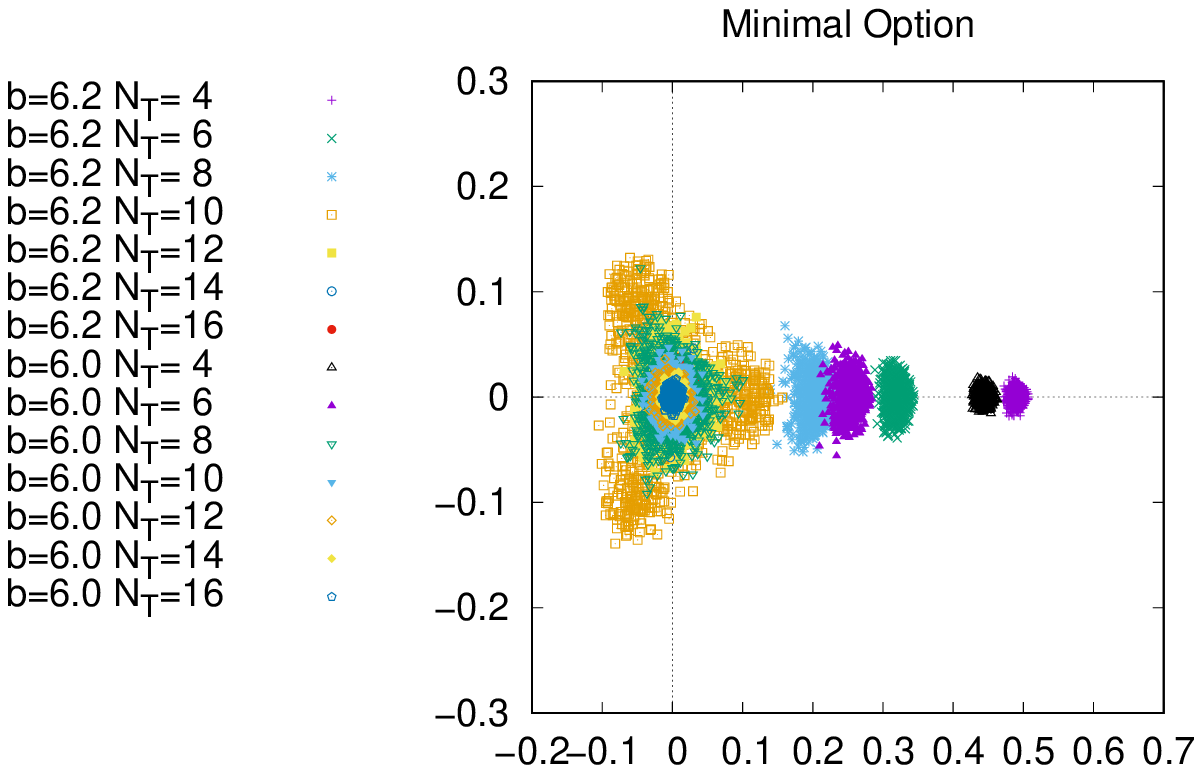}%
\includegraphics[height=52mm, viewport= 115 0 3600 252 ,clip]
{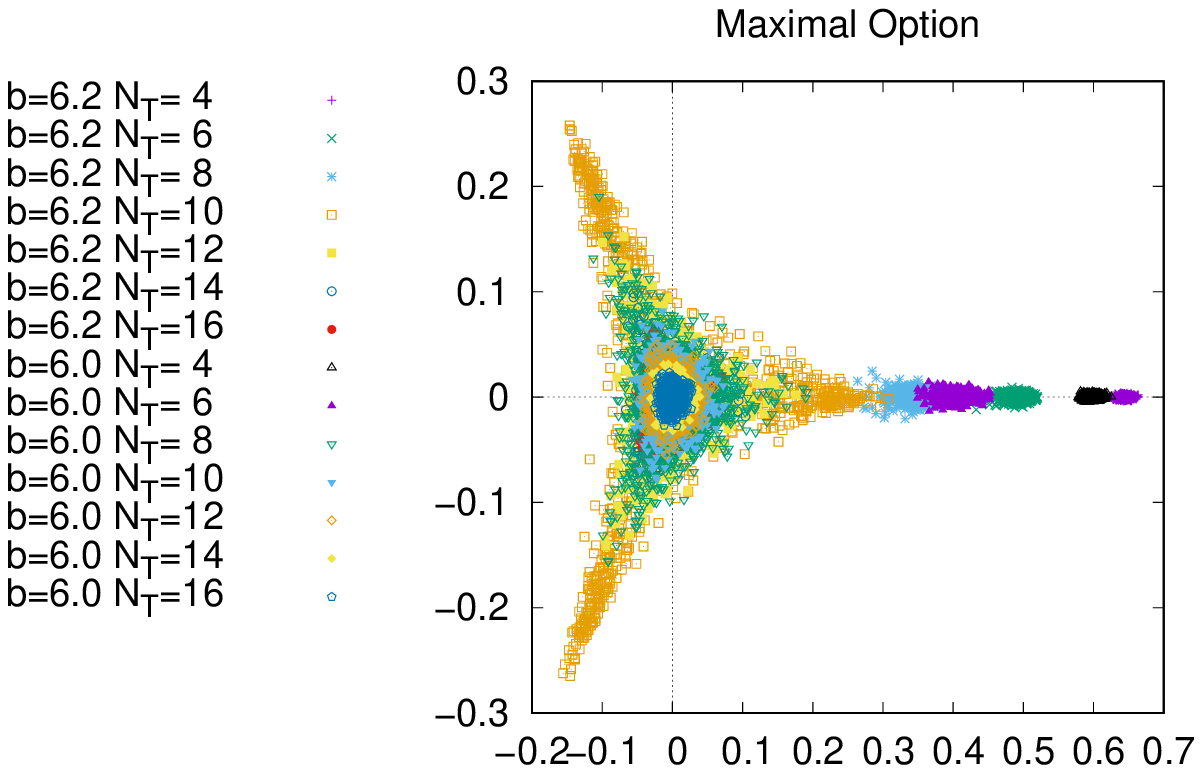}%
\caption{
Distributions of the space-averaged Polyakov loop on the complex plane for the data set II:
for $N_T=4,6,8,10,12,14,16$ at $\beta=6.2$ and $\beta=6.0$.
Plots are given from left to right for the original Yang-Mills field $P_{U}^{YM}$, the restricted field in the minimal option  $P_{V}^{\text{min}}$ and the restricted field  in  the maximal option  $P_{V}^{\text{max}}$.
} \label{Fig:PLP3}%
\end{figure*}%

First, we investigate the Polyakov loop obtained by the original gauge field
configurations $\{U_{x,\mu}\}$ and the restricted gauge field configurations
$\{ V_{x,\mu}^{\min} \}$ $\{ V_{x,\mu}^{\max} \}$ in the minimal and maximal
options, in order to clarify the role of the restricted fields at finite
temperature. The Polyakov loop average is the order parameter associated with
the center symmetry breaking. Note that the Polyakov loop average of the
Yang-Mills field is conventionally used as a criterion for the confinement and
deconfinement phase transition.

For the original gauge field configurations $\{U_{x,\mu}\}$, the restricted
gauge field configurations in the minimal and maximal options $\{V_{x,\mu
}^{\min}\}$ and $\{V_{x,\mu}^{\max}\}$, we define the respective Polyakov loop
$P_{\ast}(\vec{x}\mathbf{)}$ ( $\ast=U,V^{\min},V^{\max}$ ) by%
\begin{subequations}
\begin{align}
P_{U}() &  :=\frac{1}{3}\text{\textrm{tr}}\left(  P\prod\limits_{t=1}^{N_{T}%
}U_{(\vec{x},t),4}\right)  ,\\
\ P_{V^{\min}}(\vec{x}) &  :=\frac{1}{3}\text{\textrm{tr}}\left(
P\prod\limits_{t=1}^{N_{T}}V_{(\vec{x},t),4}^{\min}\right)  ,\\
P_{V^{\max}}(\vec{x}) &  :=\frac{1}{3}\text{\textrm{tr}}\left(  P\prod
\limits_{t=1}^{N_{T}}V_{(\vec{x},t),4}^{\max}\right)  ,
\end{align}
\label{eq:Polyakov-dist}%
\end{subequations}
and define the space-averaged Polyakov loop $P_{\ast}$ , i.e., the value of
the Polyakov loop averaged over the spatial volume by%
\begin{subequations}%
\begin{align}
P_{U} &  =\frac{1}{L^{3}}\sum_{\vec{x}}P_{U}(\vec{x}),\\
P_{V^{\min}} &  =\frac{1}{L^{3}}\sum_{\vec{x}}P_{V^{\min}}(\vec{x}),\ \\
P_{V^{\max}} &  =\frac{1}{L^{3}}\sum_{\vec{x}}P_{V^{\max}}(\vec{x}),
\end{align}
\label{eq:Ploop-Space-ave1}%
\end{subequations}
for each configuration in the complex plane.

FIG.~\ref{Fig:PLP} and FIG. \ref{Fig:PLP3} show the distributions of the
space-averaged Polyakov loop $P_{\ast}$ on the complex plane obtained from the
data set I and II respectively.
In each figure, the three panels are arranged from left to right to show the
plots measured using a set of the original gauge field configurations $P_{U}$,
the restricted gauge field configurations in the minimal option $P_{V^{\min}}$
and maximal one $P_{V^{\max}}$, respectively.
We find that all the distributions of the space-averaged Polyakov loops
$P_{\ast}$ on the complex plane equally reflect the expected center symmetry
$Z(3)$ of $SU(3)$, although they take different values option by option.
Notice that the Polyakov loop average is in general complex-valued for the
$SU(3)$ group. In our simulations, especially, we have used the cold start and
obtained the real-valued Polyakov loop average at high temperature, as shown
in FIG.~\ref{Fig:PLP} and FIG.~\ref{Fig:PLP3}.

Then, we measure the Polyakov-loop average $\left\langle P_{\ast}\right\rangle
$  $(\ast=U,V^{\min},V^{\max})$ by simply averaging the space-averaged
Polyakov loop given in FIG.~\ref{Fig:PLP} and FIG.~\ref{Fig:PLP3} over the
total sets of the original field configurations and the restricted field
configurations in the minimal and maximal options. In what follows, the symbol
$\left\langle \mathcal{O}\right\rangle $ denotes the average of the operator
$\mathcal{O}$ over the ensemble of the configurations.
%

\begin{figure}[hbt] \centering
\includegraphics[width=0.48\textwidth]
{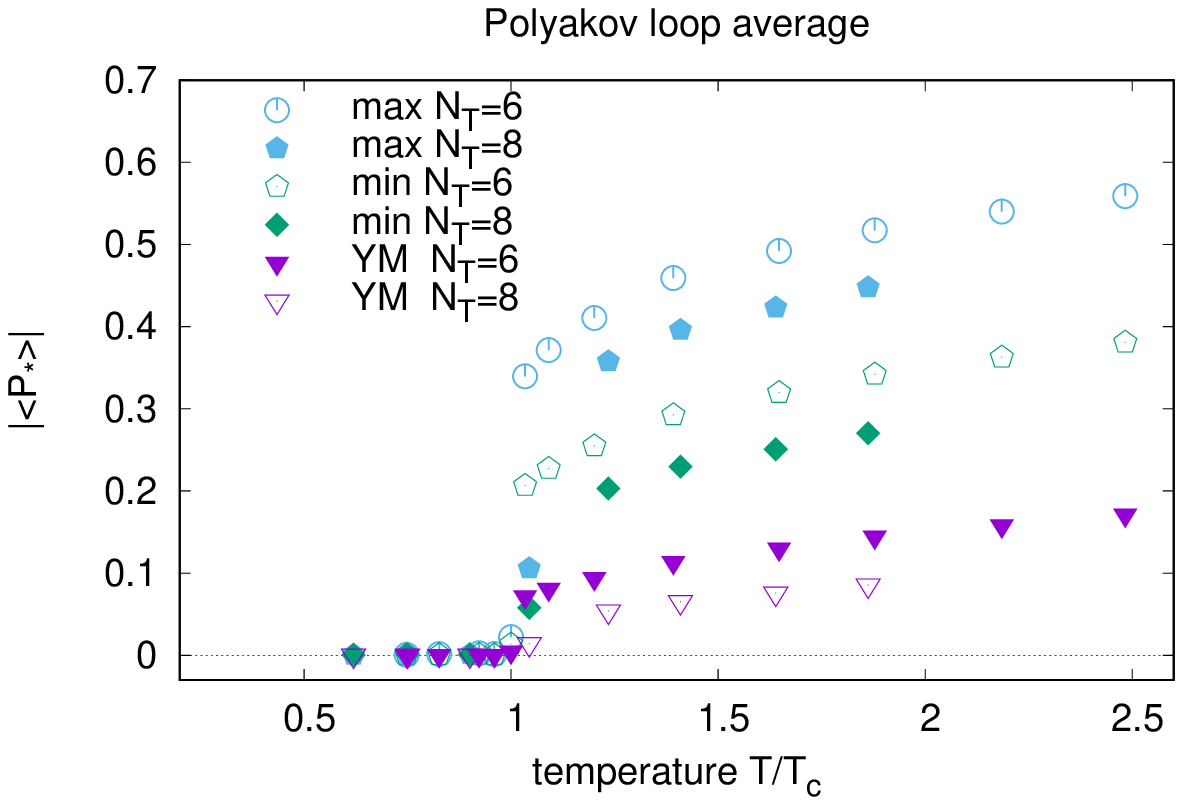}
\includegraphics[width=0.48\textwidth]
{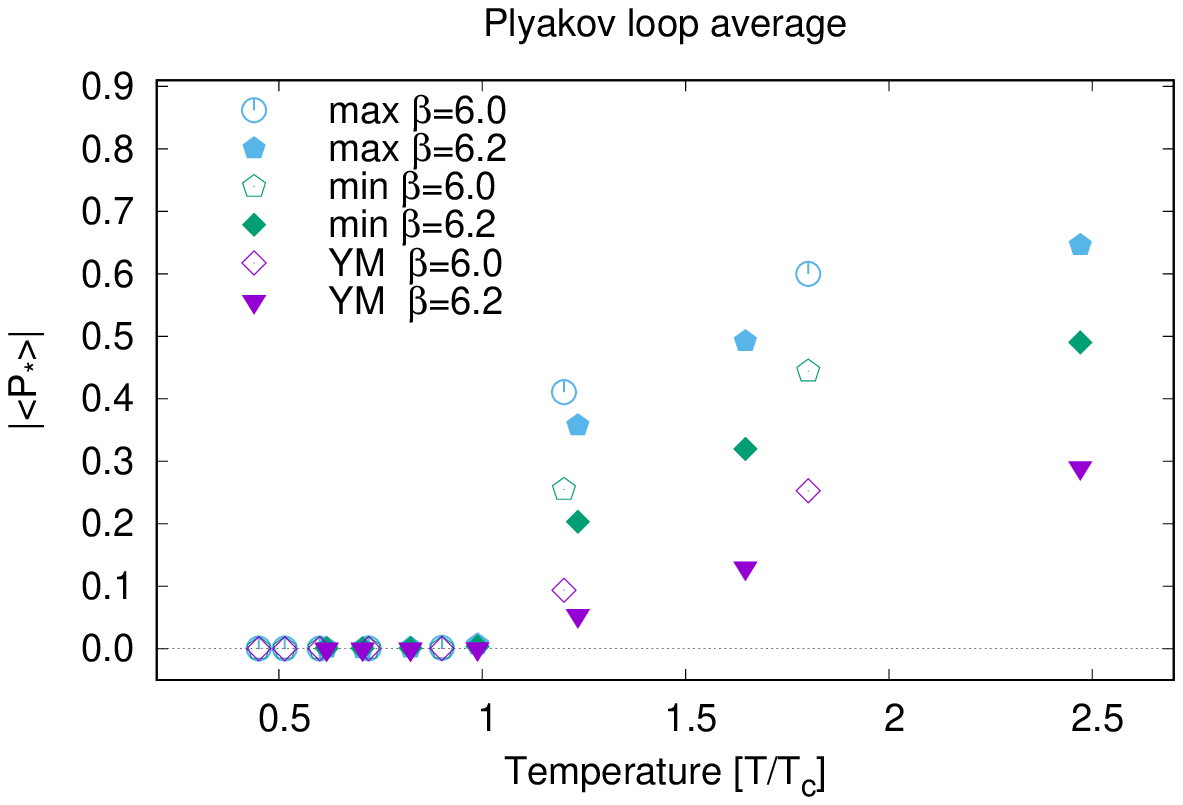}%
\caption{
The Polyakov-loop averages versus temperature:
The three plots for the Polyakov loop averages $\left< P_{*}\right>$ $(*=U,V^{\min},V^{\max})$ at a temperature $T$ represent from bottom to top the Polyakov loop averages for
the original gauge field $U$,  the restricted field $V^{\min}$ in the minimal option and the restricted field $V^{\max}$ in the maximal one, respectively.
(upper panel) The plot for the data set I.
(lower panel) The plots for  the data set II.
} \label{fig:PLPave}%
\end{figure}%

FIG.~\ref{fig:PLPave} shows the Polyakov loop average versus the temperature
determined by using Table~\ref{tbl:physscale} and \ref{tbl:physscale2}.
We find that the Polyakov loop averages $\left\langle P_{\ast}\right\rangle $
$(\ast=U,V^{\min},V^{\max})$ take different values option by option, namely,
$\ast=U,V^{\min},V^{\max}$.
It is observed that among the Polyakov loop averages $\left\langle P_{\ast
}\right\rangle $ at a temperature $T$ the original gauge field $U$ gives the
smallest value and that the restricted fields $V$ give larger values:
$V^{\max}$ in the maximal option gives the largest value, while the restricted
field $V^{\min}$ in the minimal option gives the value in between. However,
the three Polyakov loop averages, $\left\langle P_{U}\right\rangle $,
$\left\langle P_{V^{\min}}\right\rangle $ and $\left\langle P_{V^{\max}%
}\right\rangle ,$ give the same critical temperature for the phase transition
separating the low-temperature confined phase characterized by the vanishing
Polyakov loop average $\left\langle P_{U}\right\rangle =\left\langle
P_{V^{\min}}\right\rangle =\left\langle P_{V^{\max}}\right\rangle =0$ from the
high-temperature deconfined phase characterized by the non-vanishing Polyakov
loop average $\left\langle P_{U}\right\rangle \neq0$, $\left\langle
P_{V^{\min}}\right\rangle \neq0$, and $\left\langle P_{V^{\max}}\right\rangle
\neq0$. We also find that the critical temperature determined from the data
set I and II agrees with each other.%

\begin{figure*}  [t] \centering
\includegraphics[height=60mm, viewport=40 0 290 252, clip]
{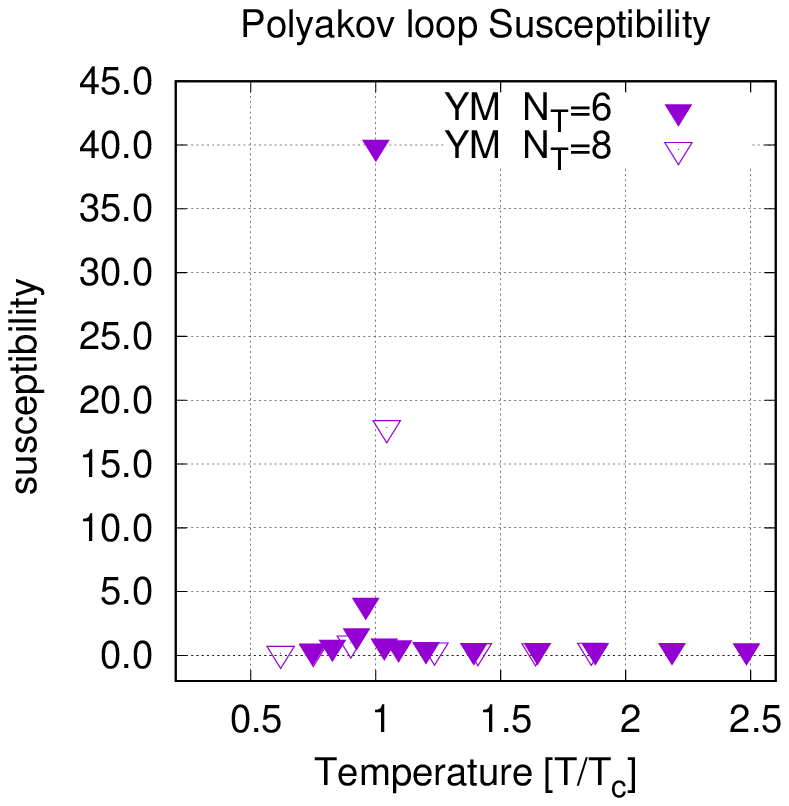}\includegraphics[height=60mm, viewport=40 0 290 252, clip ]
{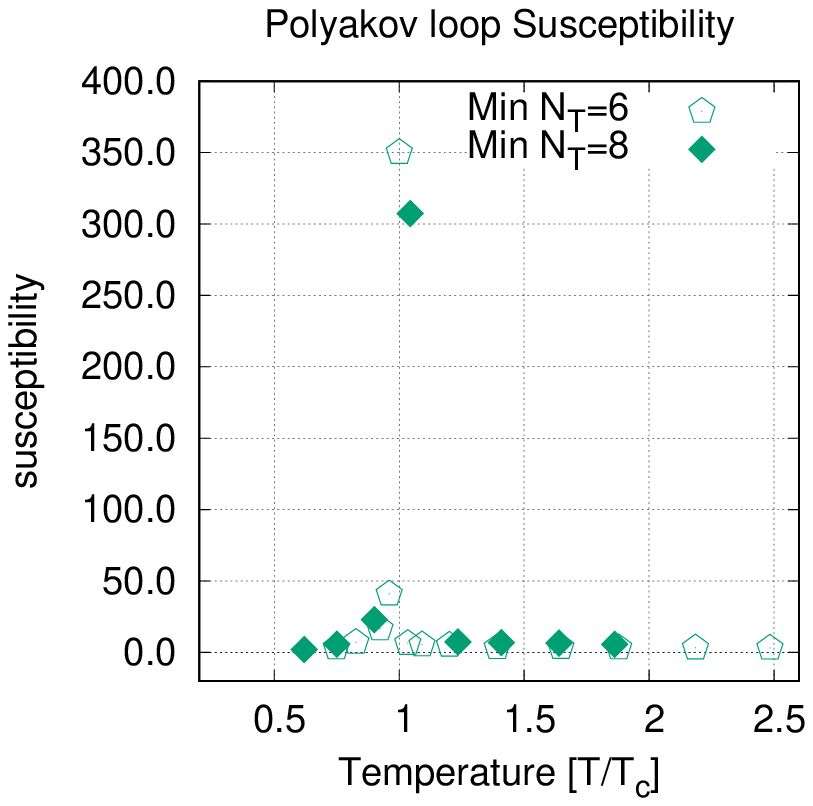}\includegraphics[height=60mm, viewport=40 0 290 252, clip ]
{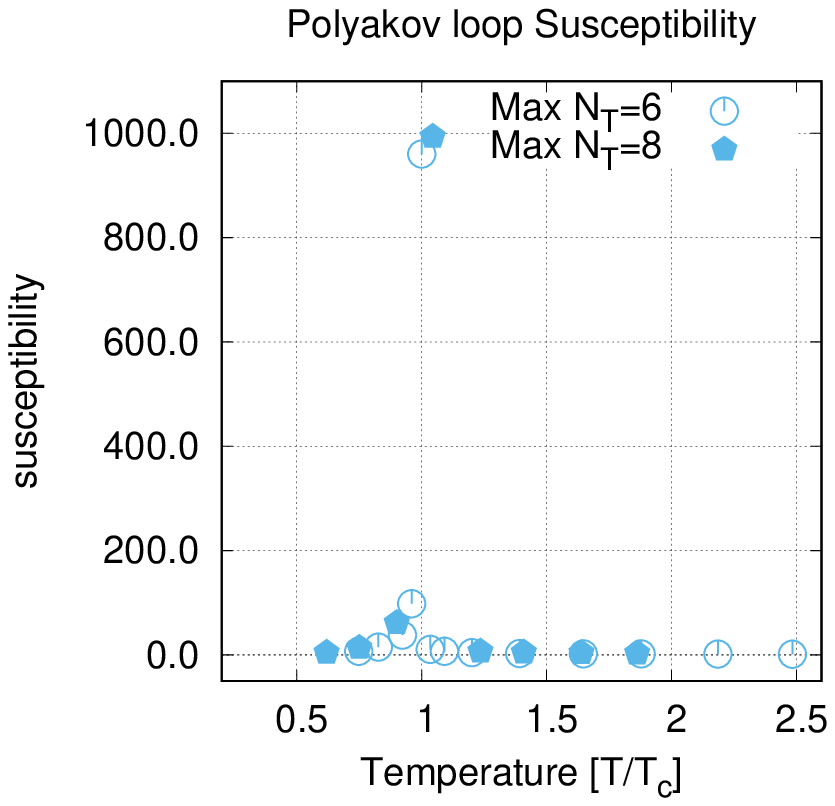}%
\linebreak

\includegraphics[height=60mm, viewport=40 0 290 252, clip]
{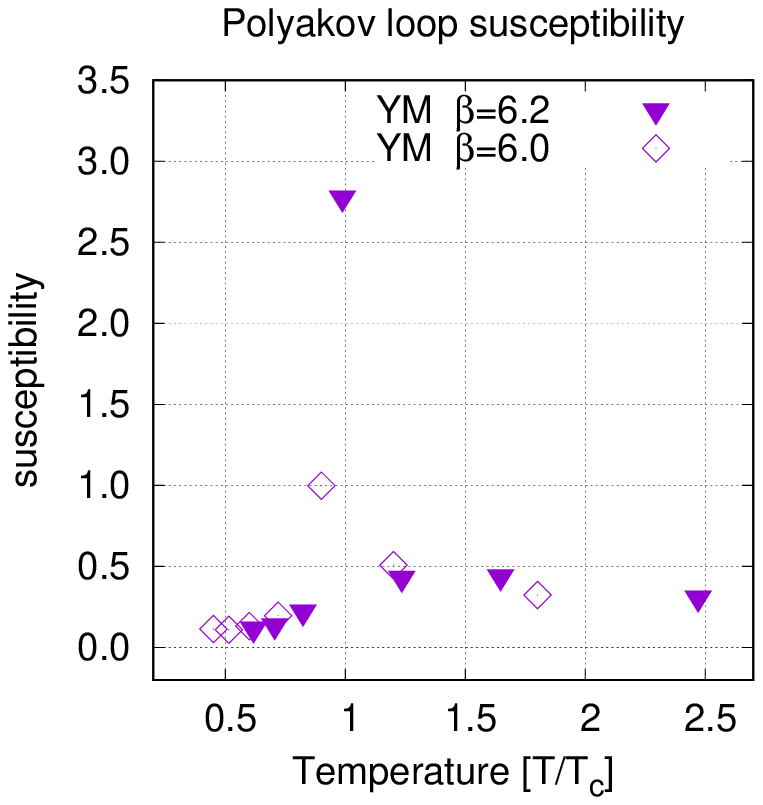}\includegraphics[height=60mm, viewport=40 0 290 252, clip ]
{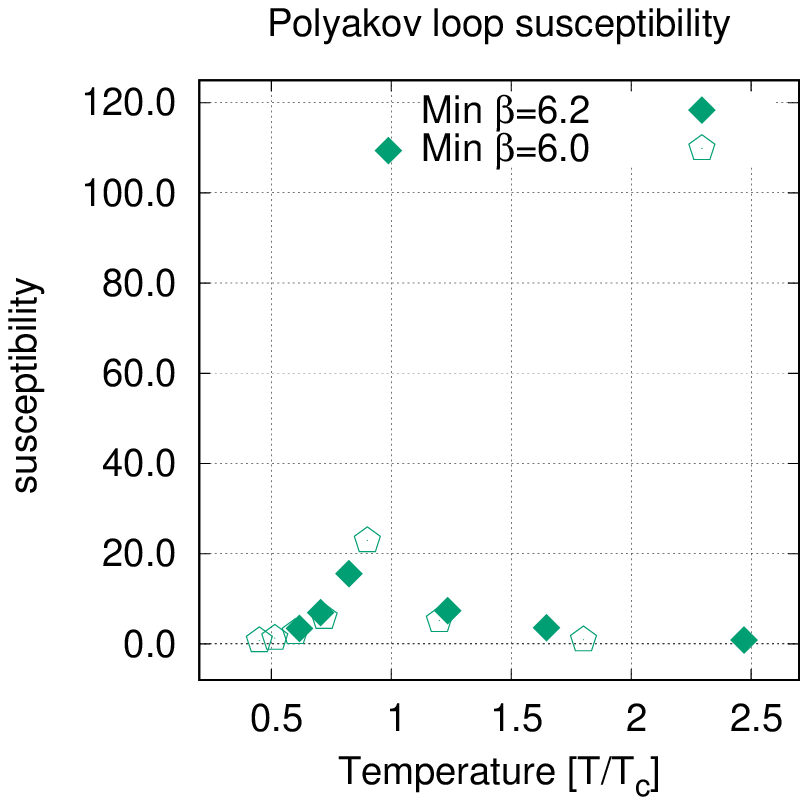}\includegraphics[height=60mm, viewport=40 0 290 252, clip ]
{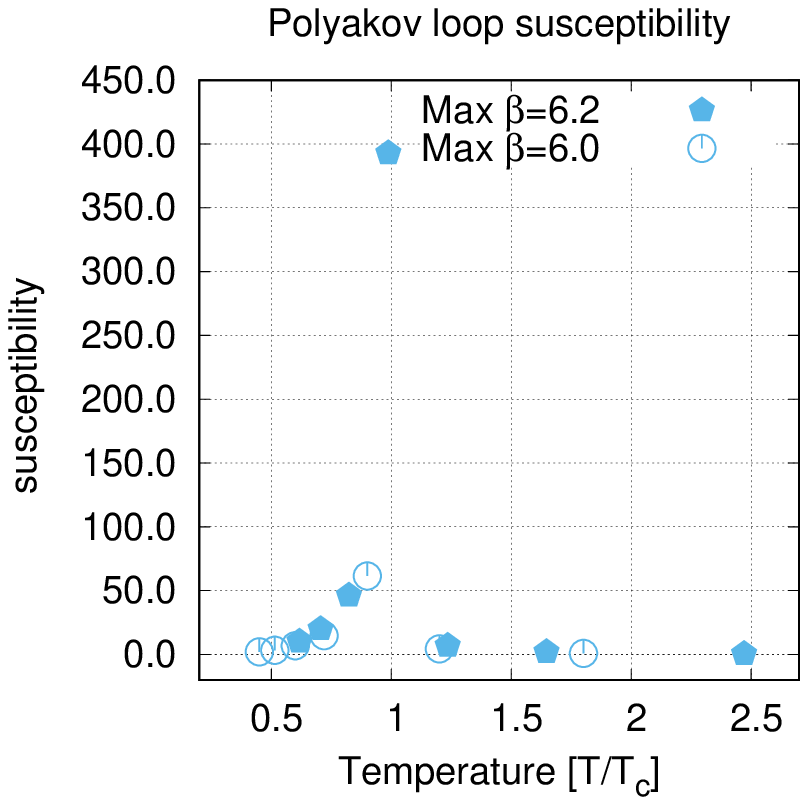} \caption{
The  Polyakov loop susceptibility versus temperature:
(upper row) data set I.
(lower row) data set II.
The three panels in each row represent from left to right the susceptibilities calculated from the Polyakov loops of the original Yang-Mills field  $P_{U}^{\text{YM} } $, the restricted field in the minimal option $P_{V}^{min}$, and the restricted field in the maximal option $P_{V}^{max}$.
} \label{Fig:PLP_sus}%
\end{figure*}%

In order to determine a precise location of the transition temperature, we
introduce the susceptibility $\chi_{\ast}$ $(\ast=U,V^{\min},V^{\max})$ of the
Polyakov loop by
\begin{equation}
\chi_{\ast}:=\frac{1}{L^{3}}\left[  \left\langle \left\vert \sum_{\vec{x}%
}P_{\ast}(\vec{x})\right\vert ^{2}\right\rangle -\left\vert \left\langle
\sum_{\vec{x}}P_{\ast}(\vec{x})\right\rangle \right\vert ^{2}\right]  .
\label{eq:ploop-susciptbility}%
\end{equation}
FIG.\ref{Fig:PLP_sus} gives the susceptibility versus temperature obtained
from the data set I and II.
They are calculated from the space-averaged Polyakov loop over the total sets
of the original field configurations and the restricted field configurations
in the minimal and maximal options, given in FIG.~\ref{Fig:PLP} and
FIG.~\ref{Fig:PLP3}, respectively. These data clearly shows that both the
minimal and maximal options reproduce the critical point of the original
Yang-Mills field theory. Thus, the three Polyakov-loop averages give the
identical critical temperature as
\begin{equation}
T_{c}=0.6285\text{ [in the unit }\sqrt{\sigma}\text{]}=276.5\text{ \ MeV}.
\label{eq:Tc}%
\end{equation}

\subsection{Static quark--antiquark potential at finite temperature}%

\begin{figure*}[tb] \centering
\includegraphics[height=50mm , clip, viewport= 0 10 510  550 ]
{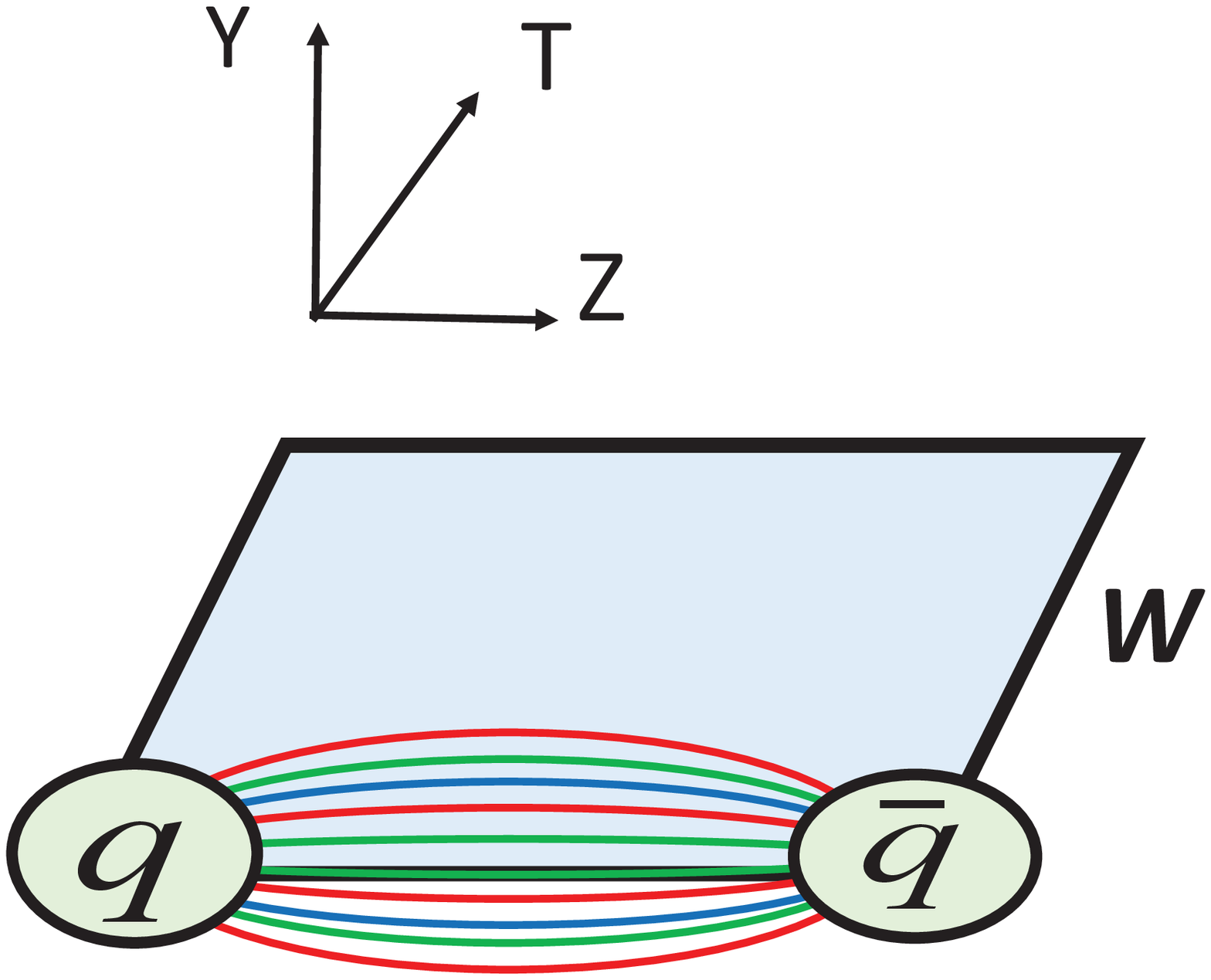}\includegraphics[height=35mm] {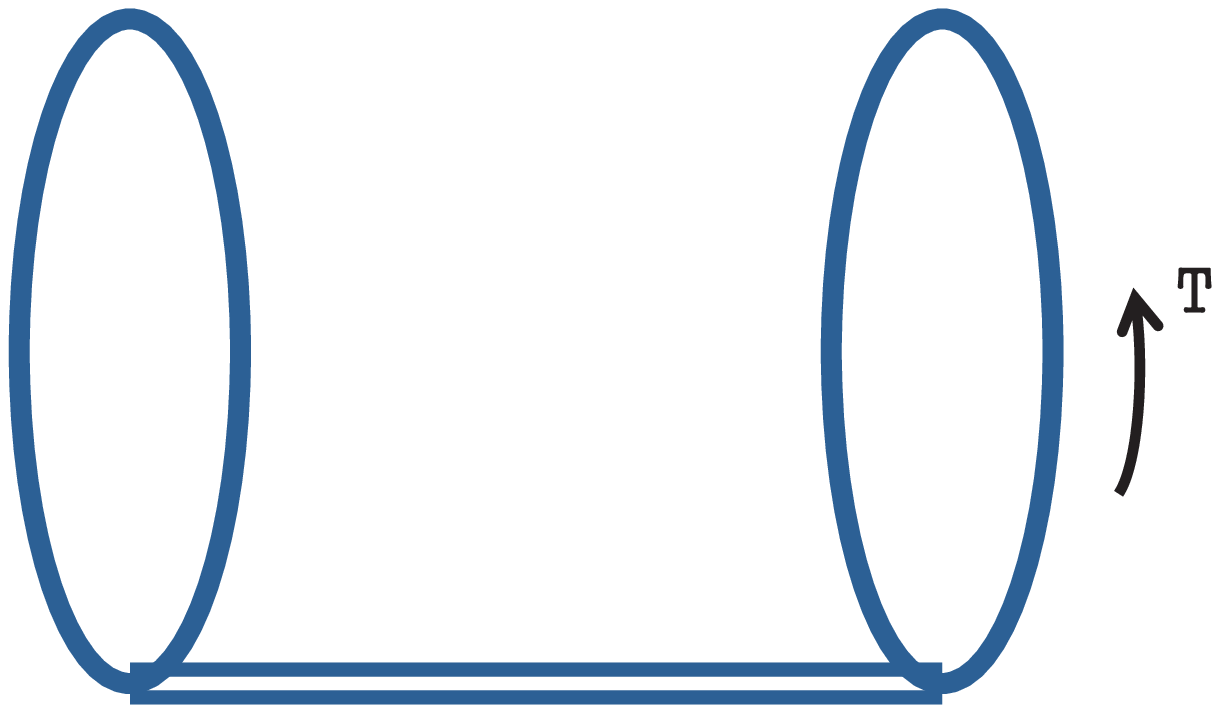}\includegraphics[height=35mm]{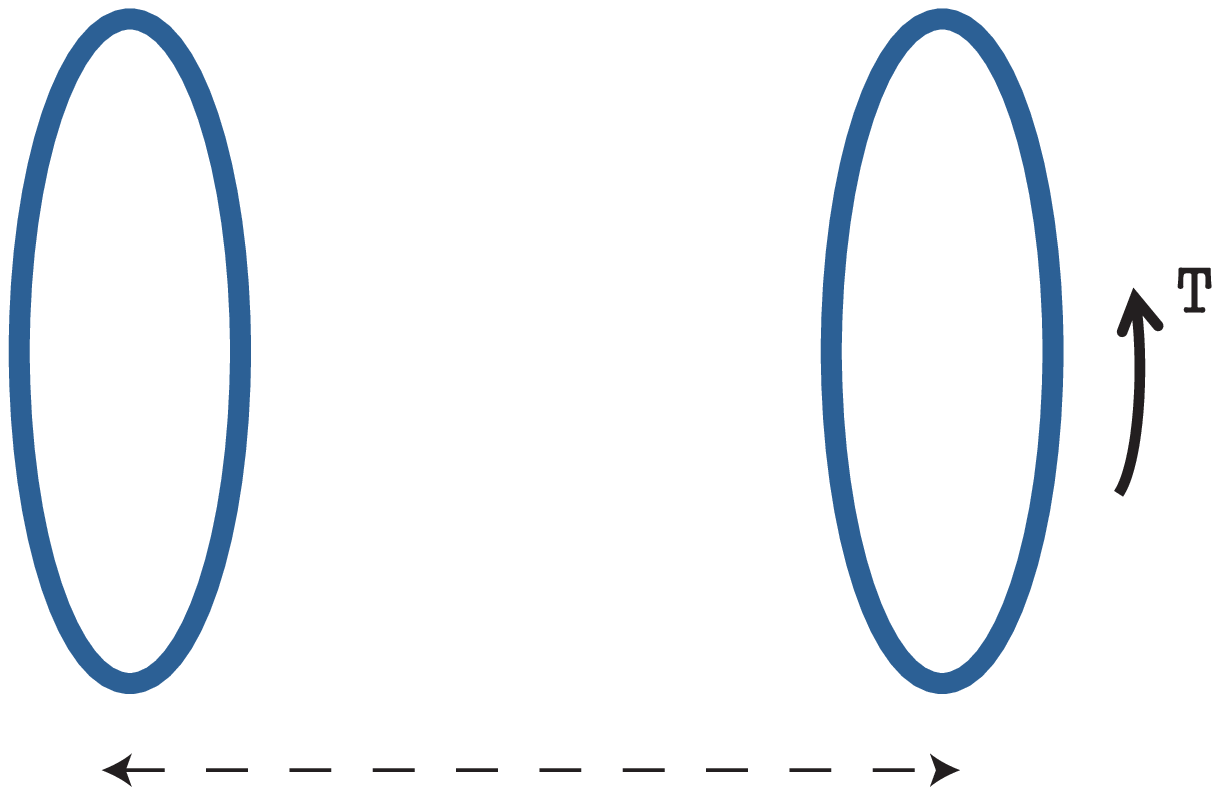}
\caption{ The set up for the measurement of the static quark-antiquark
potential at finite temperature: (Left) the Wilson loop, (Center) maximally extended Wilson loop,
(Right) a pair of a Polyakov loop and the anti-Polyakov loop}\label{fig:Op-potential}%
\end{figure*}%
%

\begin{figure*}[tb] \centering
\includegraphics
[height=53mm , clip, viewport= 0 0 340 252 ]{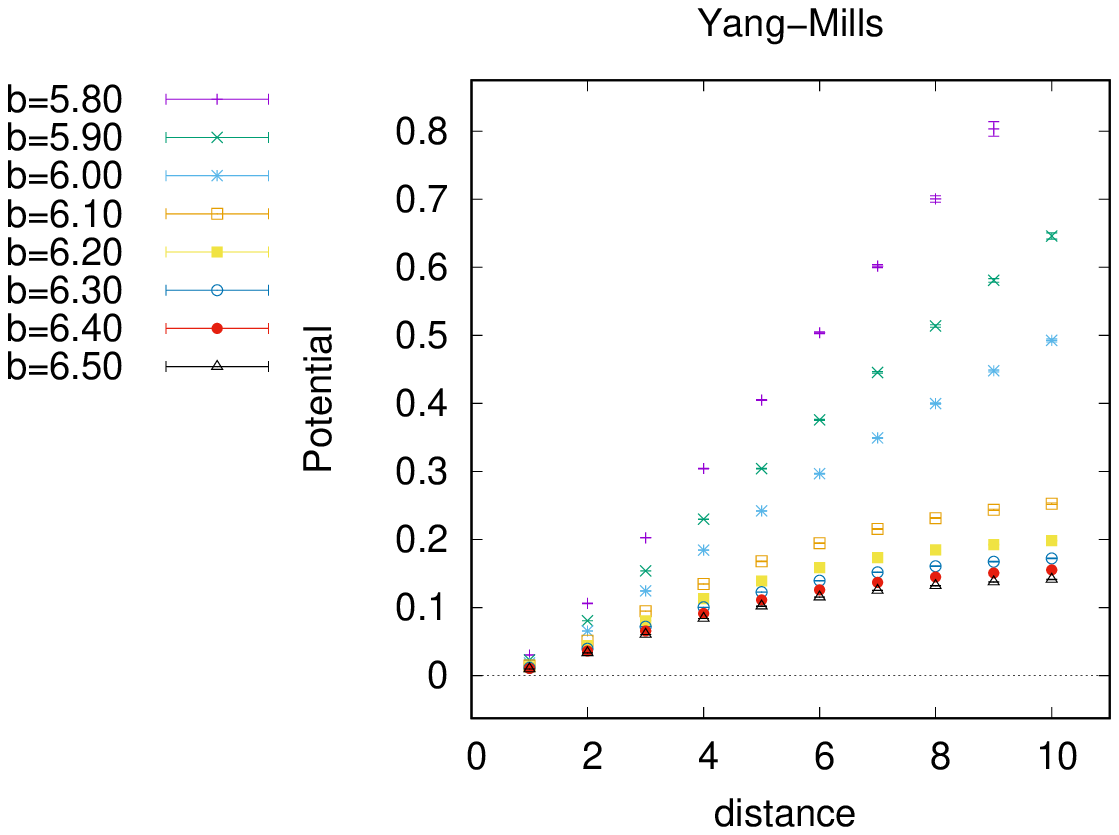}%
\includegraphics
[height=53mm, clip, viewport=90 0 340 252 ]{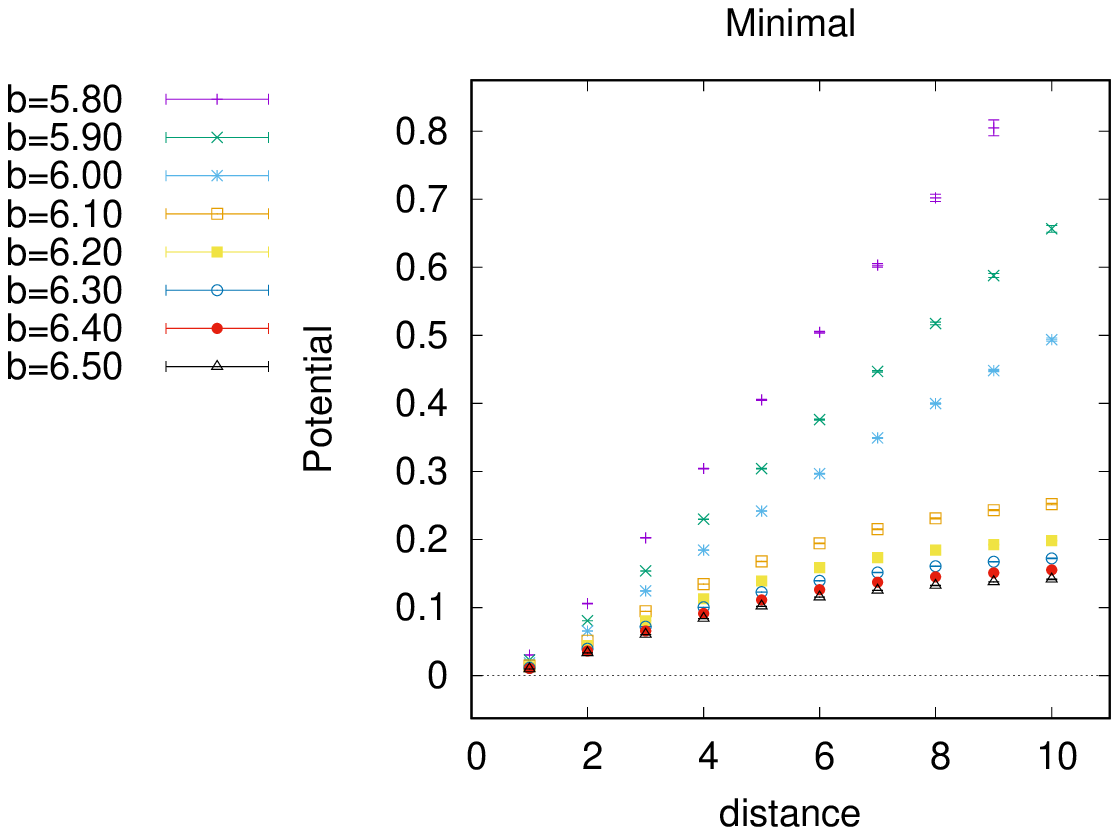}%
\includegraphics
[height=53mm , clip, viewport= 90 0 340 250 ]{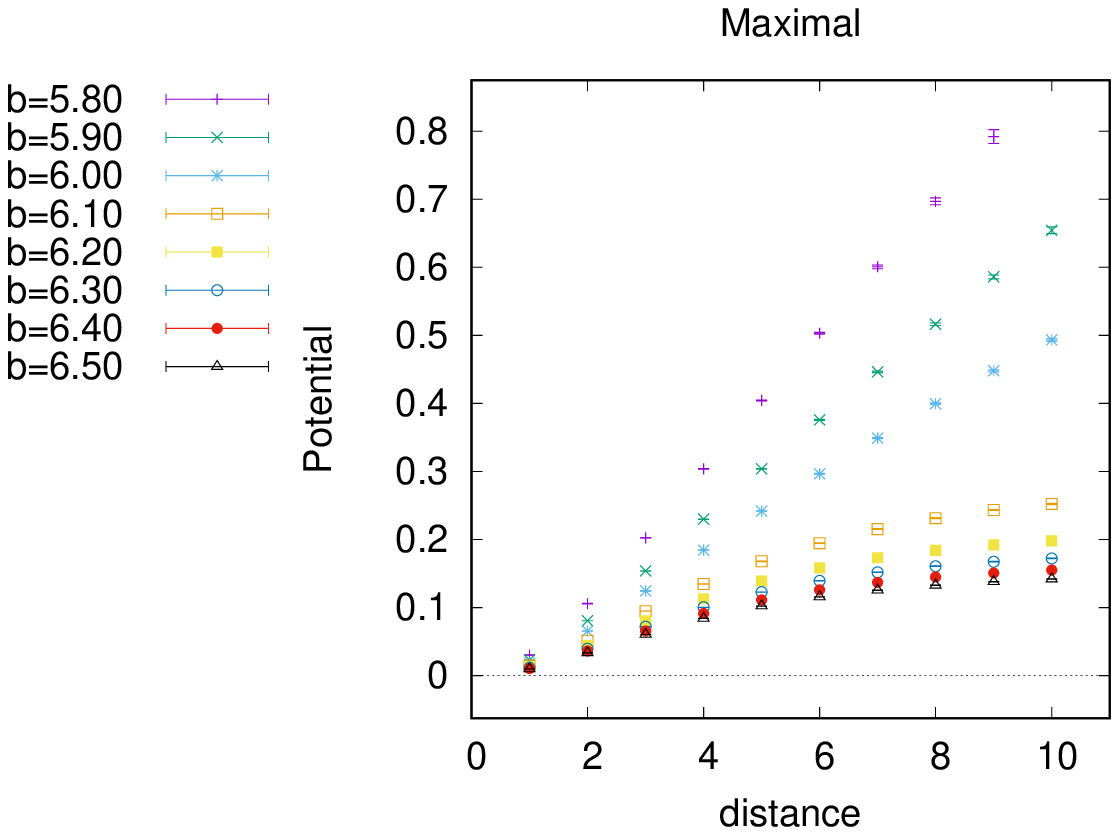}%
\linebreak
%

\includegraphics
[height=53mm , clip, viewport= 0 0 340 252 ]{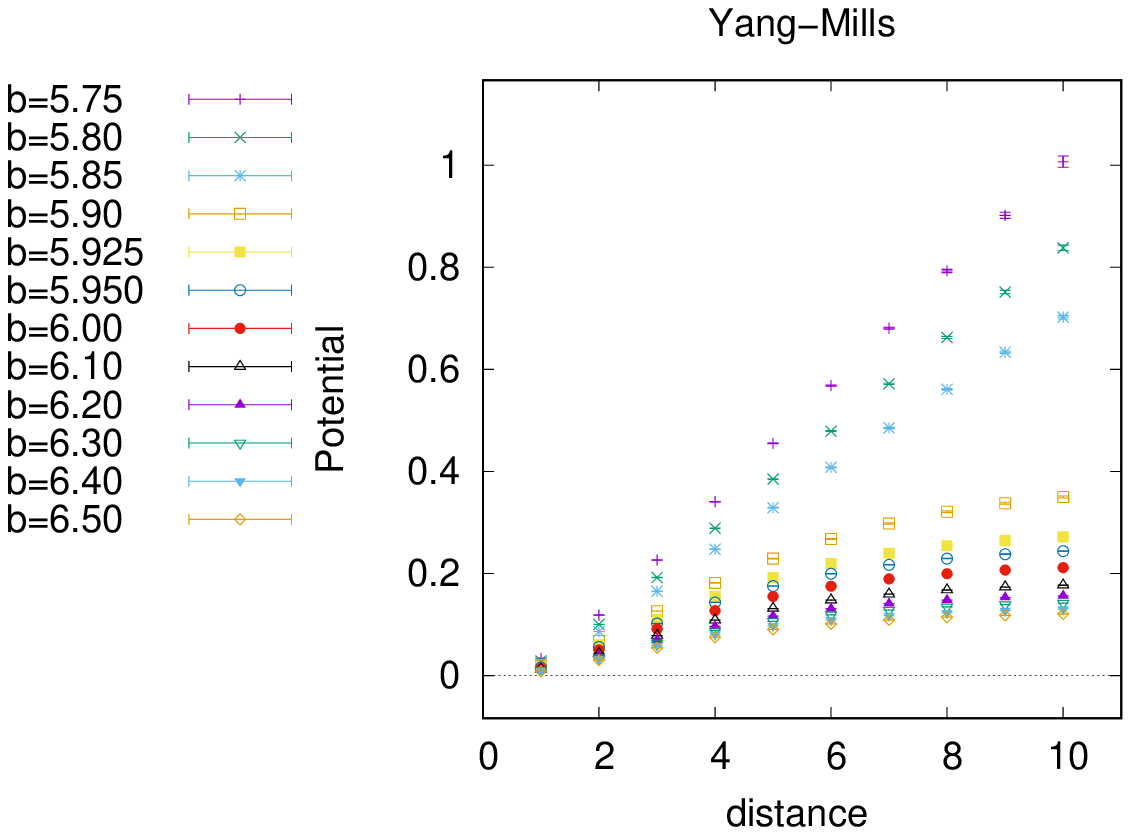}%
\includegraphics
[height=53mm , clip, viewport= 90 0 340 252 ]{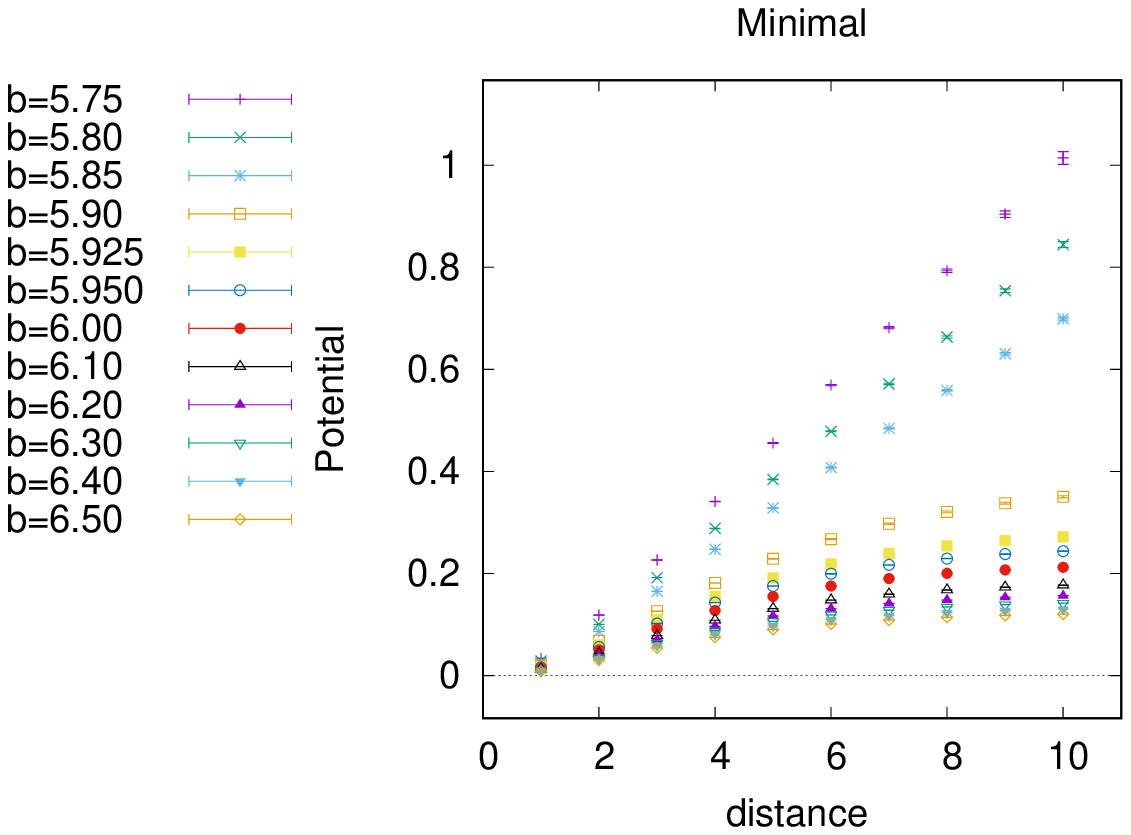}%
\includegraphics
[height=53mm , clip, viewport= 90 0 340 252 ]{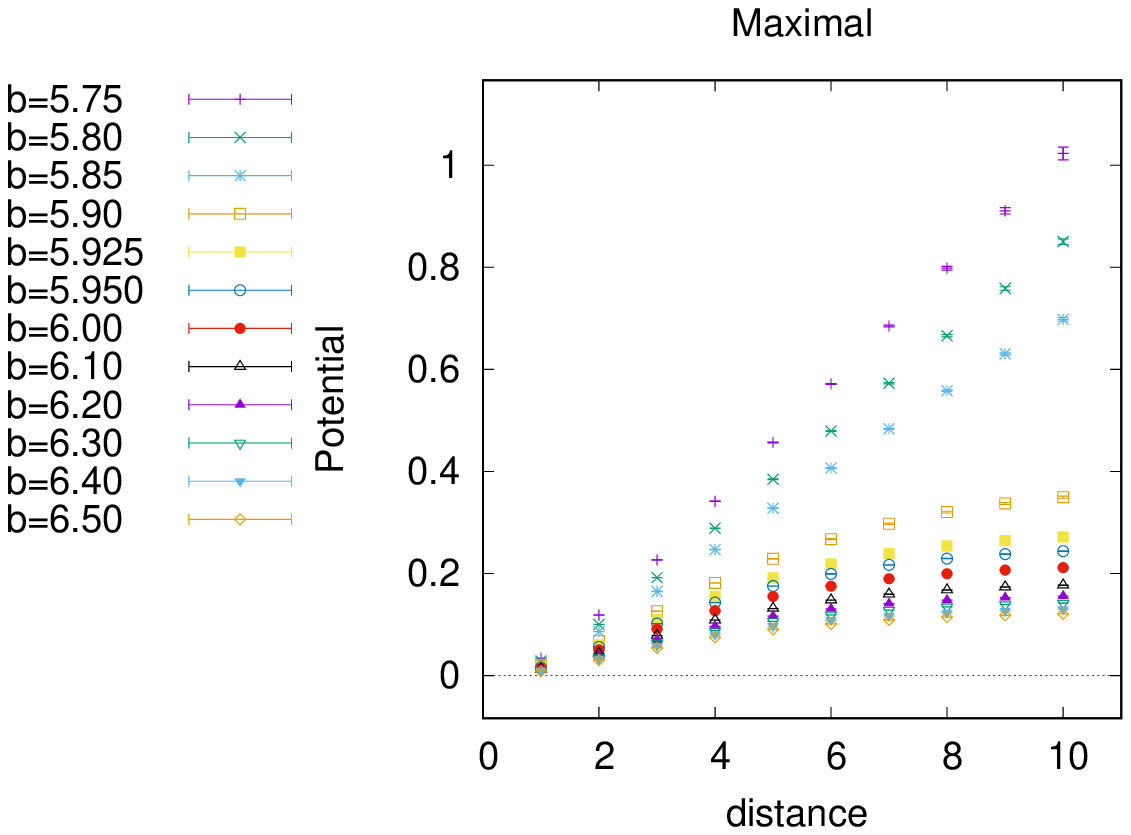}%
\caption{
The static quark-antiquark potentials $V$ obtained from the maximally extended Wilson loop (\ref{eq:wloopPotential}) using the data set I in the case of  (upper row) $N_T=6$ and  (lower row) $N_T=8$.
The three panels in each row represent from left to right the original Yang-Mills field, the restricted field in the minimal option, and the restricted field in the maximal option.
}\label{fig:Wpotential-Nt}%
\end{figure*}%
%

\begin{figure*}[tb] \centering
\includegraphics[height=54mm, clip, viewport=0 0 350 252 ]
{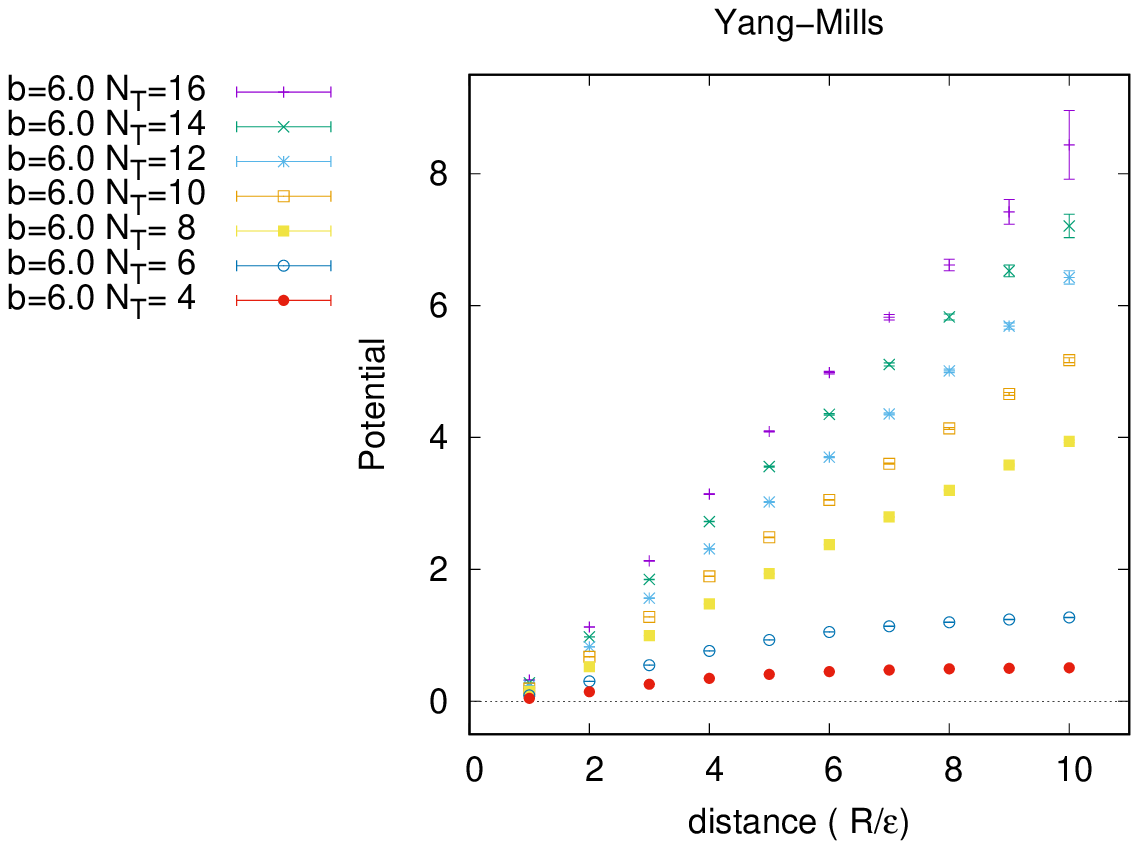}%
\includegraphics[height=53mm,  clip, viewport= 110 0 350 252 ]
{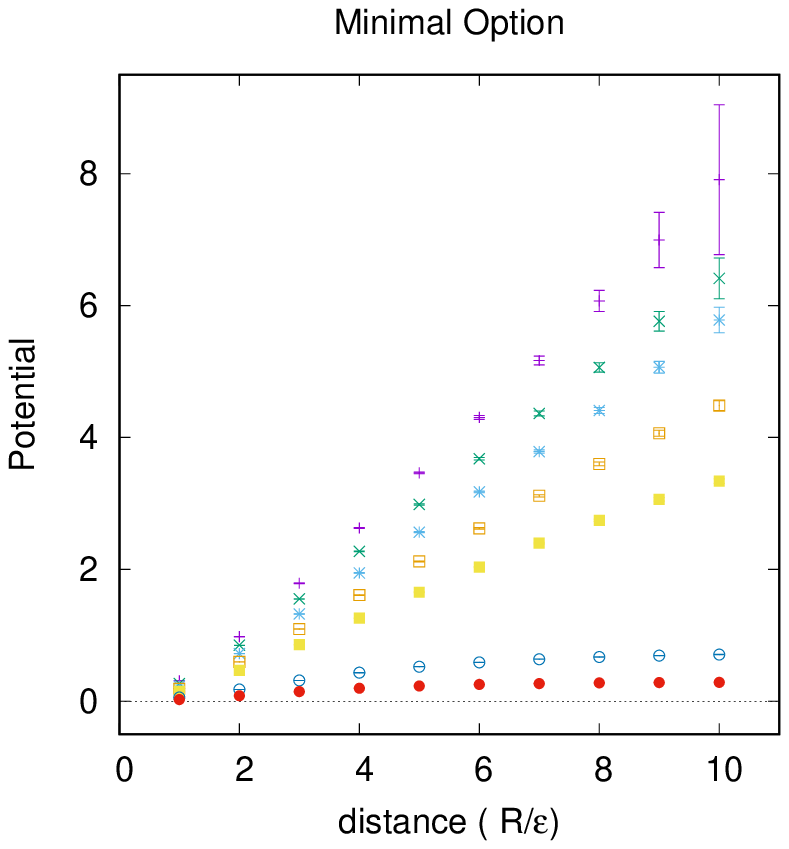}%
\includegraphics[height=54mm, clip, viewport= 110 0 350 252 ]
{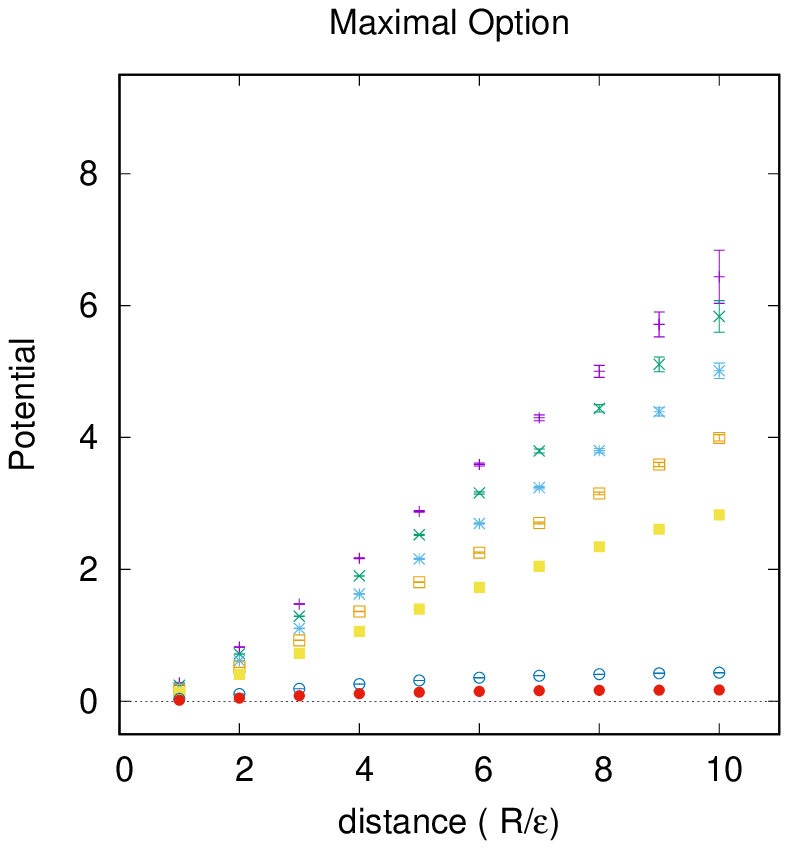}%
\linebreak
%

\includegraphics[height=54mm]{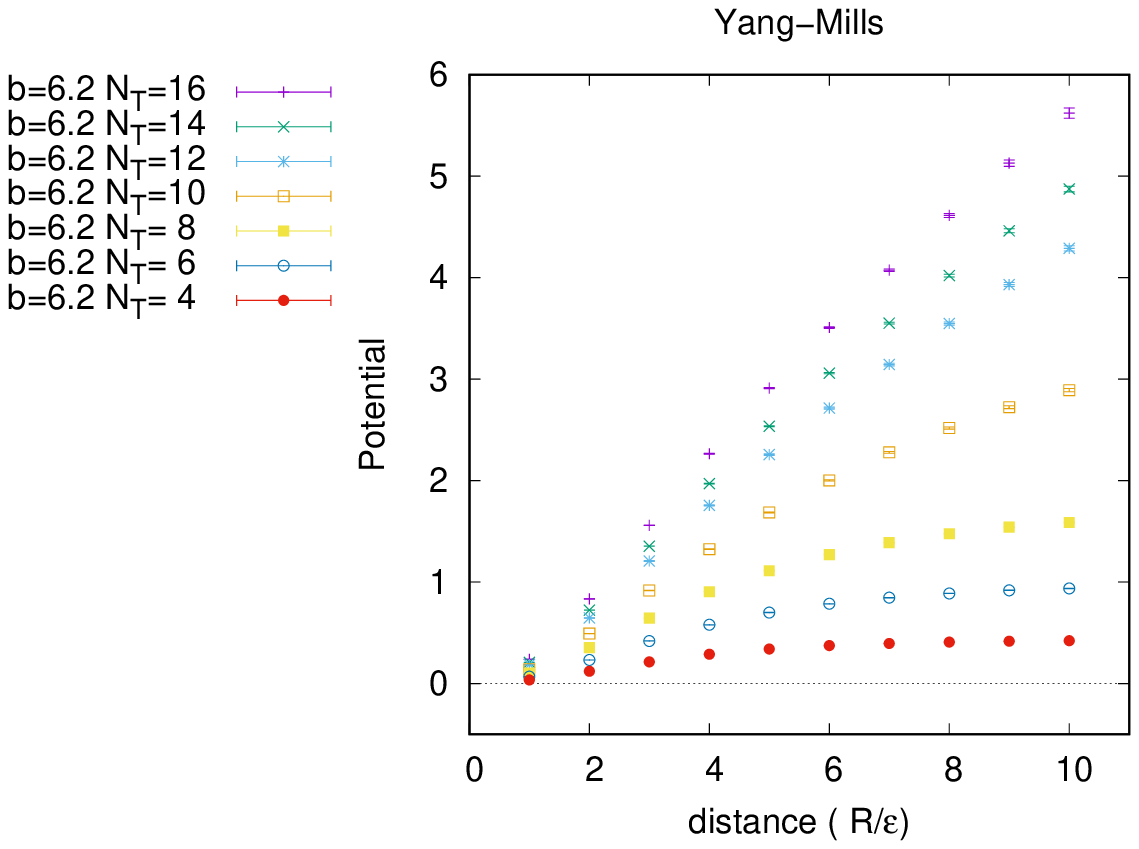}%
\includegraphics
[height=54mm, clip, viewport= 110 0 350 252 ]{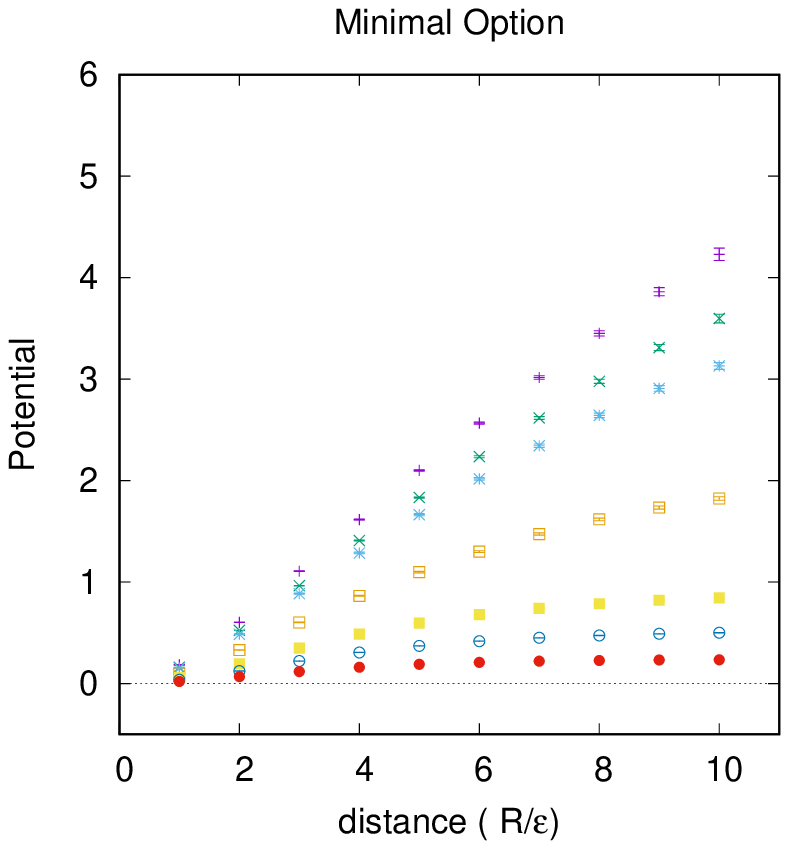}%
\includegraphics
[height=54mm ,clip , viewport= 110 0 350 252 ]{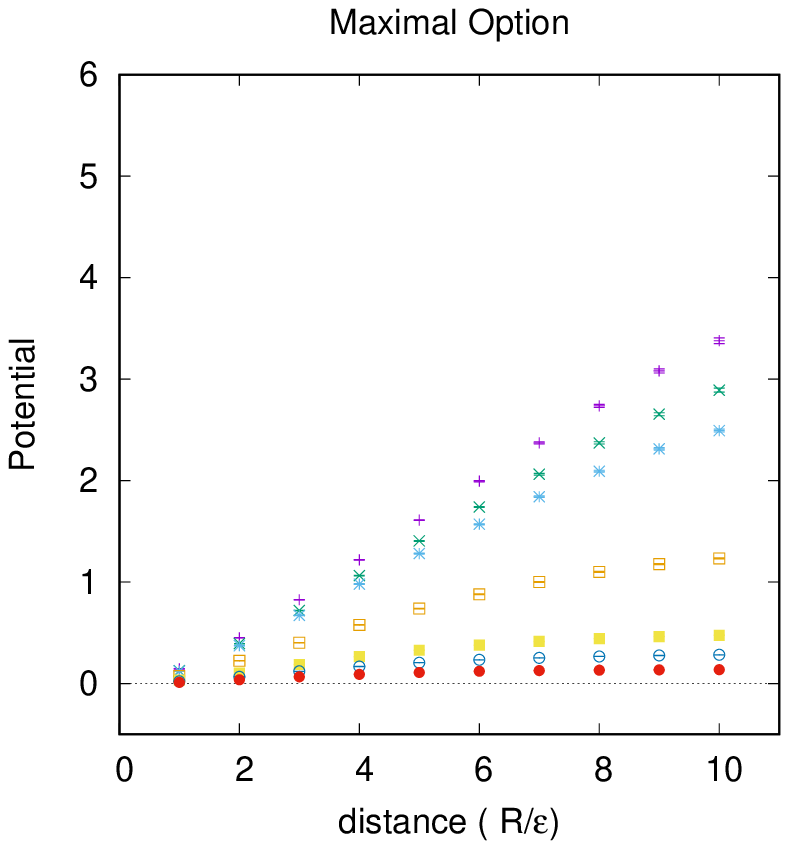}%
\caption{
The static quark-antiquark potentials $V$ obtained from the maximally extended Wilson loop (\ref{eq:wloopPotential}) using the data set  II in the case of (upper row) $\beta=6.0$ and  (lower row) $\beta=6.2$.
The three panels in each row represent from left to right  the original Yang-Mills field, the restricted field in the minimal option, and the restricted field in the maximal option.
} \label{fig:Wpotential-b}%
\end{figure*}%

Second, we investigate the static quark--antiquark potential at finite
temperature. To obtain the static potential at finite temperature, we adopt
the Wilson loop operator defined for the rectangular loop $C$ with the spatial
length $R$ and the temporal length $\tau$ which is maximally extended in the
temporal direction, i.e., $\tau=1/T$. According to the standard argument, for
large $\tau$ i.e., small $T$, the static potential is obtained from the
original gauge field $U$ and the restricted field $V$:
\begin{subequations}
\label{eq:wloopPotential}%
\begin{align}
V(R;U) &  :=-\frac{1}{\tau}\log\left\langle \left\langle W_{U}\right\rangle
\right\rangle =-T\log\left\langle \left\langle W_{U}\right\rangle
\right\rangle ,\quad\label{eq:wloopPotential-U}\\
V(R;V) &  :=-\frac{1}{\tau}\log\left\langle \left\langle W_{V}\right\rangle
\right\rangle =-T\log\left\langle \left\langle W_{V}\right\rangle
\right\rangle .\label{eq:wloopPotential-V}%
\end{align}
In what follows, the symbol $\left\langle \left\langle \mathcal{O}%
\right\rangle \right\rangle $ denotes the average of the operator
$\mathcal{O}$ over the space-time and \ ensemble of the configurations.

According to the imaginary time formalism of the quantum field theory at
finite temperature $T$, the temporal (imaginary time) direction $\tau$ of the
space-time has a finite length of extension $1/T$. Therefore, if the
rectangular closed loop is maximally extended in the temporal direction, then
it eventually reaches the boundaries at $\tau=0$ and $\tau=1/T$ which are to
be identified by the periodic boundary condition. See the left and middle
panels of FIG.~\ref{fig:Op-potential}.

This definition (\ref{eq:wloopPotential}) of the static potential is based on
an analogy with that of the Wilson loop average at zero temperature, and
agrees with the static potential at zero temperature in the limit
$T\rightarrow0$. Therefore, this definition of the static potential at finite
temperature is valid for low temperatures or not-so-high temperatures, but it
will be problematic to use in extremely high temperatures above the critical temperature.

FIG.\ref{fig:Wpotential-Nt} and FIG. \ref{fig:Wpotential-b} show the static
potentials at various temperatures calculated according to the definition
(\ref{eq:ploopPotential}). The panels in the upper and lower rows of FIG.
\ref{fig:Wpotential-Nt} show the potentials calculated respectively from
configurations with $N_{t}=6$ and $N_{t}=8$ of the data set I . While, the
panels in the upper and lower rows of FIG. \ref{fig:Wpotential-b} show the
results calculated respectively from configurations with $\beta=6.0$ and
$\beta=6.2$ of the data set II . The three static potentials are obtained by
the original gauge field (Yang-Mills field), the restricted fields
($V$-fields) in the minimal and maximal options. By comparing these results,
we find that the restricted fields in both options reproduce the string
tension of the original Yang-Mills field. Therefore, we have shown the
restricted field ($V$-field) dominance in the string tension for both options
at finite temperature.%

\begin{figure*}[tb] \centering
\includegraphics[width=\textwidth]{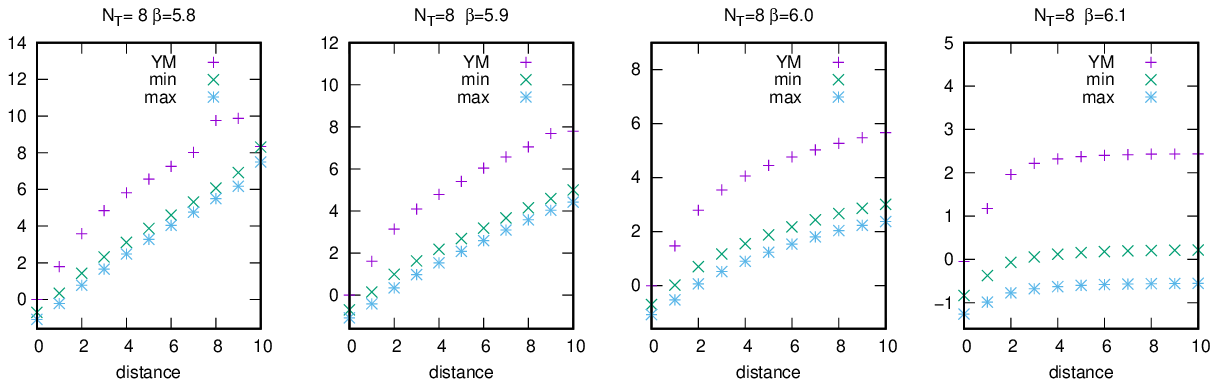}
\linebreak
\includegraphics[width=\textwidth] {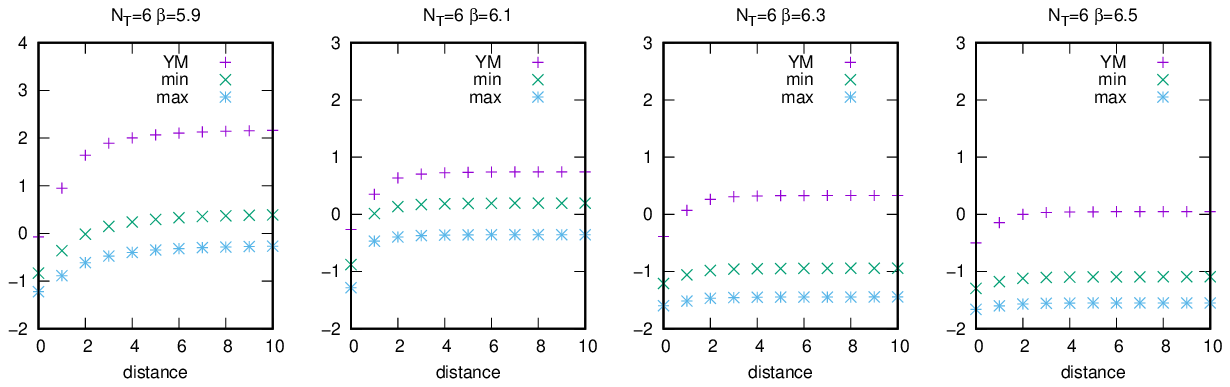}%
\caption{
The static quark-antiquark potentials $\tilde V$ obtained from the Polyakov loop correlator using the data set II in the case of  (upper row) $N_T=6$ for a low temperature $T< T_c$, and (lower row) $N_T=8$ for a high temperature $T> T_c$.
The three potentials in each panel represent from top to bottom the original Yang-Mills field, the restricted field in the minimal option, and the restricted field in the maximal option.
}\label{fig:PP-corretation}%
\end{figure*}%

The Wilson loop maximally extended in the temporal direction looks like a pair
of the Polyakov loop $P(\vec{x})$ and the anti-Polyakov loop $P(\vec{x}%
)^{\ast}$ (the Polyakov loop defined by the path-ordering in the reverse
direction, i.e., the complex conjugate of the Polyakov loop), see the right
panel of FIG~\ref{fig:Op-potential}. However, it should be remarked that the
potential obtained from the maximally extended Wilson loop is different from
the \textquotedblleft potential\textquotedblright\ $\tilde{V}(R;U)$ defined
from the correlation function for a pair of the Polyakov loop $P(\vec{x})$ and
anti-Polyakov loop $P^{\ast}(\vec{y})$ separated by the spatial distance
$R=\left\Vert \vec{x}-\vec{y}\right\Vert $ as
\end{subequations}
\begin{subequations}
\label{eq:ploopPotential}%
\begin{align}
\tilde{V}(R;U) &  :=-T\log\left\langle \left\langle P_{U}(\vec{x})P_{U}^{\ast
}(\vec{y})\right\rangle \right\rangle ,\label{eq:ploopPotential-U}\\
\quad\tilde{V}(R;V) &  :=-T\log\left\langle \left\langle P_{V}(\vec{x}%
)P_{V}^{\ast}(\vec{y})\right\rangle \right\rangle ,\label{eq:ploopPotential-V}%
\end{align}
where each Polyakov loop is defined by the corresponding closed loop obtained
by identifying the end points at $\tau=0$ and $\tau=1/T$ due to the periodic
boundary condition. FIG. \ref{fig:PP-corretation} gives the result for the
measurement of $\tilde{V}$ at various temperatures.

In fact, it is known \cite{YHM05,BW79} that the Polyakov loop correlation
function which is related to the partition function in the presence of a quark
at $\vec{x}$ and an anti-quark at $\vec{y}$ is decomposed into the singlet and
the adjoint combinations in the color space:
\end{subequations}
\begin{align}
&  \left\langle \left\langle P_{U}(\vec{x})P_{U}^{\ast}(\vec{y})\right\rangle
\right\rangle \simeq e^{-F_{q\bar{q}}/T}\nonumber\\
&  =\frac{1}{N_{c}^{2}}e^{-F^{(S)}/T}+\frac{N_{c}^{2}-1}{N_{c}^{2}}%
e^{-F^{(A)}/T}, \label{PP}%
\end{align}
where $F^{(S)}$ and $F^{(A)}$ are the free energies in the singlet channel and
the adjoint channel respectively. The left-hand side of (\ref{PP}) is
gauge-invariant for arbitrary $\vec{x}$ and $\vec{y}$ by construction.
However, the decomposition in the right-hand side of (\ref{PP}) is
gauge-dependent and thus should be taken with care. In sharp contrast to
$\tilde{V}$, the potential $V$ obtained from the Wilson loop is color singlet
and gauge-independent object.

\subsection{Chromo-flux tube at finite temperature}%

\begin{figure*}[htb] \centering
\includegraphics[height=50mm]
{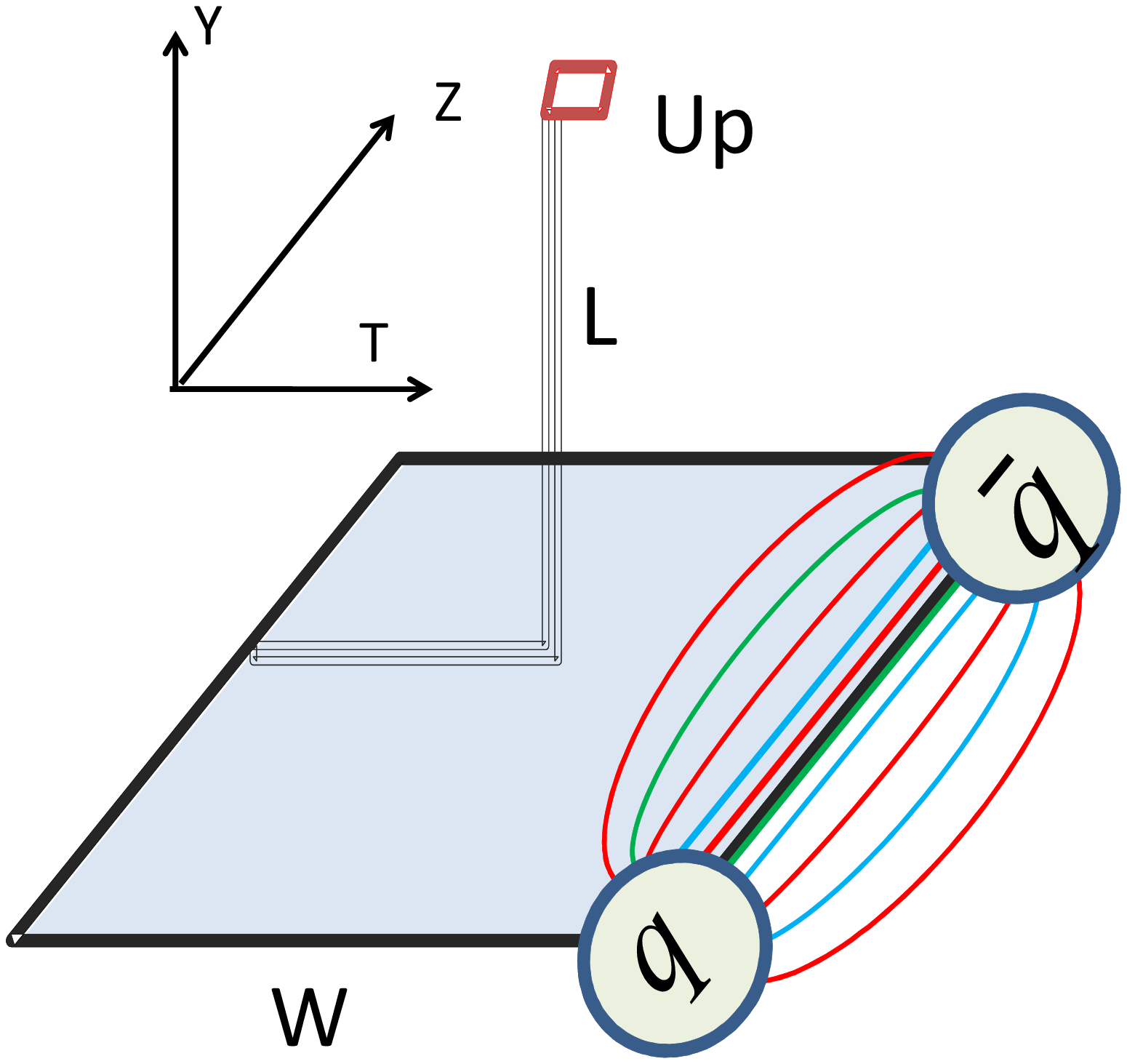} \ \includegraphics[height=40mm]
{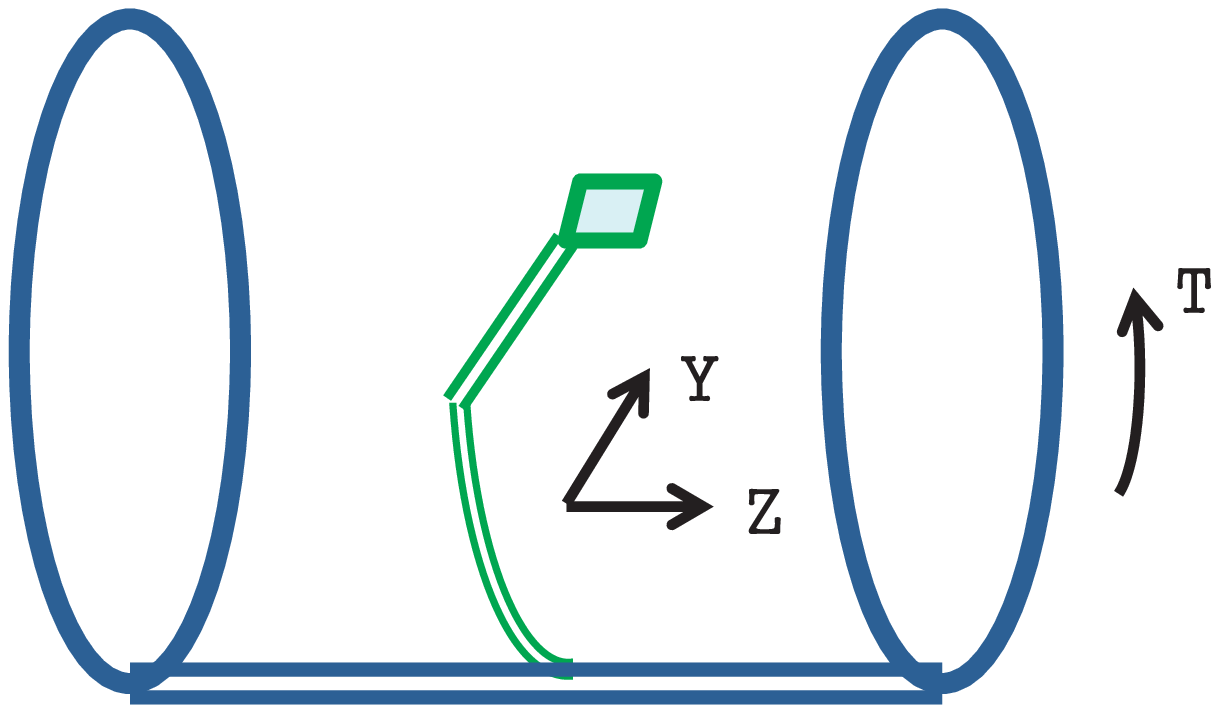}
\caption{ The set up for the measurement of the chromo-fluxes.
(Left)  The graphical representation of the operator
$\text{tr}\left( WLU_{p}L^\dagger  \right )$.
(Right) The operator $\text{tr}\left( WLU_{p}L^\dagger  \right )$ with  the maximally extended Wilson loop $W$.
} \label{fig:Op-flux}%
\end{figure*}%
%

\begin{figure*}[tbp] \centering
\includegraphics[height=45mm, clip, viewport= 80 0 318 242 ]
{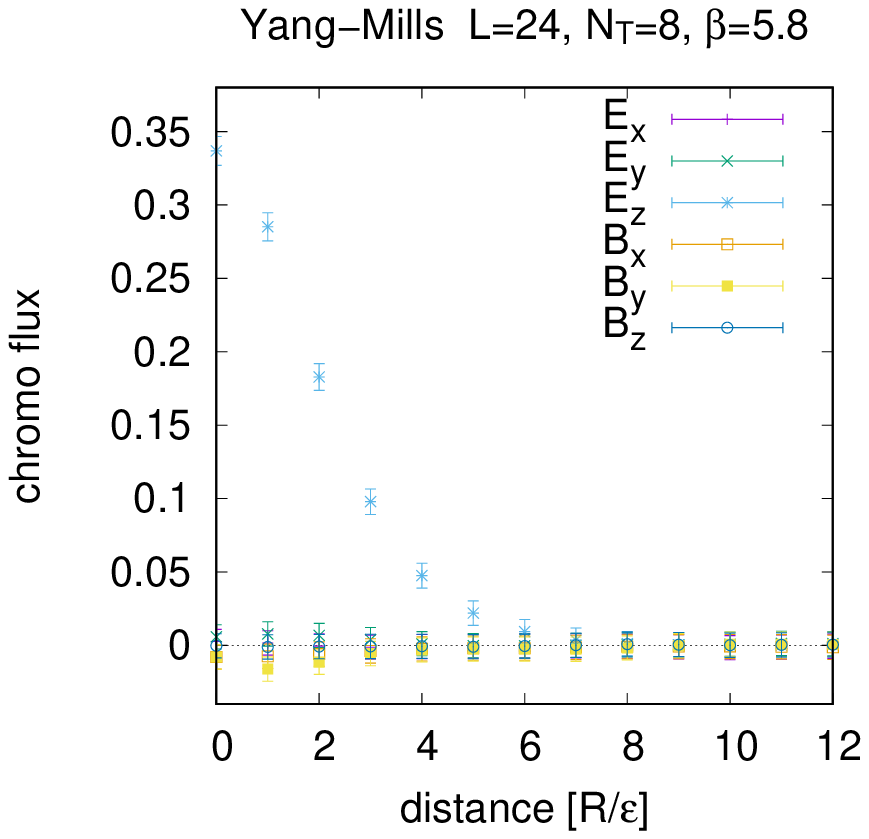}\includegraphics[height=45mm ,clip, viewport=80 0 318 242 ]
{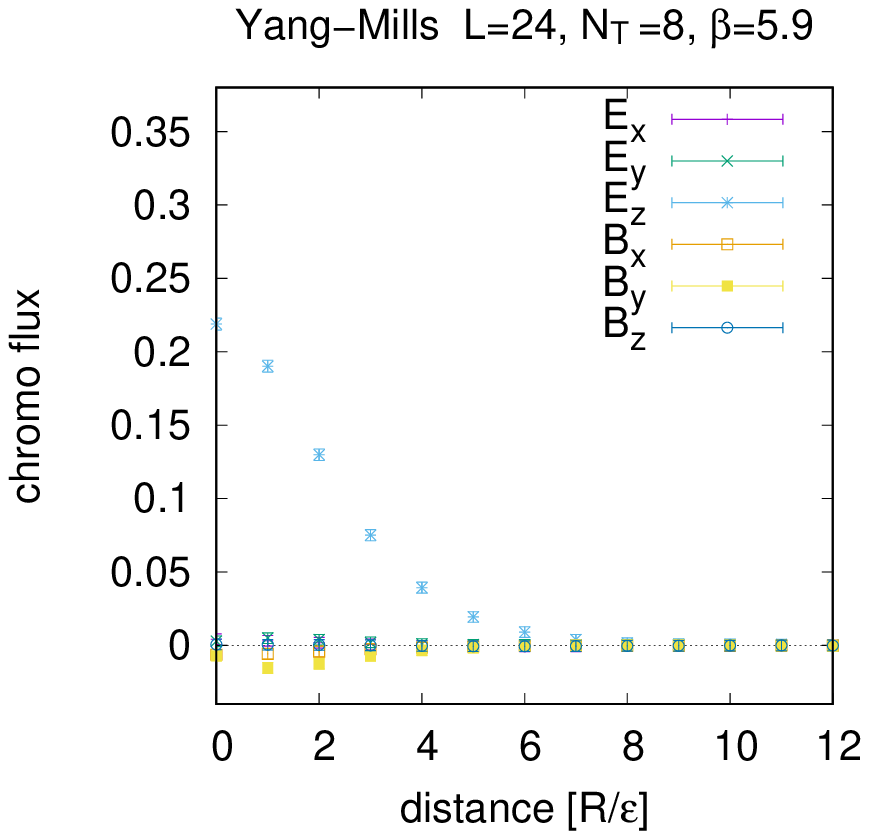}\includegraphics[height=45mm , clip, viewport=80 0 318 242 ]
{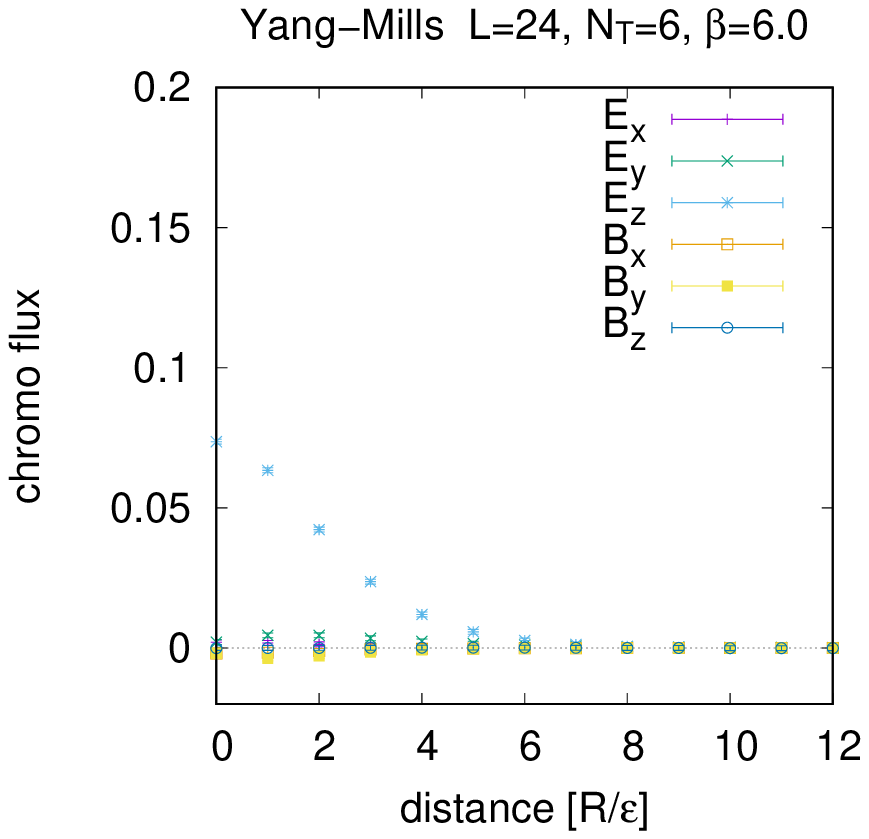}\includegraphics[height=45mm, clip, viewport= 80 0 318 242 ]
{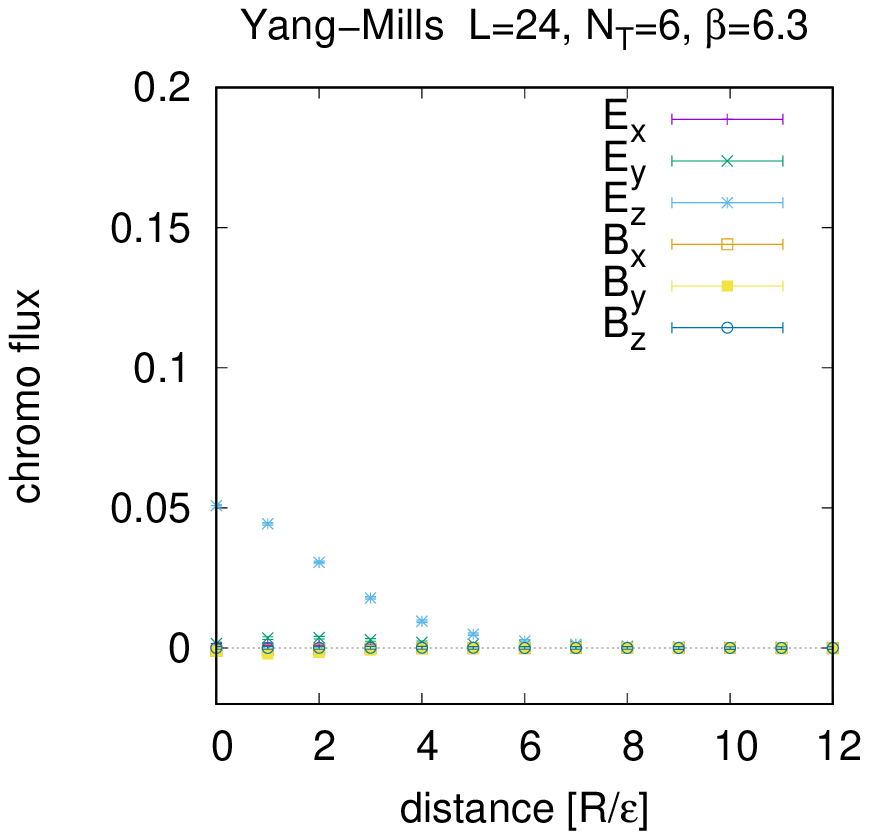}
\linebreak
\includegraphics[height=45mm , clip, viewport=80 0 318 242 ]
{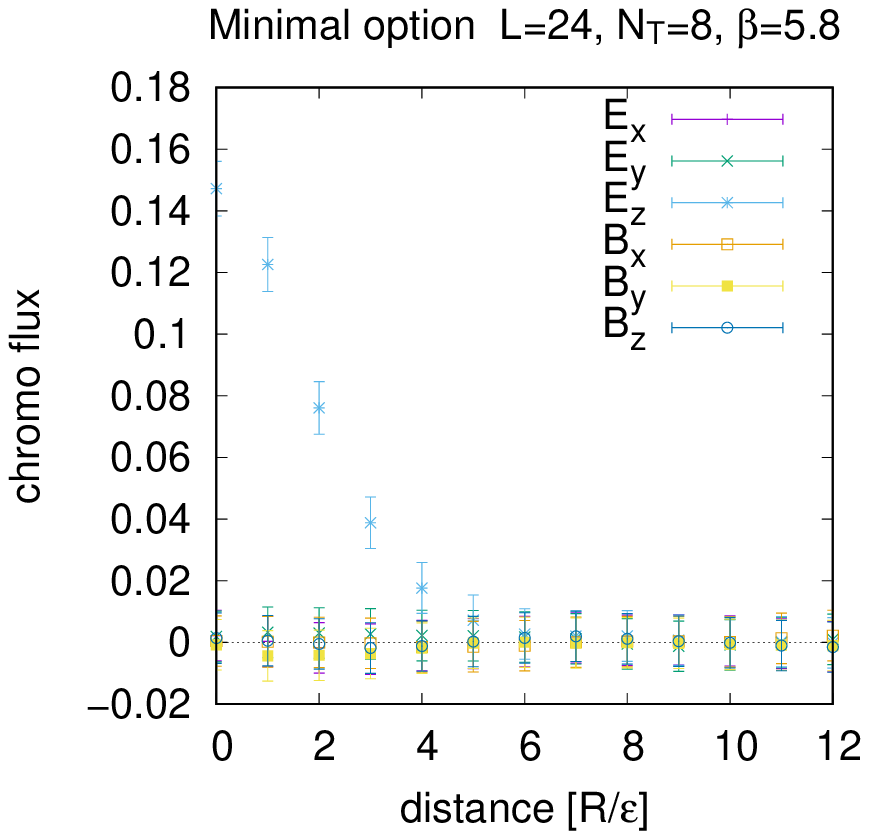}\includegraphics[ height=45mm , clip, viewport= 80 0 318 242]
{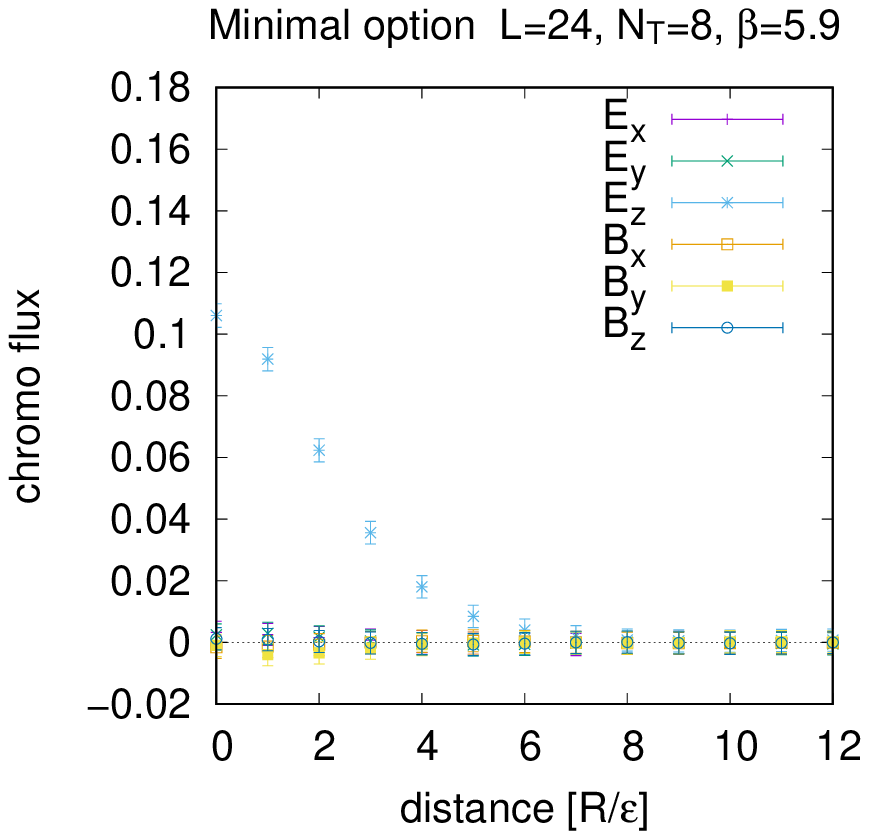}\includegraphics[height=45mm, clip, viewport=80 0 318 242 ]
{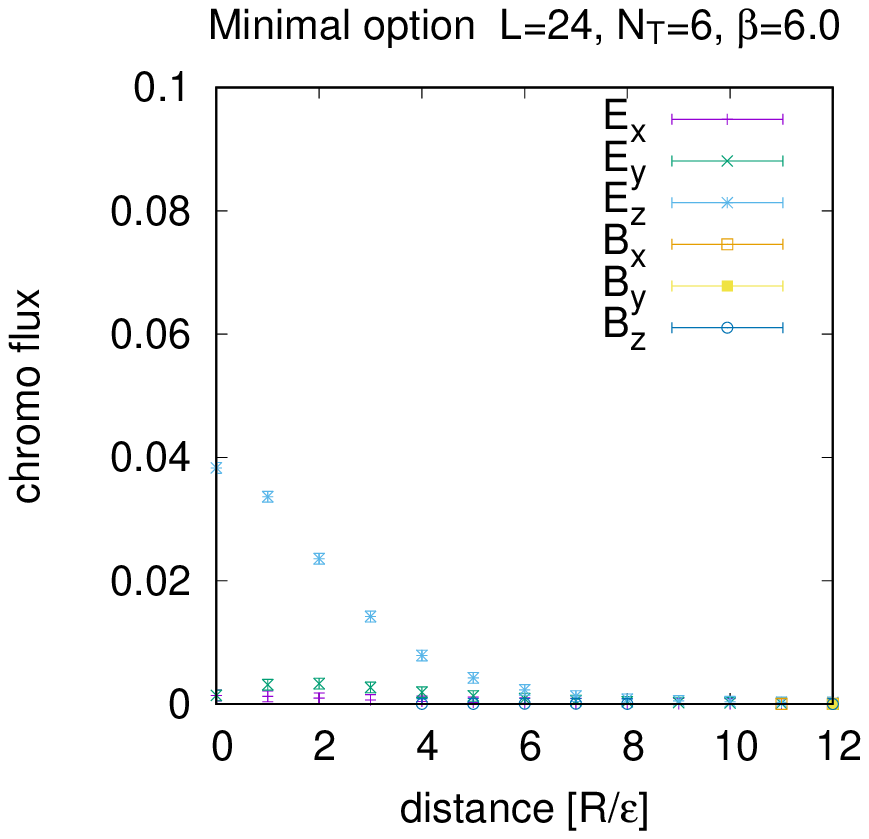}\includegraphics [height=45mm , clip, viewport= 80 0 318 242 ]
{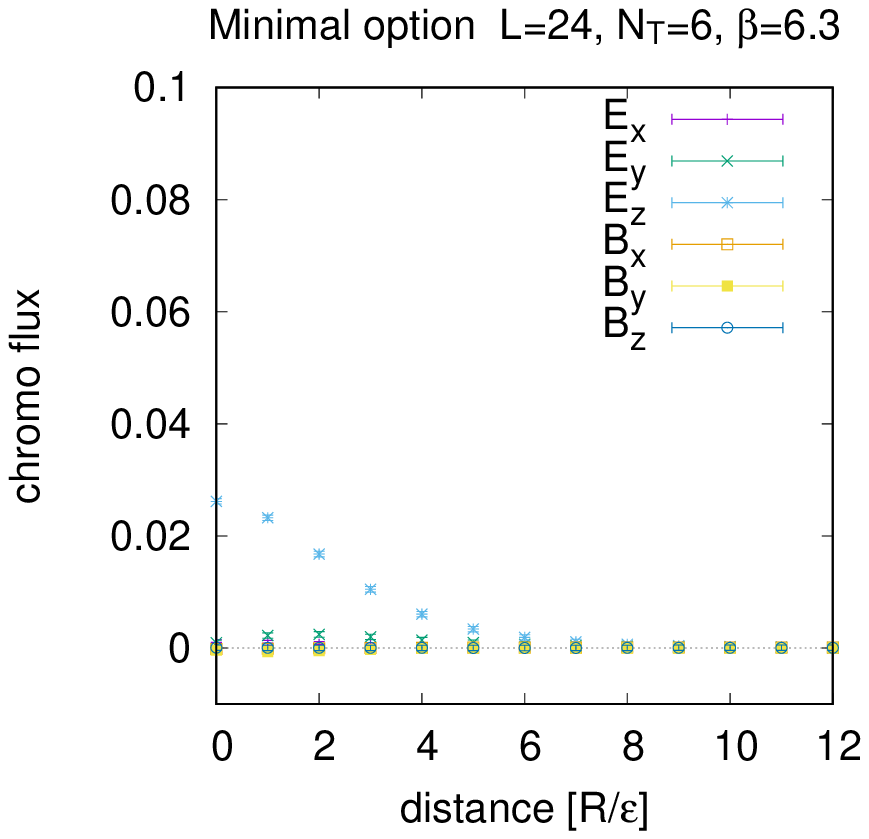}
\linebreak
\includegraphics[height=45mm, clip, viewport=80 0 320 242 ]
{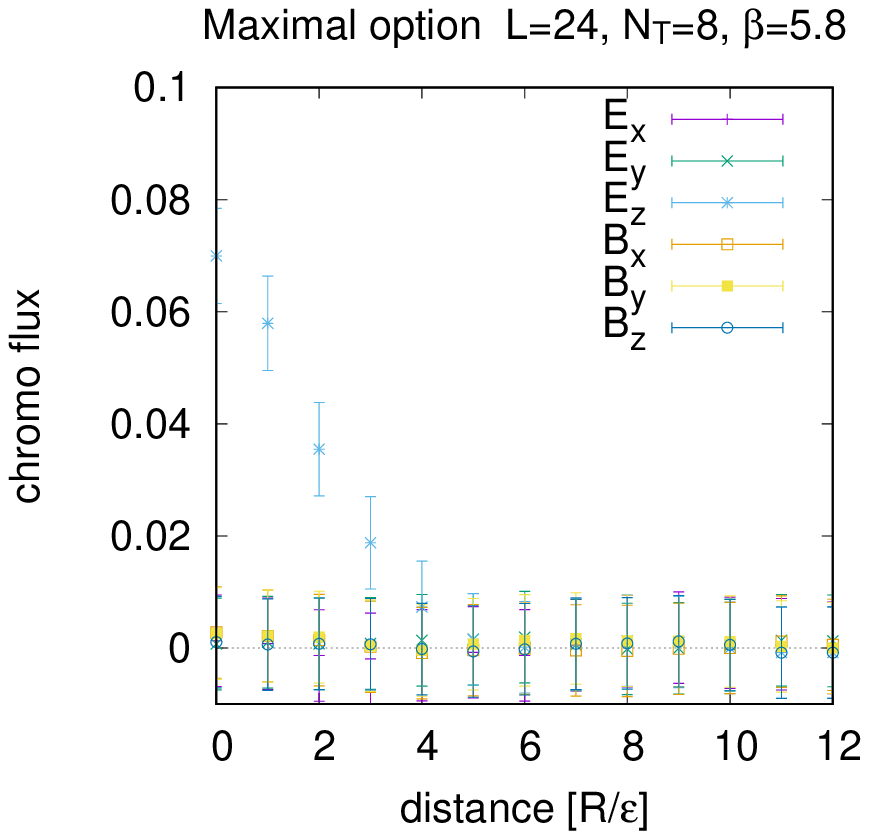}\includegraphics[height=45mm, clip, viewport= 80 0 320 242 ]
{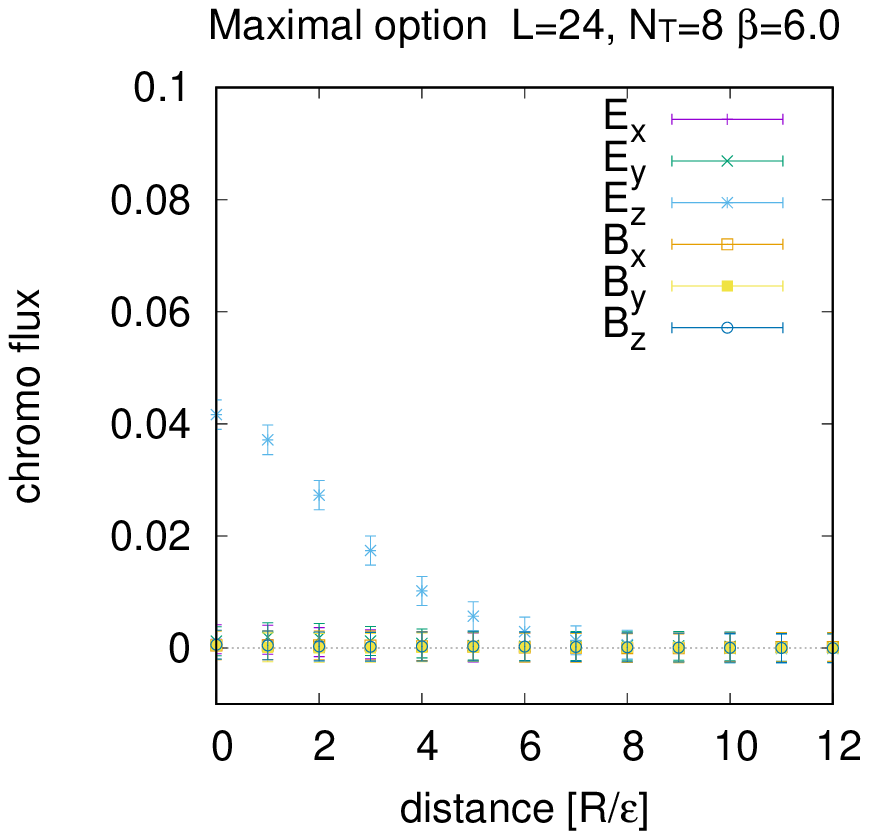}\includegraphics[height=45mm , clip, viewport= 80 0 320 242]
{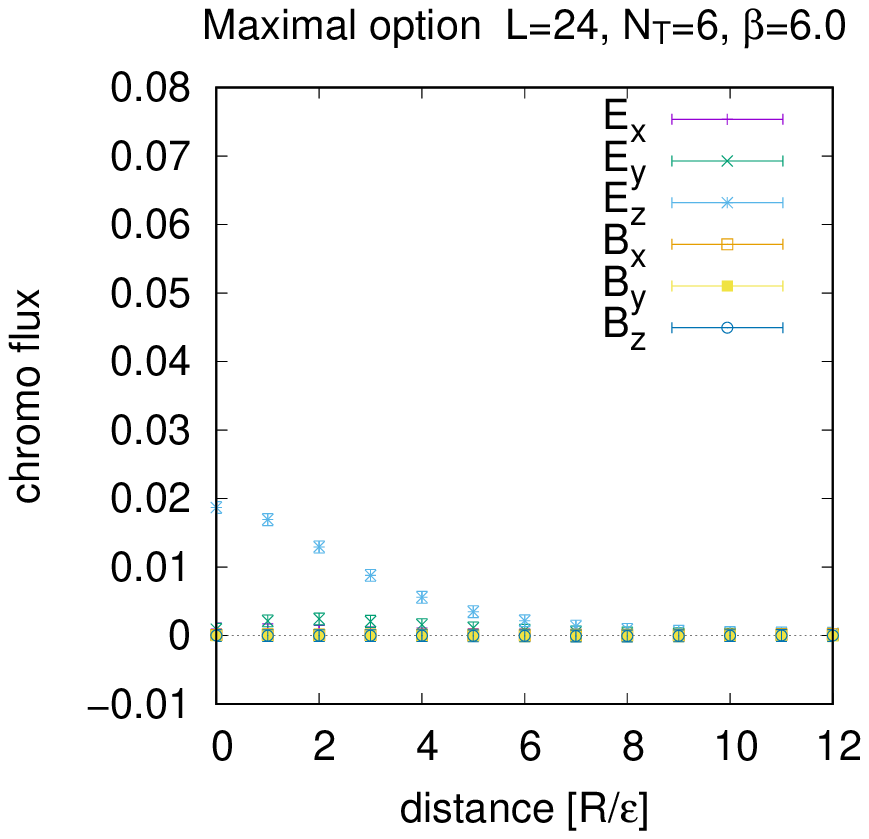}\includegraphics [ height=45mm, clip , viewport= 80 0 320 242 ]
{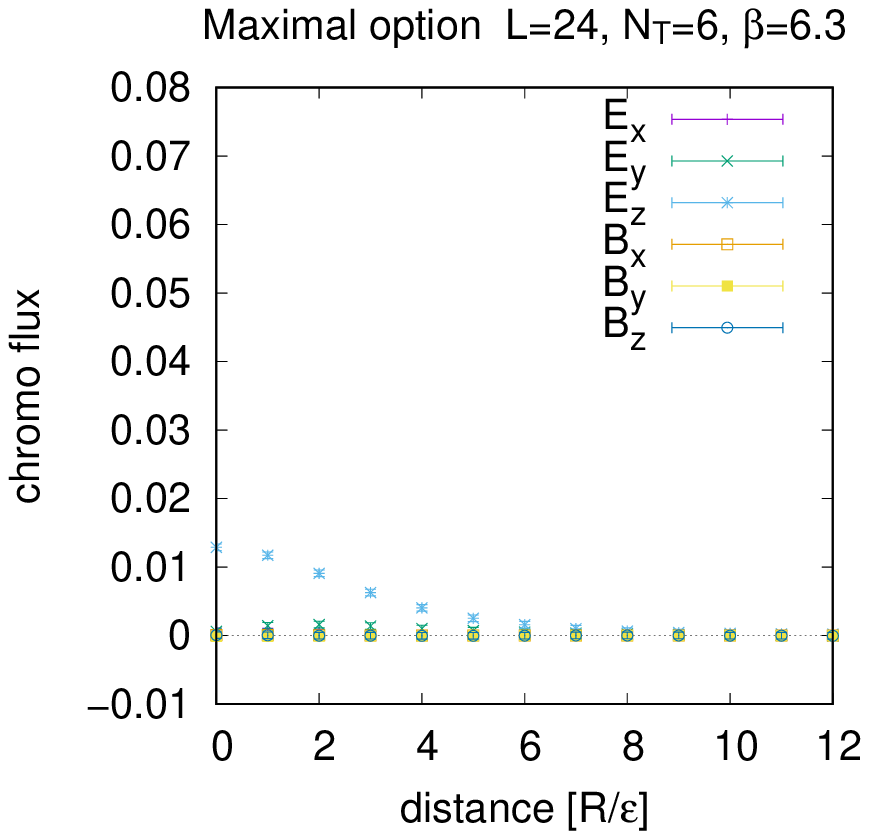} \caption{
The components of the chromo flux obtained from the data set I:
$E_x=F^{q\bar{q}}_{14}$, $E_y=F^{q\bar{q}}_{24}$, $E_z=F^{q\bar{q}}_{34}$,
$B_x=F^{q\bar{q}}_{23}$, $B_y=F^{q\bar{q}}_{31}$, $B_z=F^{q\bar{q}}_{12}$.
The three panels in the same column correspond to the different options: (top row) the original Yang-Mills field,
(mid row) the restricted field in the minimal option,
and (bottom row) the restricted field in the maximal option.
In the four panels in each row, the left two  are  at low temperatures in the confinement phase $T < T_c$, while  the right two are at high temperatures in the deconfinement phase $T_c <T$.
} \label{fig:flux-phase}%
\end{figure*}%

We proceed to investigate the dual Meissner effect at finite temperature. For
this purpose, we use the gauge-invariant correlation function which is the
same as that used at zero temperature \cite{GMO90}. We measure the chromo-flux
created by a quark-antiquark pair, which is represented by the maximally
extended Wilson loop $W$ given in the center panel of
FIG.\ref{fig:Op-potential}. The chromo-field strengths, i.e., the field
strengths of the chromo-flux created by the Wilson loop of the original
Yang-Mills field, the restricted field in the minimal and maximal options are
measured by using the three kind of the probes defined by
\begin{subequations}
\label{eq:Op-flux-YM}%
\begin{align}
F_{\mu\nu}^{\text{YM}}(y)  &  =\sqrt{\frac{\beta}{6}}\rho_{_{U_{P}}}(y)\text{,
\ }\label{eq:Op-flux-YM0}\\
\rho_{_{U_{P}}}^{\text{YM}}(y)  &  :=\frac{\left\langle \left\langle
\mathrm{tr}\left(  WL[U]U_{p}(y)L^{\dag}[U]\right)  \right\rangle
\right\rangle }{\left\langle \left\langle \mathrm{tr}\left(  W\right)
\right\rangle \right\rangle }\nonumber\\
&  -\frac{1}{N_{c}}\frac{\left\langle \left\langle \mathrm{tr}\left(
U_{p}(y)\right)  \mathrm{tr}\left(  W\right)  \right\rangle \right\rangle
}{\left\langle \left\langle \mathrm{tr}\left(  W\right)  \right\rangle
\right\rangle }\text{ }, \label{eq:Op-flux-YM1}%
\end{align}%
\end{subequations}
\begin{subequations}
\begin{align}
\text{\ }F_{\mu\nu}^{\text{min}}(y)  &  =\sqrt{\frac{\beta}{6}}\rho
_{_{V_{P}^{\min}}}(y)\text{, }\label{eq:Op-flux-min0}\\
\rho_{V_{p}}^{\min}(y)  &  :=\frac{\left\langle \left\langle \mathrm{tr}%
\left(  WL[V^{\min}]V_{p}^{\min}(y)L^{\dag}[V^{\min}]\right)  \right\rangle
\right\rangle }{\left\langle \left\langle \mathrm{tr}\left(  W\right)
\right\rangle \right\rangle }\nonumber\\
&  -\frac{1}{N_{c}}\frac{\left\langle \left\langle \mathrm{tr}\left(
V_{p}^{\min}(x)\right)  \mathrm{tr}\left(  W\right)  \right\rangle
\right\rangle }{\left\langle \left\langle \mathrm{tr}\left(  W\right)
\right\rangle \right\rangle }, \label{eq:Op-flux-min1}%
\end{align}
\end{subequations}
\begin{subequations}
\label{eq:Op-flux-max}%
\begin{align}
\text{\ \ \ }F_{\mu\nu}^{\max}(y)  &  =\sqrt{\frac{\beta}{6}}\rho
_{_{V_{P}^{\max}}}(y)\text{, }\label{eq:OP-flux-max0}\\
\rho_{_{V_{P}}}^{\max}(y)  &  :=\frac{\left\langle \left\langle \mathrm{tr}%
\left(  WL[V^{\max}]V_{p}^{\max}(y)L^{\dag}[V^{\max}]\right)  \right\rangle
\right\rangle }{\left\langle \left\langle \mathrm{tr}\left(  W\right)
\right\rangle \right\rangle }\nonumber\\
&  -\frac{1}{N_{c}}\frac{\left\langle \left\langle \mathrm{tr}\left(  V^{\max
}(x)\right)  \mathrm{tr}\left(  W\right)  \right\rangle \right\rangle
}{\left\langle \left\langle \mathrm{tr}\left(  W\right)  \right\rangle
\right\rangle }, \label{eq:Op-flux-max1}%
\end{align}
where $U_{p}$, $V_{p}^{\min}$, and $V_{p}^{\max}$ represent plaquette
variables for the field strength made of the Yang-Mills fields $U$, the
restricted fields in the minimal and maximal options, $V^{\min}$ and $V^{\max
}$, respectively, and $L$ is the Schwinger line connecting the source $W$ and
the probes, $U_{p}$ ,$\ V_{p}^{\min}$ , and $V_{p}^{\max}$, which is
introduced to guarantee the gauge-invariance. FIG.\ref{fig:Op-flux} shows the
graphical representation of the connected correlation Eqs.(\ref{eq:Op-flux-YM}%
)--(\ref{eq:Op-flux-max}). We prepare the maximally extended Wilson loop of
the size $R\times N_{T}$ with $\ R=7$ and $N_{T}$ being the size of the
temporal direction where the Wilson loop is placed at $z-t$ plane (see
FIG.\ref{fig:Op-flux}), and ae quark and an anti-quark are placed along the
temporal direction ($t$-direction) at the distance $R$ in the $z$-direction.
When we introduce the coordinates relative to the Wilson loop, the quark and
antiquark are respectively represented by the segments, $z=0,$ $0\leq t$ $\leq
N_{T}$ and $z=R$, $0\leq t$ $\leq N_{T}$ in the Wilson loop, respectively. The
coordinates $y$ in Eqs.(\ref{eq:Op-flux-YM})--(\ref{eq:Op-flux-max})
represents the amount of shift from the Wilson loop. We measure the chromo
flux at the midpoint of the Wilson loop, $t_{\text{mid}}=$ $N_{T}/2-1$ and
$z_{\text{mid}}=3$, i.e., the Schwinger line $L$ ($L^{\dag}$) is connected at
the point $z=z_{mid}$, $t=0$ in the $z-t$ plane and extended to the center of
the Wilson loop, $z=z_{mid}$, $t=t_{mid}$. The plaquette is displaced along
the $y$-direction from the center of the Wilson loop (the point of the
distance $0$) to the point of the distance $Y$, i.e., ($x=0,y$, $z_{mid}$,
$t_{mid})$, $0\leq y\leq Y\,$.

The three kind of the chromo fluxes for the Yang-Mills fields and the
restricted fields in the minimal and maximal options are distinguished by
using different probes made of the Yang-Mills fields and the restricted fields
in the minimal and maximal options, $L[U]U_{p}L^{\dag}[U]$, $L[V^{\min}%
]V_{p}^{\min}L^{\dag}[V^{\min}]$, and $L[V^{\max}]V_{p}^{\max}L^{\dag}%
[V^{\max}]$, \ respectively. Indeed, in the naive continuum limit, the
connected correlations are reduced to
\end{subequations}
\begin{align}
&  \rho_{_{U_{P}}}^{\text{YM}}\overset{\varepsilon\rightarrow0}{\simeq
}g\varepsilon^{2}\left\langle \mathcal{F}_{\mu\nu}[U]\right\rangle _{q\bar{q}%
}\nonumber\\
&  :=\frac{\left\langle \left\langle \mathrm{tr}\left(  g\varepsilon
^{2}\mathcal{F}_{\mu\nu}[U]L[U]^{\dag}WL[U]\right)  \right\rangle
\right\rangle }{\left\langle \left\langle \mathrm{tr}\left(  W\right)
\right\rangle \right\rangle }+O(\varepsilon^{4}),
\end{align}%
\begin{align}
&  \rho_{_{V_{P}}}^{\min}\overset{\varepsilon\rightarrow0}{\simeq}%
g\varepsilon^{2}\left\langle \mathcal{F}_{\mu\nu}[V^{\min}]\right\rangle
_{q\bar{q}}\nonumber\\
&  :=\frac{\left\langle \left\langle \mathrm{tr}\left(  g\varepsilon
^{2}\mathcal{F}_{\mu\nu}[V^{\min}]L[V^{\min}]^{\dag}WL[V^{\text{min}}]\right)
\right\rangle \right\rangle }{\left\langle \left\langle \mathrm{tr}\left(
W\right)  \right\rangle \right\rangle }\nonumber\\
&  +O(\varepsilon^{4}),
\end{align}%
\begin{align}
&  \rho_{_{V_{P}}}^{\max}\overset{\varepsilon\rightarrow0}{\simeq}%
g\varepsilon^{2}\left\langle \mathcal{F}_{\mu\nu}[V^{\max}]\right\rangle
_{q\bar{q}}\nonumber\\
&  :=\frac{\left\langle \left\langle \mathrm{tr}\left(  g\varepsilon
^{2}\mathcal{F}_{\mu\nu}[V^{\max}]L[V^{\max}]^{\dag}WL[V^{\max}]\right)
\right\rangle \right\rangle }{\left\langle \left\langle \mathrm{tr}\left(
W\right)  \right\rangle \right\rangle }\nonumber\\
&  +O(\varepsilon^{4})\text{ }.
\end{align}

FIG.\ref{fig:flux-phase} exhibits the chromo fluxes measured by using
Eqs.(\ref{eq:Op-flux-YM})--(\ref{eq:Op-flux-max}) for the data set I. Note
again that the physical unit (the lattice spacing) is different $\beta$ by
$\beta$ and we cannot directly compare measured values between the different
lattice, because the temperature varies by changing $\beta$ for the fixed size
of the lattice and the lattice spacing is a function of $\beta$. The three
panels from top to bottom in each column show the option dependence of the
chromo fluxes obtained for the Yang-Mills field, the restricted field in the
minimal option and the restricted field in the maximal options with the
lattice size and the gauge coupling being fixed. The four panels from left to
right in each row show the temperature dependence of the chromo fluxes at
different temperatures in the same option. At a low-temperature in the
confinement phase, $T<T_{c}$, we observe that only the component $E_{z}$ of
the chromoelectric flux in the direction connecting a quark and antiquark pair
is non-vanishing, while the other components take vanishing values, see the
left two panels in each row of FIG. \ref{fig:flux-phase}. At a
high-temperature in the deconfinement phase, $T>T_{c}$, we observe that the
component $E_{z}$ of the chromoelectric flux becomes much smaller as the
temperature increases and that the other components still take vanishing
values, see the right two panels in each row of FIG. \ref{fig:flux-phase}. The
magnitude of the chromoelectric field decreases as the temperature increases,
and rapidly falls off when the temperature exceeds the critical temperature.
Therefore, the flux tube gradually disappears above the critical temperature.

Thus, the results of FIG.\ref{fig:flux-phase} give the numerical evidence for
the disappearance of the dual Meissner effect in the high-temperature
deconfinement phase. These results should be compared with similar analyses
for the Yang-Mills field at finite temperature by using a pair of Polyakov
loops \cite{Cea2016,Cea2018}. Our results are consistent with them.

\subsection{Magnetic--monopole current and dual Meissner effect at finite
temperature}%

\begin{figure}[hbt] \centering
\includegraphics[height=52mm]
{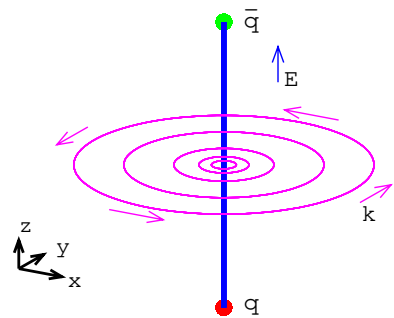}\caption{
The magnetic--monopole current $k$ induced around the chromo-flux tube created by a quark-antiquark pair.
}\label{fig:M_current}%
\end{figure}
\begin{figure*}[tbp] \centering
\includegraphics[height=52mm, clip, viewport= 0 0 350 250  ]{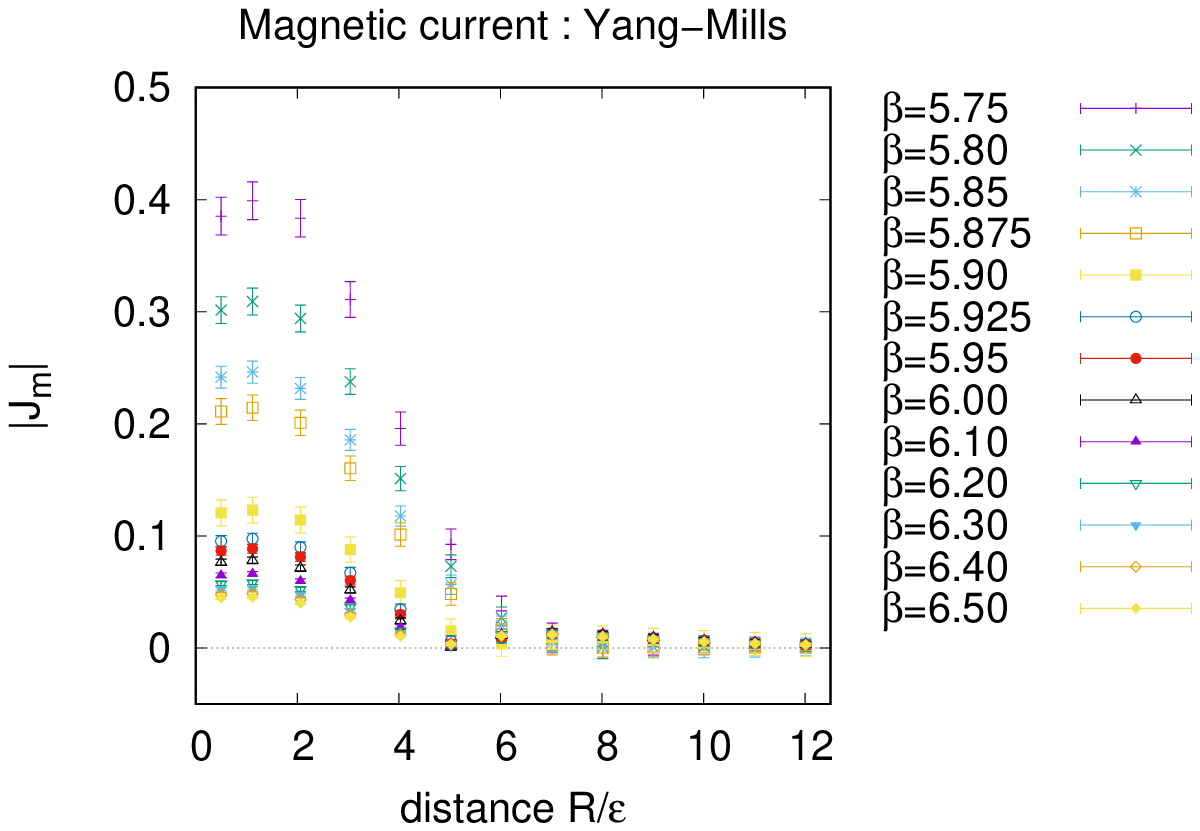}\includegraphics[height=52mm, clip, viewport= 20 0 250 250  ]{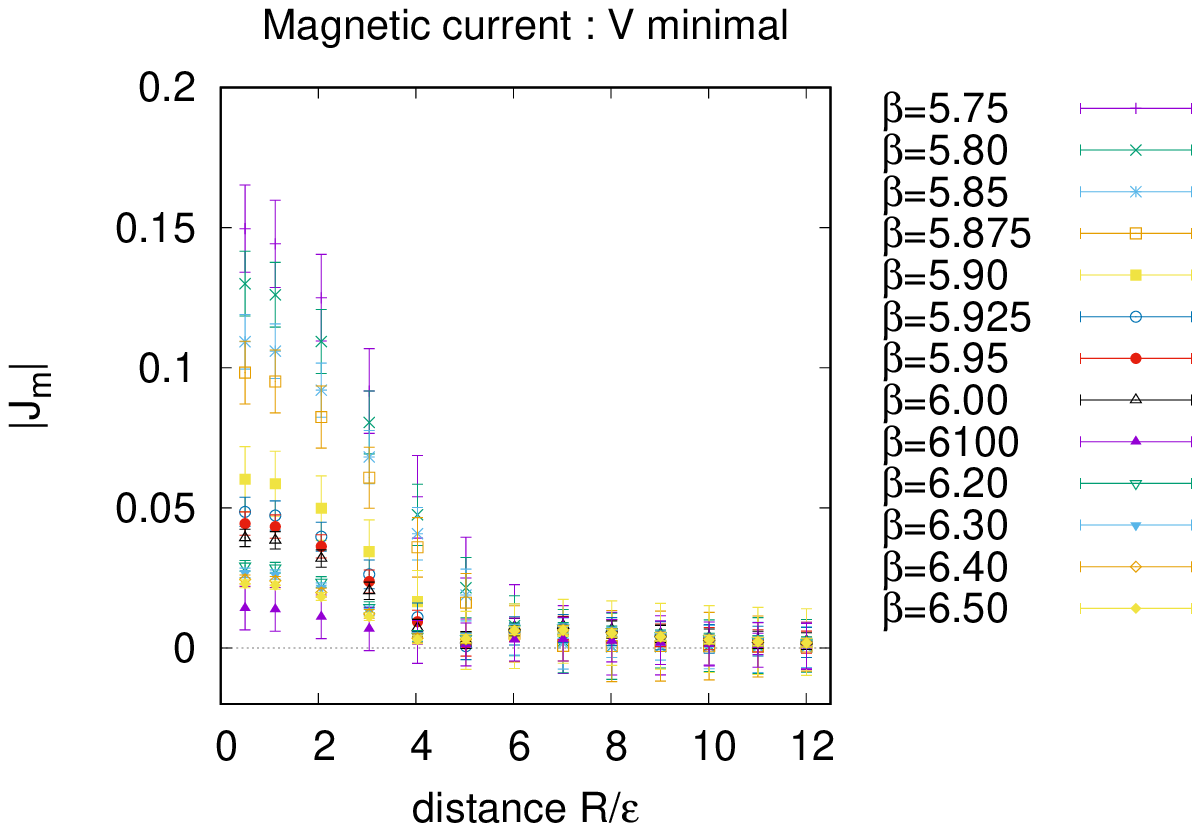}\includegraphics[height=52mm, clip, viewport= 20 0 250 250  ]{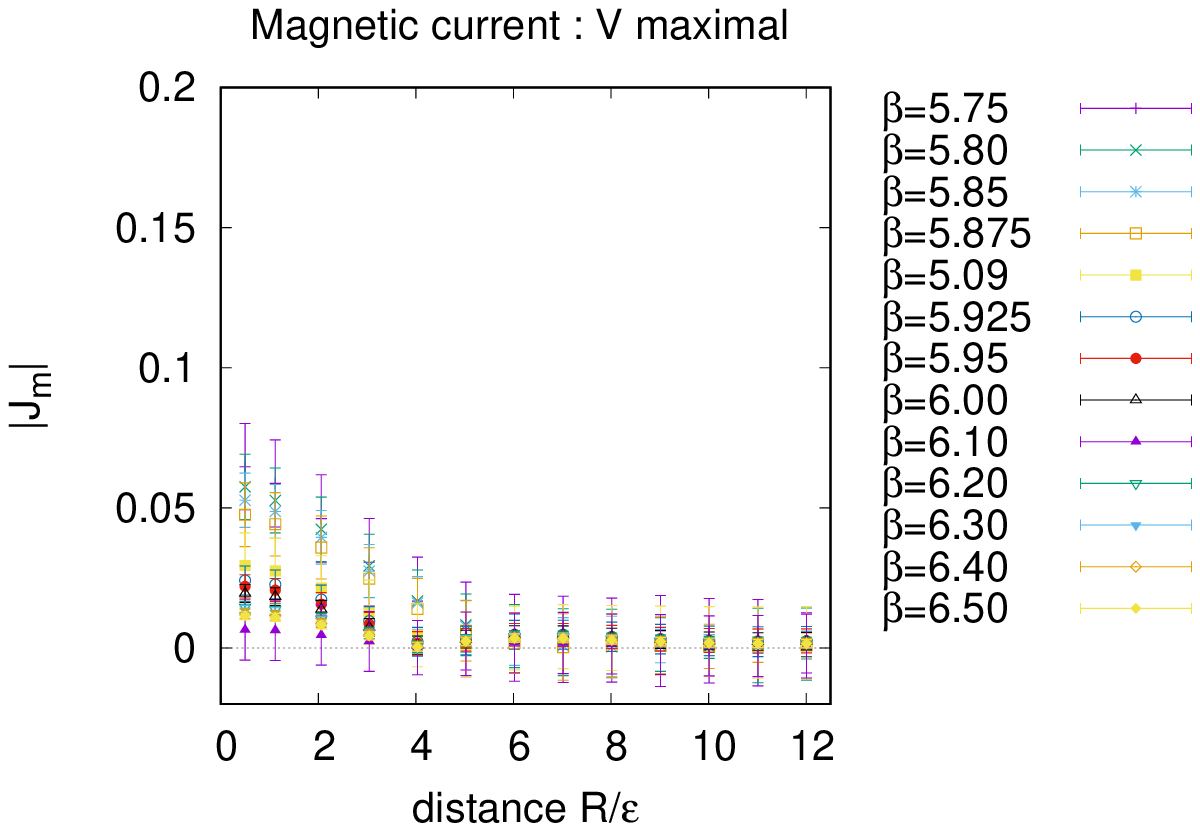}
\linebreak
\includegraphics[height=50mm, clip, viewport= 0 0 350 250  ]{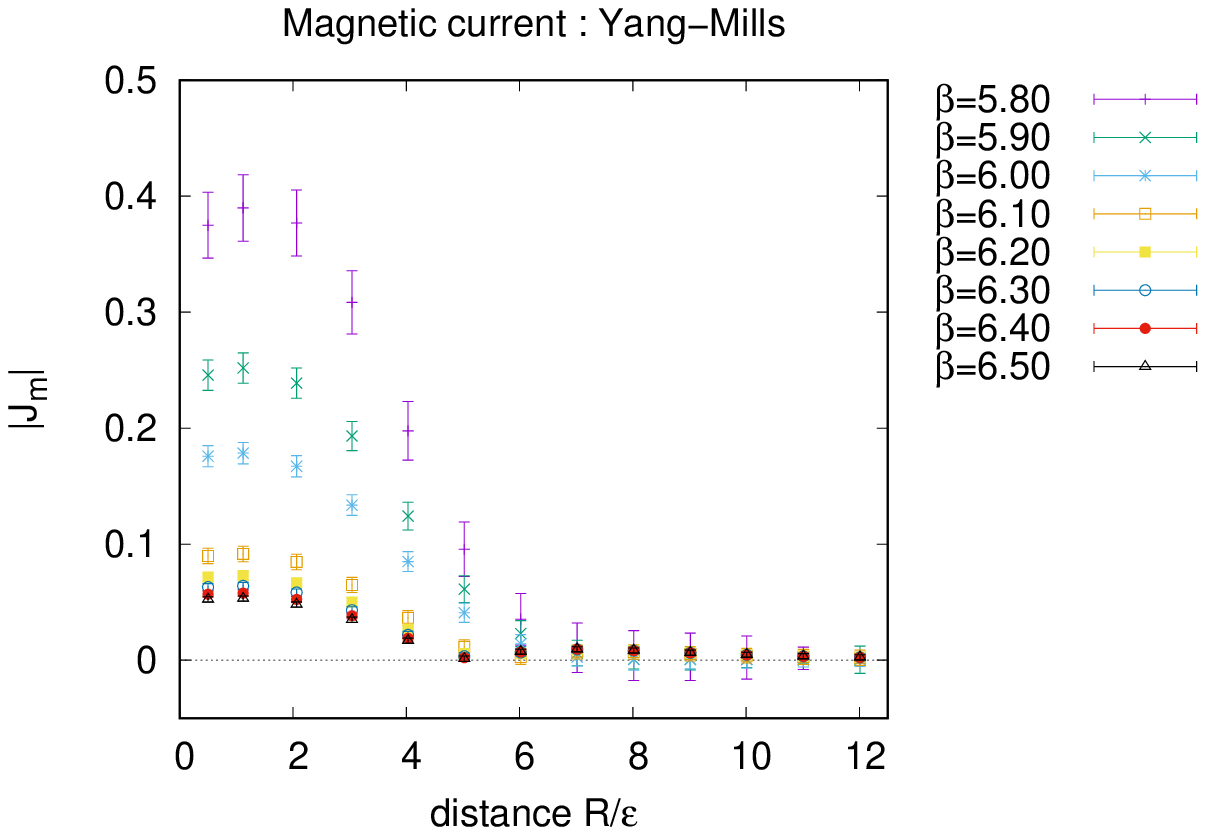}
\includegraphics[height=50mm, clip, viewport= 20 0 260 250  ]{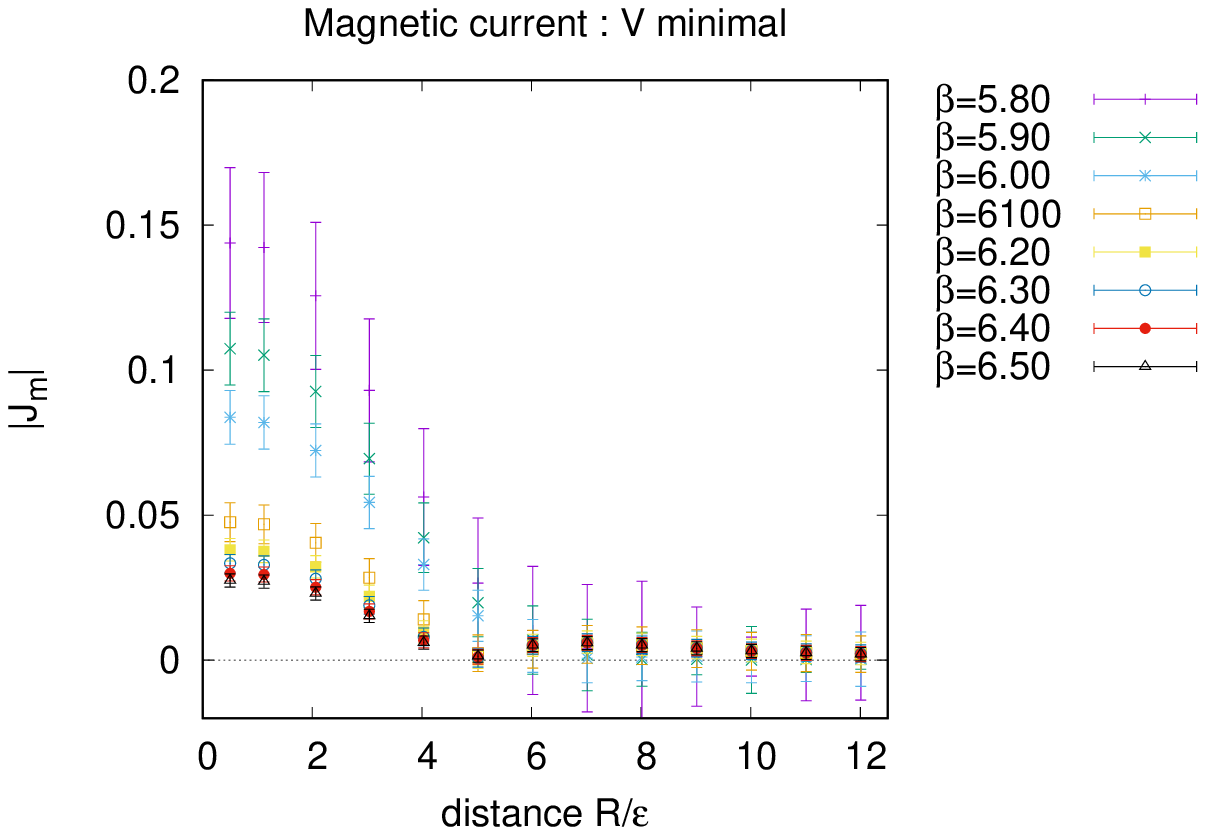}
\includegraphics[height=50mm, clip, viewport= 20 0 260 250 ]{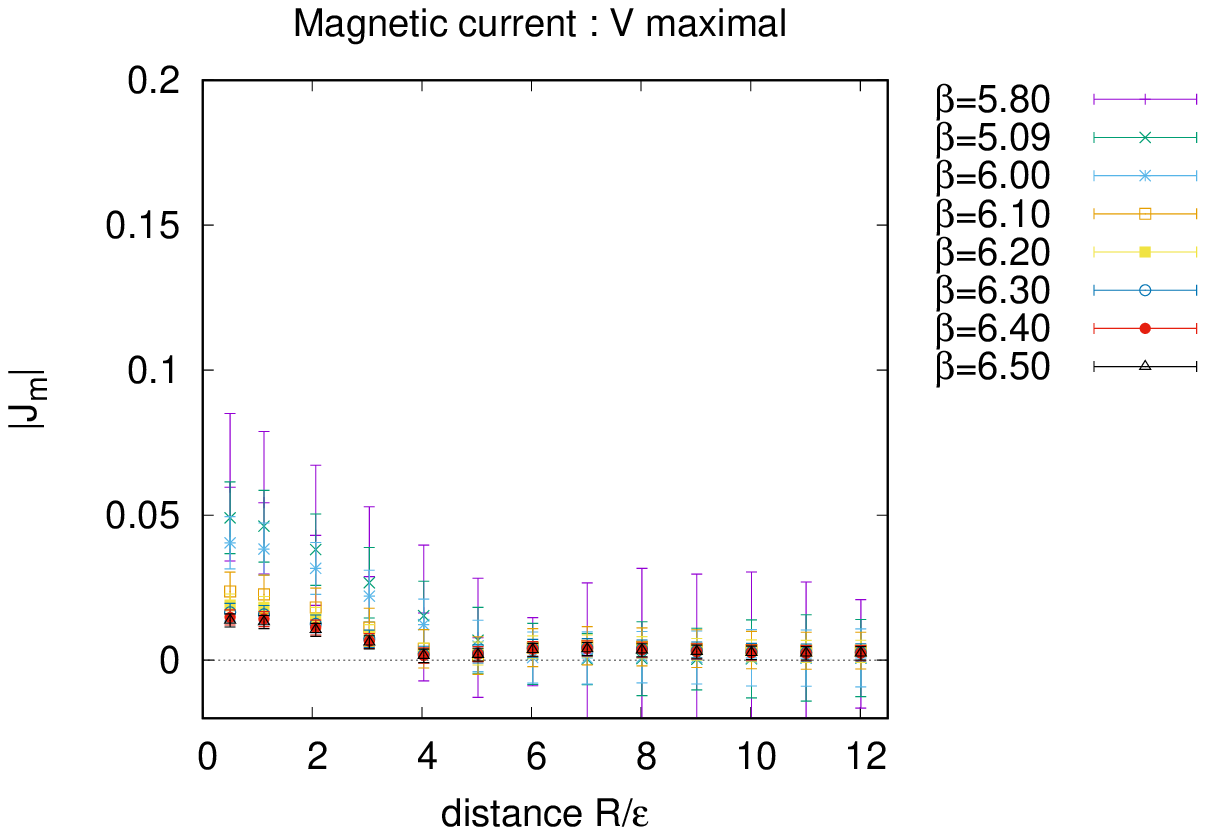}
\caption{
The magnitude $|J_m| = \sqrt{k_\mu k_\mu} $ of the induced magnetic current $k_{\mu}$ ($J_{m,\mu}$)
around the flux tube connecting a quark-antiquark pair as a function of the distance $y$ from the $z$ axis for
various values of $\beta$.
(Upper row) The results for $N_T=6$ from data set I.
(Upper row) The results for $N_T=8$ from data set I.
The three panels in each row represent from left to right  the original Yang-Mills field, the restricted field in the minimal option, and the restricted field in the maximal option.
} \label{fig:m_current}%
\end{figure*}%

Finally, we investigate the dual Meissner effect by measuring the
magnetic--monopole current $k$ induced around the chromo-flux tube created by
a quark-antiquark pair, see FIG.\ref{fig:M_current}. We use the
magnetic--monopole current $k$ defined by%
\begin{subequations}
\begin{align}
k_{\mu}^{\text{YM}}(x)  &  =\frac{1}{2}\varepsilon_{\mu\nu\alpha\beta}\left(
F_{\alpha\beta}^{YM}(x+\hat{\nu})-F_{\alpha\beta}^{YM}(x)\right)
,\label{eq:m_current_ym}\\
k_{\mu}^{\text{min}}(x)  &  =\frac{1}{2}\varepsilon_{\mu\nu\alpha\beta}\left(
F_{\alpha\beta}^{\min}(x+\hat{\nu})-F_{\alpha\beta}^{\min}(x)\right)
,\label{eq:m_current_min}\\
k_{\mu}^{\text{max}}(x)  &  =\frac{1}{2}\varepsilon_{\mu\nu\alpha\beta}\left(
F_{\alpha\beta}^{\max}(x+\hat{\nu})-F_{\alpha\beta}^{\max}(x)\right)  ,
\label{eq:m_current_nax}%
\end{align}
\label{eq:m_current}%
\end{subequations}
where $F_{\mu\nu}^{(\ast)}$ are the field strength\ defined by
Eqs.(\ref{eq:Op-flux-YM})--(\ref{eq:Op-flux-max}). This definitions satisfy
the conserved current, i.e., $\partial_{\mu}k_{\mu}^{(\ast)}(x):=\sum_{\mu
}\left(  k_{\mu}^{(\ast)}(x+\hat{\mu})-k_{\mu}^{(\ast)}(x)\right)  \equiv0$.
There is a general belief that the magnetic--monopole current
(\ref{eq:m_current}) must vanish due to the Bianchi identity as far as there
exist no singularities in the gauge potential $A$. In fact, if the field
strength $F^{(\ast)}$ was written in the exact form $F^{(\ast)}=dA^{(\ast)}$
using the differential form, the magnetic--monopole current would vanish
$k^{(\ast)}:={}^{\ast}dF={}^{\ast}ddA=0$ due to $dd \equiv0$. This is not the
case in the Yang-Mills theory. We show that the magnetic--monopole current
defined in this way (\ref{eq:m_current}) can be the order parameter for the
confinement/deconfinement phase transition, as suggested from the dual
superconductivity hypothesis. FIG.~\ref{fig:m_current} shows the result of the
measurements of the magnitude $\sqrt{k_{\mu}k_{\mu}}$ of the induced magnetic
current $k_{\mu}$ obtained according to (\ref{eq:m_current}) for various
temperatures ($\beta$). The current decreases as the temperature becomes
higher and eventually vanishes above the critical temperature for both
options. We observe respectively the appearance and disappearance of the
magnetic monopole current in the low temperature phase and high temperature phase.


\section{Conclusion}

Using a new formulation of the Yang-Mills theory on a lattice, we have
investigated the confinement/deconfinement phase transition at finite
temperature in the $SU(3)$ Yang-Mills theory from the viewpoint of the dual
superconductivity. We have compared the two possible realizations of the dual
superconductivity at finite temperature by adopting the two possible options
in the $SU(3)$ Yang-Mills theory, i.e., the non-Abelian dual superconductivity
in the minimal option and the conventional Abelian dual superconductivity in
the maximal one. In both options, we have succeeded even at finite temperature
to extract the restricted field variable ($V$-field) from the original
Yang-Mills field variable as the dominant mode for confining quarks in the
fundamental representation, which could be called the restricted field
dominance at finite temperature. In both options, moreover, we have
investigated the dual Meissner effect to show that the chromoelectric flux
tube appears in the confining phase, but it disappears in the deconfining
phase. Thus, both options can be adopted to give the low-energy effective
description for quark confinement of the original Yang-Mills theory.

In fact, we have given the numerical evidences for the restricted field
dominance. First, we have investigated the Polyakov loop averages at finite
temperature and have shown that that the Polyakov loop average $\langle
P_{V}\rangle$ of the restricted field $V$ gives the same critical temperature
$T_{c}$ as that detected by the Polyakov loop average $\langle P_{U}\rangle$
of the original gauge field $U$: $\langle P\rangle=0$ for $T<T_{c}$ and
$\langle P\rangle\neq0$ for $T>T_{c}$.

Next, we have investigated the static quark potential at finite temperature by
using the maximally extended Wilson loop and the correlation function for a
pair of the Polyakov loop and the anti-Polyakov loop. We have found the
restricted field ($V$-field) dominance in the string tension at finite
temperature for both options: the string tension calculated from the
restricted field reproduces that from the original Yang-Mills theory. Both
measurements by using the Wilson loop and the Polyakov loop are in good
agreements. When the temperature exceeds the critical temperature, the string
tension rapidly vanishes as the temperature increases. The restricted fields
in both options reproduce the center symmetry restoration/breaking of the
original Yang-Mills theory, and give the same critical temperature of the
confining/deconfining phase transition.

However, the Polyakov loop average cannot be the direct signal of the dual
Meissner effect or magnetic monopole condensation. Therefore, it is important
to find an order parameter which enables one to detect the dual Meissner
effect directly.
In view of these, we have measured the chromoelectric and chromomagnetic flux
for both the original field and the restricted fields in the two options. In
the low--temperature confined phase $T<T_{c}$, we have obtained the numerical
evidences of the dual Meissner effect in the $SU(3)$ Yang-Mills theory,
namely, the squeezing of the chromoelectric flux tube created by a
quark-antiquark pair and the associated magnetic--monopole current induced
around the flux tube. In the high--temperature deconfined phase $T>T_{c}$, on
the other hand, we have observed the disappearance of the dual Meissner
effect, namely, no more squeezing of the chromoelectric flux tube detected by
non-vanishing component in the chromoelectric flux and the vanishing of the
magnetic-monopole current associated with the chromo-flux tube. We have
confirmed that the dual Meissner effect can be described by the restricted
field alone for both options. Therefore, we have confirmed the restricted
field dominance in the dual Meissner effect even at finite temperature. Thus,
we have given the evidences that the confinement/deconfinement phase
transition at finite temperature is caused by appearance/disappearance of the
dual superconductivity for the $SU(3)$ Yang-Mills theory.

\subsection*{Acknowledgement}

This work was supported by Grant-in-Aid for Scientific Research, JSPS KAKENHI
Grant Number (C) No. 24540252, No.15K05042, and No.19K03840. A.S. was
supported also in part by JSPS Grant-in-Aid for Scientific Research (S)
22224003. The numerical calculations are supported by the Large Scale
Simulation Program of High Energy Accelerator Research Organization (KEK):
No.12/13-20(FY2012-13), No.13/14-23(FY2013-14), No.14/15-24(FY2014-15), and No.16/17-20(FY2016-17)


\begin{thebibliography}{99}                                                                                               %


\bibitem {dualsuper}
Y. Nambu,
Phys. Rev. D\textbf{10}, 4262--4268 (1974). \newline G. 't Hooft, in: High
Energy Physics, edited by A. Zichichi (Editorice Compositori, Bologna, 1975).
\newline S. Mandelstam,
Phys. Report \textbf{23}, 245--249 (1976).



\bibitem {KMS05}




K.-I. Kondo, T. Murakami and T. Shinohara,
Prog. Theor. Phys. \textbf{115}, 201--216 (2006). [hep-th/0504107]

K.-I. Kondo, T. Murakami and T. Shinohara,
Eur. Phys. J. C\textbf{42}, 475--481 (2005). [hep-th/0504198]

K.-I. Kondo, T. Shinohara and T. Murakami,
Prog. Theor. Phys. \textbf{120}, 1--50 (2008). arXiv:0803.0176 [hep-th]

\bibitem {KKMSSI06}S. Kato, K.-I. Kondo, T. Murakami, A. Shibata, T. Shinohara
and S. Ito,
Phys. Lett. B\textbf{632}, 326--332 (2006). [hep-lat/0509069]

\bibitem {KSSMKI08}K.-I. Kondo, A. Shibata, T. Shinohara, T. Murakami, S.
Kato, S. Ito,
Phys. Lett. B\textbf{669}, 107--118 (2008). arXiv:0803.2451 [hep-lat]



\bibitem {SKS10}A. Shibata, K.-I. Kondo and T. Shinohara,
Phys. Lett. B\textbf{691}, 91--98 (2010). arXiv:0911.5294 [hep-lat]



\bibitem {KKSS15}K.-I. Kondo, S. Kato, A. Shibata and T. Shinohara,
Phys. Rep. \textbf{579}, 1--226 (2015). arXiv:1409.1599 [hep-th].





\bibitem {tHooft81}G. 't Hooft,
Nucl.Phys. B\textbf{190} [FS3], 455--478 (1981).

\bibitem {KLSW87}A. Kronfeld, M. Laursen, G. Schierholz and U.-J. Wiese,
Phys. Lett. B\textbf{198}, 516--520 (1987).

\bibitem {Kondo08}K.-I. Kondo,
Phys. Rev. D \textbf{77}, 085029 (2008). arXiv:0801.1274 [hep-th]



\bibitem {KS08}K.-I. Kondo and A. Shibata,
arXiv:0801.4203 [hep-th]


\bibitem {KSSK11}K.-I. Kondo, A. Shibata, T. Shinohara, and S. Kato,
Phys. Rev. D\textbf{83}, 114016 (2011). arXiv:1007.2696 [hep-th]



\bibitem {lattice2008}A. Shibata, S. Kato, K.-I. Kondo, T. Shinohara \ and S.
Ito, PoS(LATTICE 2008) 268, arXiv:0810.0956 [hep-lat]

\bibitem {lattice2009}A. Shibata, K.-I. Kondo, S.Kato, S. Ito, T. Shinohara
and N. Fukui, PoS LAT2009 (2009) 232, arXiv:0911.4533 [hep-lat].

\bibitem {lattice2010}A. Shibata, K.-I. Kondo, S. Kato and T. Shinohara,
PoS(Lattice 2010) 286

\bibitem {SKKS13}A. Shibata, K.-I. Kondo, S. Kato and T. Shinohara,
Phys. Rev. D\textbf{87}, 054011 (2013). arXiv:1212.6512 [hep-lat]





\bibitem {flusx:AP}
Y.~Matsubara, S.~Ejiri and T.~Suzuki,
Nucl. Phys. B. Proc. Suppl. \textbf{34}, 176--178 (1994).

\bibitem {Cardaci2011}M.~S.~Cardaci, P.~Cea, L.~Cosmai, R.~Falcone and
A.~Papa,
Phys.\ Rev.\ D \textbf{83}, 014502 (2011) [arXiv:1011.5803 [hep-lat]]




\bibitem {Cardso}
N.~Cardoso, M.~Cardoso and P.~Bicudo,
Phys.\ Rev.\ D \textbf{84}, 054508 (2011) [arXiv:1107.1355 [hep-lat]].






\bibitem {CCP12}P.~Cea, L.~Cosmai and A.~Papa,
Phys. Rev. D\textbf{86}, 054501 (2012). arXiv:1208.1362 [hep-lat]

\bibitem {CCCP14}P. Cea, L. Cosmai, F. Cuteri, and A. Papa,
Phys. Rev. D\textbf{89}, 094505 (2014). arXiv:1404.1172 [hep-lat]




\bibitem {SUN-decomp1}
Y.M. Cho,
Unpublished preprint, MPI-PAE/PTh 14/80 (1980). \newline Y.M. Cho,
Phys. Rev. Lett. \textbf{44}, 1115--1118 (1980).

\bibitem {SUN-decomp2}L. Faddeev and A.J. Niemi,
Phys. Lett. B\textbf{449}, 214--218 (1999). [hep-th/9812090]

L. Faddeev and A.J. Niemi,
Phys. Lett. B\textbf{464}, 90--93 (1999). [hep-th/9907180]

T.A. Bolokhov and L.D. Faddeev,
Theoretical and Mathematical Physics, \textbf{139}, 679--692 (2004).

\bibitem {CFNS-C}Y.M. Cho, Phys. Rev. D 21, 1080 (1980). Phys. Rev. D 23, 2415 (1981);

Y.S. Duan and M.L. Ge, Sinica Sci., 11, 1072(1979);

L. Faddeev and A.J. Niemi, Phys. Rev. Lett. 82, 1624 (1999);

S.V. Shabanov, Phys. Lett. B 458, 322 (1999). Phys. Lett. B 463, 263 (1999).

\bibitem {lattce2007}A. Shibata, S. Kato, K.-I. Kondo, T. Shinohara \ and S.
Ito, POS(LATTICE2007) 331, arXiv:0710.3221 [hep-lat]

\bibitem {CCLL14}N. Cundy, Y.M. Cho, W. Lee, and J. Leem,
Phys. Lett. B\textbf{729}, 192--198 (2014). arXiv:1307.3085 [hep-lat]
\newline N. Cundy, Y.M. Cho, and W. Lee,
PoS LATTICE2013 (2013) 471, arXiv:1311.3029 [hep-lat]
\newline N. Cundy, W. Lee, J. Leem, Y.M. Cho,
PoS LATTICE2012 (2012) 213, arXiv:1211.0664 [hep-lat]


\bibitem {GIS12}S. Gongyo, T. Iritani, and H. Suganuma,
Phys. Rev. D\textbf{86}, 094018 (2012).
e-Print: arXiv:1207.4377 [hep-lat]





\bibitem {SS14}N. Sakumichi and H. Suganuma,
Phys.Rev. D\textbf{90}, 111501 (2014), e-Print: arXiv:1406.2215 [hep-lat]

\bibitem {IKKMSS06}S. Ito, S. Kato, K.-I. Kondo, T. Murakami, A. Shibata and
T. Shinohara,
Phys. Lett. B\textbf{645}, 67--74 (2007). [hep-lat/0604016]



\bibitem {SKKMSI07}A. Shibata, S. Kato, K.-I. Kondo, T. Murakami, T. Shinohara
and S. Ito,
Phys. Lett. B\textbf{653}, 101--108 (2007). arXiv:0706.2529 [hep-lat]



\bibitem {lattice2009k}S. Kato, K.-I. Kondo, A. Shibata and T. Shinohara,
PoS(LAT2009) 228.

\bibitem {Cea1995}P. Cea and L. Cosmai, Phys. Rev. D52, 5152 (1995)

P. Cea and L. Cosmai, Phys. Lett. B349, 343 (1995).

\bibitem {Cea2012a14}P. Cea, L. Cosmai and A. Papa, Phys. Rev. D86, 054501 (2012).

P. Cea, L. Cosmai, F. Cuteri, and A. Papa, Phys. Rev.D89, 049505 (2014).

\bibitem {KKS14}S. Kato, K.-I. Kondo, and A. Shibata,
Phys. Rev. D\textbf{91}, 034506 (2015).
arXiv:1407.2808 [hep-lat]

\bibitem {NKSSK2018}A. Shibata, K.-I. Kondo, S. Nishino, T. Sasago, and S.
Kato, PoS (Confinement2018) 269, arXiv:1903.10487 [hep-lat] , CHIBA-EP-232,
KEK Preprint 2018-78

S. Nishino, K.-I. Kondo, A. Shibata, T. Sasago, and S. Kato, Eur.Phys.J. C79
(2019) no.9, 774, \ arXiv:1903.10488 [hep-lat], CHIBA-EP-236, KEK Preprint 2018-82

\bibitem {SY90}T. Suzuki and I. Yotsuyanagi,
Phys. Rev. D\textbf{42}, 4257--4260 (1990).



\bibitem {SS94}H. Shiba and T. Suzuki,
Phys. Lett. B\textbf{333}, 461--466 (1994).
[hep-lat/9404015]

\bibitem {SNW94}J.D. Stack, S.D. Neiman and R. Wensley,
Phys. Rev. D\textbf{50}, 3399--3405 (1994).
[hep-lat/9404014]





\bibitem {MSKK2019}R. Matsudo, A. Shibata, S. Kato, K.-I. Kondo, Phys. Rev.
D\textbf{100}, 014505, arXiv:1904.09388 [hep-lat]

A. Shibata , R. Matsudo, S. Kato, K.-I. Kondo, PoS LATTICE2018 (2018) 254,
arXiv:1812.05827 [hep-lat]

\bibitem {lattice2012}A. Shibata, K.-I. Kondo, S. Kato and T. Shinohara, PoS
LATTICE2012 (2012) 215, arXiv:1212.2835 [hep-lat]

\bibitem {lattice2013}A. Shibata, K.-I. Kondo, S. Kato, T.Shinohara, PoS
LATTICE2013 (2014), arXiv:1403.3809 [hep-lat]

\bibitem {lattice2014}A. Shibata, K.-I. Kondo, S. Kato, T.Shinohara, PoS
LATTICE2014 (2015) 340, arXiv:1501.06271 [hep-lat]

\bibitem {lattice2015}A. Shibata, K.-I. Kondo, S. Kato, T.Shinohara, PoS
LATTICE2015 (2016) 320, arXiv:1512.03695 [hep-lat]

\bibitem {lattice2016}A. Shibata, K.-I. Kondo, S. Kato, and T. Shinohara, PoS
LATTICE2016 (2017) 345, :arXiv:1701.02442 [hep-lat]

\bibitem {SCGT15}A. Shibata, K.-I. Kondo, S. Kato, and T. Shinohara,
Proceedings of Origin of Mass and Strong Coupling Gauge Theories (SCGT15),
168-174 (2018) ; \ Int.J.Mod.Phys. A32 (2017) no.36, 1747016, :
arXiv:1511.04155 [hep-lat]

\bibitem {confinementX}A. Shibata, K.-I. Kondo, S. Kato and T. Shinohara, PoS
ConfinementX (2012) 052, arXiv:1302.6865 [hep-lat]

\bibitem {confinement2018}A. Shibata, K.-I. Kondo, S. Kato, PoS
Confinement2018 (2019) 061, arXiv:1812.06797 [hep-lat]

\bibitem {Edward98}R.G. Edwards, Urs. M. Heller and T.R. Klassen, Nucl. Phys.
B517 (1998) 377-392

\bibitem {Albanese87}M.~Albanese et al. (APE Collaboration),
Phys. Lett. B\textbf{192}, 163--169 (1987).



\bibitem {YHM05}K. Yagi, T. Hatsuda and Y. Miake, \emph{Quark-Gluon Plasma}
(Cambridge Univ. press, Cambridge, 2005).

\bibitem {BW79}L.S. Brown and W.I. Weisberger,
Phys.Rev. D20 (1979) 3239




\bibitem {GMO90}A.~Di Giacomo, M.~Maggiore and S.~Olejnik,
Nucl.\ Phys.\ B\textbf{347}, 441--460 (1990).
\newline A.~Di Giacomo, M.~Maggiore and S.~Olejnik,
Phys.\ Lett.\ B\textbf{236}, 199--202 (1990).


\bibitem {Cea2016}P.~Cea, L.~Cosmai, F.Cuteri,~A. Papa,\ JHEP 1606 (2016) 033

\bibitem {Cea2018}P.Cea, L.Cosmai, F.Cuteri, A.Papa, EPJ Web Conf. 175 (2018)
12006

\end{thebibliography}
\end{document}